%% file: pi0_review.tex
\documentclass[rmp,aps,preprint]{revtex4}
\usepackage{epsfig}
\usepackage{amsmath,amssymb}
\usepackage{graphics}

\begin{document}

\title{Neutral Pion Lifetime Measurements and the
QCD Chiral Anomaly}
\author{A. M. Bernstein\\
Physics Department and Laboratory for Nuclear Science\\
 M.I.T., Cambridge Mass., 02139 USA \\
 Barry R. Holstein\\ Department of Physics LGRT\\
 University of Massachusetts, Amherst. Mass, 01003 USA}

\begin{abstract}

A  fundamental property of QCD is the presence of the chiral anomaly, which is the dominant component of the $\pi^0\rightarrow\gamma\gamma$ decay rate.  Based on this anomaly and its small ($\simeq$ 4.5\%) chiral correction, a  prediction of the $\pi^0$ lifetime can be used as a test of QCD at confinement scale energies.  The interesting experimental and theoretical histories of the $\pi^0$ meson are reviewed, from discovery to the present era. Experimental results are in agreement with the theoretical prediction, within the current ($\simeq$ 3\%) experimental error; however, they are not yet sufficiently precise to test the chiral corrected result, which is a firm QCD prediction and is known to $\simeq$ 1\% uncertainty. At this level there exist experimental inconsistencies, which require attention.  Possible future work to improve the present precision is suggested.
\end {abstract}

\maketitle
\tableofcontents

\input{section_Introduction}
\input{section_Early_Experimental_History}
\input{section_Barrymod12}

\input{section_PDB}
\input{section_PrimEx}

\input{section_End}
\bibliographystyle{apsrmp}
\bibliography{pi0_review.bib}

 \end{document}

%% file: section_Introduction.tex
\section{Introduction}

An important feature of Quantum Chromodynamics (QCD)\cite{Gross:2005,Wilczek:2005}, the accepted theory of the strong interactions, is the existence of the chiral anomaly \cite{bej,sdl}.  An anomaly is said to occur when a symmetry of the classical Lagrangian is {\it not} a symmetry of the full quantum mechanical theory. In the case which is the focus of this article the conservation of the third isospin component of the axial current, which is present in the classical chiral (massless) version of the QCD Lagrangian, is lost upon quantization due to the fluctuations of the gauge fields. The $\pi^{0} \rightarrow \gamma \gamma$ decay is perhaps the best example of a process that proceeds primarily via the chiral anomaly. The lifetime predicted by the anomaly is exact in the chiral limit (when the light quarks are massless) and has no free parameters. As is discussed in  section \ref{sec:Theory} this is the leading order (LO) term of the chiral series which formally starts at order $q^{4}$. At higher order (HO)  a small ($4.5\pm$ 1\%) increase has been calculated in chiral perturbation theory\cite{bgh,kam,anm}(see section \ref{sec:HO}). Considering the fundamental nature of the subject and the 1\% accuracy which has been reached in the theoretical lifetime prediction, it is important for future experiments to aim for a comparable level of precision.

Precision measurements of this quantity thus serve as a stringent probe of the validity of QCD itself.  The $\pi^0$ lifetime represents a particularly interesting test, since at low energies QCD is very difficult to solve, because the quarks and gluons interact very strongly and predictions must in general  be made by the use of effective field theories such as Chiral Perturbation Theory (ChPT)\cite{Weinberg,chl,chl1} or lattice calculations.  The history of the $\pi^{0}$ lifetime is not only of fundamental importance, but also represents an interesting story of the ingenuity of experimental and theoretical physicists.

The existence of the $\pi$ meson was first postulated by Yukawa in 1935 in order to explain the short range and large magnitude of the nucleon-nucleon interaction\cite{Yukawa:1935}.  Initially, the newly discovered $\mu$ meson was thought to be Yukawa's particle, but the muon turned out to participate in only weak and electromagnetic interactions.  The charged $\pi$ meson was finally discovered in a 1947 cosmic ray experiment\cite{Lattes:1947}.  This was followed in 1950 by a series of experiments that observed the $\pi^{0}$ meson\cite{Bjorklund:1950,Panofsky:1950,Panofsky:1950a,Steinberger:1950,Carlson:1950}, and its primary decay mode into two gamma rays.  This last feature is closely connected with the chiral symmetry of QCD\cite{Nambu:2009,dgh}, which makes  $\pi$ mesons the lightest hadrons\cite{PDB}.

During the 1950's it was discovered that the pion family is an isotriplet with spin = 0 and negative parity $J^\pi =0^-.$\footnote{For a brief history of the experiments leading to the measurement of $J^{\pi} = 0^{-}$ for the pion family, and the connection with the two photon decay of positronium see Perkins\cite{Perkins:1982}} The pseudoscalar nature of the pions\cite{PDB} was interpreted by Nambu\cite{Nambu:2009} as being due to the breaking of the  underlying chiral symmetry of nature. In modern terms, the QCD Lagrangian is chiral symmetric in the limit where the light quark masses vanish\cite{dgh}. If this symmetry were to be manifested in the conventional Wigner-Weyl fashion, each quantum state, such as the proton, would have a nearly degenerate opposite-parity partner particle.  Since this is not the case experimentally, Nambu realized that the axial symmetry is instead realized via the appearance of massless pseudoscalar mesons (now called Nambu-Goldstone Bosons) so that, {\it e.g.}, the opposite parity partner of the proton is a  state containing the proton and a massless "pion".  This conjecture was put on a stronger theoretical basis by Goldstone\cite{Goldstone:1961}.  Of course, in the real world pions  have small but nonvanishing mass due to the explicit breaking of chiral symmetry, since the masses of the up and down quarks are small, but non-zero\cite{Leutwyler:2009,Leutwyler:1996,PDB}.  The modern picture of pions  is that they are Nambu-Goldstone Bosons in addition to being Yukawa's mesons and are the source of the longest-range component of the nucleon-nucleon interaction.  They play this role by having relatively weak interactions with nucleons in the s-wave (vanishing in the chiral limit when the masses of the light quarks vanish) but strong interactions in the p-wave channel.

Electromagnetic effects make the charged pions 4.6 MeV heavier than the neutral pion.  This means that the  $\pi^{0}$ primarily decays in  the two gamma mode or the relatively weak($\simeq$ 1.2\%) $\gamma e^{+}e^{-}$ Dalitz decay mode\cite{Dalitz:1951}.  This decay, like the two-photon decay of positronium, requires that the two photons are E1 and M1, in order to carry away the negative parity of the $J^{\pi} = 0^{-}$ state\cite{Perkins:1982}.  This means that the electric field vectors of the two photons are orthogonal, as has been experimentally demonstrated in the double Dalitz $\pi^{0} \rightarrow e^{+}e^{-}e^{+}e^{-}$ decay\cite{Plano:1959,Abouzaid:2008}.

 Since the $\pi^{0}$ lifetime $\tau(\pi^{0})$ is $\simeq 10^{-16}$ s it is far too short to measure by electronic means. The conceptually simplest technique is to measure the mean distance that the $\pi^{0}$ meson travels before it decays. By measuring the upper limit to the decay distance $d(\pi^{0})$ in low energy reactions it was realized that  $\tau(\pi^{0}) < 5 \times 10^{-14}$ s within the first year of its discovery. The difficulty with  this technique is the small magnitude of  $d(\pi^{0})=\beta\gamma c \tau(\pi^{0})$ where $\beta$ is the $\pi^{0}$ velocity relative to the velocity of light c, $\gamma=1/\sqrt{1-\beta^2}$ is the relativistic boost factor, and $c \tau(\pi^{0}) \simeq 2.4\times 10^{-6}$ cm  using the currently accepted value of  $\tau(\pi^{0}) \simeq 0.8\times 10^{-16}$ s. This short decay distance $d(\pi^{0})$ is hard to measure unless the pion is accelerated to high energies, where $\beta$ approaches unity and $\gamma$ is large, and is why the early series of low energy direct decay distance measurements obtained only upper limits.  This effort was not concluded until  1963 with the first definitive, high energy, measurement, which utilized an 18 GeV proton beam at CERN with the result that $\tau(\pi^{0}) = (0.95 \pm 0.15)\times 10^{-16} $ s\cite{vonDardel:1963}.\footnote{This is the corrected value presented in \cite{Atherton:1985}}.

\begin{figure}
\begin{center}
\epsfig{figure=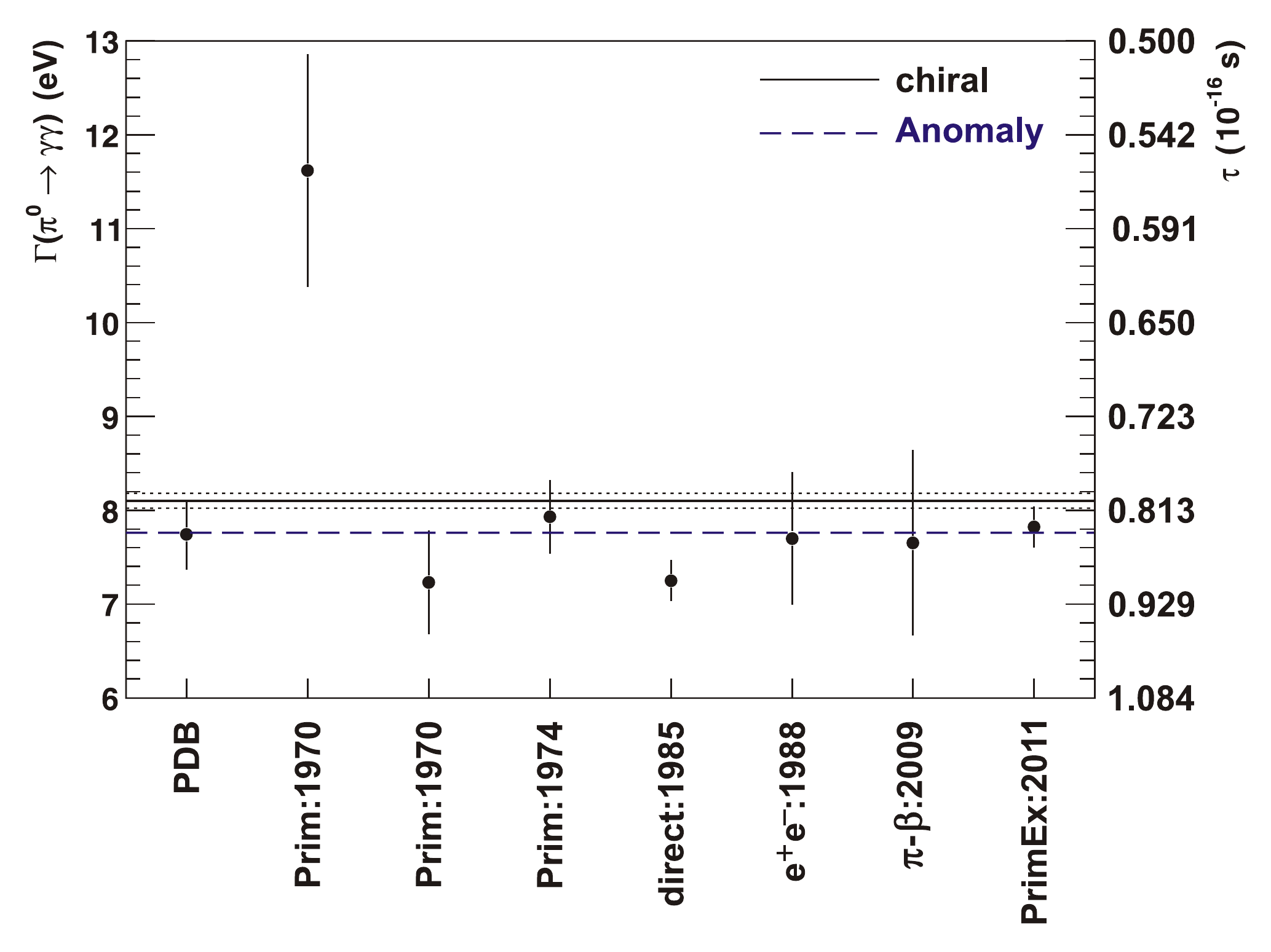,width=14cm,height=12cm}
\end{center}
\vspace{-0.5cm}
\caption{$\pi^{0} \rightarrow \gamma \gamma$ decay width
in eV (left scale) and  $\tau(\pi^{0})$, the mean $\pi^{0}$ lifetime in units of $10^{-16}$ s(right scale). The experimental results
with errors and publication dates are from left to right: 1) 2011 particle data book average\cite{PDB-online};
2,3,4) Primakoff experiments(1970-1974)\cite{Bellettini:1970,Kryshkin:1970,Browman:1974}; 5) direct method(1985)\cite{Atherton:1985}; 6)
$e^{+}e^{-}$(1988)\cite{Williams:1988}; 7) $\pi \beta$ experiment(2009)\cite{Bychkov:2009} ; 8) new Primakoff measurement(2011)\cite{Larin:2010}. All of these experiments with the exception of the last one are the basis of the particle data book average.
The lower dashed line is  the LO prediction of the chiral anomaly\cite{bej,sdl} ($\Gamma(\pi^{0} \rightarrow \gamma \gamma) = 7.760$ eV, $\tau(\pi^{0}) = 0.838\times 10^{-16}$ s). The upper solid line is  the HO chiral prediction and the dotted lines show the estimated 1\% error\cite{bgh,kam,anm}($\Gamma(\pi^{0} \rightarrow \gamma \gamma)= 8.10$ eV, $\tau(\pi^{0})= 0.80\times 10^{-16}$ s).
 ~For the relationship between $\Gamma(\pi^{0} \rightarrow \gamma \gamma)$ and $\tau(\pi^{0})$ see Eq. \ref{eq:Gamma-tau}. \label{fig:width}}
\end{figure}

The results for the $\pi^{0}$ lifetime (and decay width) are shown in Fig. \ref{fig:width}. There have been four different experimental methods  which have been utilized to measure the $\pi^{0}$ lifetime.  The first is the direct technique, discussed above. Fig. \ref{fig:width} shows the result obtained by the latest and most accurate direct measurement performed at  CERN  with much higher energy protons (450 GeV)\cite{Atherton:1985}. The second experimental procedure utilizes the Primakoff\cite{Primakoff:1951} effect in which an incident photon interacts with the Coulomb field of a nucleus to produce the $\pi^{0}$ meson.  A measurement of the cross section combined with detailed balance yields the value of $\tau(\pi^{0})$.  Measurements using this technique were carried out from 1965 through 1974 (see section \ref{sec:PDB}).
The third method, published in 1988, involves measurement of the cross section for the purely electromagnetic two photon $e^{+}e^{-} \rightarrow \gamma \gamma \rightarrow \pi^{0}$ process\cite{Williams:1988}.  In these last two methods either one or two of the photons are not real, but are off shell. However as will be shown in section \ref{sec:PDB} the off shell nature of these processes does not alter the results significantly from those obtained with real photons. The fourth (indirect) technique measured radiative pion decay  $\pi^{+} \rightarrow e^{+}\nu \gamma$ in the PIBETA  experiment\cite{Bychkov:2009}.  Using isospin invariance, the weak polar-vector form factor contributing to this decay channel is related by a simple isospin rotation to the amplitude for $\pi^0\rightarrow\gamma\gamma$, and in this way one additional experimental number for the $\pi^{0}$ lifetime has been obtained (see section \ref{sec:pi-beta} for further discussion including the corrections for isospin breaking).  These measurements complete the information on which the 2011 Particle Data Book (PDB)average is based\cite{PDB,PDB-online}, and the results of these experiments are shown in Fig. \ref{fig:width}, along with the newly performed Primakoff measurement\cite{Larin:2010}.  With the exception of one major outlier\cite{Bellettini:1970}, these results are in reasonable agreement with each other.  At a more precise level, looking towards a test of the theoretical predictions\cite{bgh,kam,anm} at the 1\% level, there exist differences between the two most accurate measurements\cite{Atherton:1985,Larin:2010}. The 2011 PDB average is $\tau(\pi^{0}) = (0.84\pm 0.04)\times 10^{-16} $ s\cite{PDB,PDB-online}. As will be discussed in section \ref{Subsection:Summary}, however, we believe that this error may be understated by a significant factor.

Before the axial anomaly was understood in 1969, a standard way to calculate the $\pi^0\rightarrow\gamma\gamma$ amplitude was to utilize the partially conserved axial-vector, isovector, current (PCAC) condition, which relates the pion field to the divergence of the axial current via
\begin{equation}
\partial^\mu J_{5\mu}^a=F_\pi m_\pi^2\phi^a_\pi ,
\end{equation}
where $J_{5\mu}^{a}$ is the axial-vector current,
$\phi_{\pi}$ represents the pion field, and  $F_{\pi}= 92.21 \pm 0.02 \pm 0.14$ MeV\cite{PDB} is the pion decay constant measured via the $\pi^{+} \rightarrow \mu^{+} \nu_{\mu}$ decay rate.  However, the use of PCAC yields $\tau(\pi^{0}) \approx  10^{-13}$ s, a lifetime approximately three orders of magnitude too long (for details see section \ref{sec:Theory}).  Note that in this procedure the $\pi^0$ amplitude vanishes in the chiral limit, when the masses of the up and down quarks and the pion are set to zero.

It was the discovery of the chiral anomaly that resolved this theoretical conundrum\cite{bej,sdl}.  The existence of the anomaly requires an additional term in the divergence of the third component of the axial current
\begin{equation}
\partial^{\mu}J_{5\mu}^{3} = F_\pi m_\pi^2\phi_{\pi^0}+(\alpha/\pi) \vec{E} \cdot \vec{B} ,
 \end{equation}
 where $\alpha$ is the fine structure constant. From this additional term it can be seen that the $\pi^{0} \rightarrow \gamma \gamma$ decay is via E1 and M1 photons, as indicated by experiment\cite{Plano:1959,Abouzaid:2008}.  Note also that this additional term survives in the chiral limit, and is exact therein.  In fact the anomaly term is the {\it dominant} contribution to the  $\pi^{0}\rightarrow \gamma \gamma$ decay rate, which (in the chiral limit) has no adjustable parameters\cite{bej,sdl}---
\begin{eqnarray}
\Gamma(\pi^{0} \rightarrow \gamma \gamma)&=&(m_{\pi^{0}}/4 \pi)^{3} (\alpha/F_{\pi})^{2}
= 7.760\,\,eV \nonumber \\
\tau &=& \hbar/\Gamma_{tot}(\pi^{0}) = 0.838 \times 10^{-16}\,\,{\rm s} \nonumber \\
  \Gamma_{tot}(\pi^{0}) &=& \Gamma(\pi^{0} \rightarrow \gamma \gamma)+ \Gamma(\pi^{0} \rightarrow \gamma e^{+}e^{-}) \nonumber \\
  \Gamma(\pi^{0} \rightarrow \gamma \gamma) \tau(\pi^{0}) &=& BR(\pi^{0} \rightarrow \gamma \gamma)\hbar
\label{eq:Gamma-tau}
\end{eqnarray}
where $BR(\pi^{0} \rightarrow \gamma \gamma)$ = 0.9882\cite{PDB} is the $\pi^{0} \rightarrow \gamma \gamma$ branching ratio.
However, Eq. \ref{eq:Gamma-tau} is exact only in the chiral limit, {\sl i.e.}, when the $u$ and $d$~quark masses vanish.  In the real world, there exist modifications and the dominant chiral corrections are due to the small masses of the up and down quarks and their difference\cite{Leutwyler:2009,Leutwyler:1996,PDB}, which mixes the I (isotopic spin) = 0  $\eta$ and $\eta'$ mesons into the  I = 1 $\pi^{0}$ wave function.  As discussed below, this chiral symmetry breaking produces a $\simeq$ 4.5\% increase in $\Gamma(\pi^{0} \rightarrow \gamma \gamma)$ to 8.10\,\,{\rm eV} ($\tau(\pi^{0}) = 0.80\times 10^{-16}$ s)\footnote{For the remainder of this review both the value of $\Gamma(\pi^{0} \rightarrow \gamma \gamma)$ and $\tau(\pi^{0})$, which are related by Eq.\ref{eq:Gamma-tau}. will usually be quoted.} with an estimated  uncertainty of less than 1\%\cite{bgh,kam,anm} and it is an important goal of modern experiments to test this firm QCD prediction.

The 2011 average experimental value for
$\Gamma(\pi^0\to\gamma\gamma) =7.74 \pm 0.37\,{\rm eV}  (\tau(\pi^{0}) = (0.84 \pm  0.04 )\times 10^{-16}\,{\rm s.}$) given by the Particle Data Group\cite{PDB,PDB-online} is in reasonable agreement with this predicted value\cite{bgh,kam,anm}. This number primarily represents an average of several experiments, all of which were performed before 1988(see section \ref{sec:PDB} for a complete discussion).  The quoted error of 5\%  is most likely too low, since many of the  experiments appears to have understated their errors (see section \ref{Subsection:Summary} for a discussion.) Even at the 5\% level, however, the precision is not sufficient for a test of such a fundamental quantity, particularly for the new higher order (HO) calculations which take the finite quark masses into account and are accurate at the 1\% level.  All of the previous experiments were performed with experimental equipment which by now has greatly improved.

In order to begin to improve this situation a modern experiment (PrimEx) was performed at Jefferson Lab using the Primakoff effect technique\cite{Larin:2010, Bernstein:2009,PrimEx}.  This experiment utilized tagged photons for the first time and incorporated many accelerator and detector improvements developed over the years. The improvements  included a cw (continuous wave)  accelerator which provides high duty cycles, and greatly improved beam focusing and  angular and energy resolution for the outgoing pion.  Such improvements enabled a significantly better measurement, with a 2.8\% overall error as shown in Fig. \ref{fig:width}, and yielded a result consistent with the chiral prediction.  Still, even with this improved Primakoff measurement there is still considerable room for experimental improvements.  A second experiment using the Primakoff effect has also been performed by the PrimEx group and the data analysis is in progress\cite{PrimEx}.\footnote{Note added in proof. The 2012 particle data book average, which includes the PrimEx experiment and has followed our suggestions about which of the older Primakoff experiments not to use (Sec. IV), gives an average value of $\Gamma(\pi^0\to\gamma\gamma) =7.63 \pm 0.16\,{\rm eV}  (\tau(\pi^{0}) = (0.852 \pm  0.018 )\times 10^{-16}\,{\rm s.}$) \cite{Beringer:2012}. For comments on the averaging procedure see Sec. VII.A.}

An interesting aspect of the history of the $\pi^{0}$ lifetime is the degree of independence of experiment and theory. In most of the experimental papers on which the Particle Data Book average is based there is no comparison of the experimental results with theory. This is even more remarkable since one of the early pioneers in the discovery, properties, and early theory of the $\pi^{0}$ was Jack Steinberger, who performed the first accurate lifetime calculation and then went on to become one of the early experimental leaders. It is only in the past decade that the PrimEx experiment\cite{Larin:2010} was designed to test QCD via the LO predictions of the anomaly plus the HO chiral corrections\cite{bgh,kam,anm}. The chiral predictions in turn were stimulated by the prospect of the PrimEx experiment.

The chiral anomaly represents quantum mechanical symmetry breaking by the electromagnetic field of the chiral symmetry associated with the third isospin component of the axial current\cite{bej,sdl}. The $\pi^{0}$ decay provides the most sensitive test of this phenomenon of symmetry breaking due to the quantum fluctuations of the quark fields in the presence of a gauge field. Considering the fundamental nature of the subject, and the 1\% accuracy which has been reached in the theoretical lifetime prediction, it is important for future experiments to aim for the same level of precision.

With this interplay of theory and experiment in mind we below review both the theoretical and experimental approaches to $\pi^0\rightarrow\gamma\gamma$
decay.  We begin in section II by examining the 1950 discovery of the neutral pion and its decay into two photons together with the early lifetime measurements which gradually converged towards $10^{-16}$ s by 1963.  In section III we review the theoretical evolution which led to our current understanding of this process.  In section IV we examine the experiments  that are used by the PDG in computing their average, and in section V, we look at the new PrimEx experiment performed during the last few years at JLab.
In section VI we briefly examine some related experiments. Finally, in section VII we summarize our findings and speculate on future improvements. 

%% file: section_Early_Experimental_History.tex
\section{Early Experimental History}\label{early-exp}

In 1947 the charged pion was discovered with photographic emulsions
exposed to cosmic rays at mountain altitudes\cite{Lattes:1947}, and its
dominant, weak, muon neutrino decay mode $\pi^{+} \rightarrow
\mu^{+} + \nu_{\mu}$ was observed.  In 1950 the neutral pion was
observed at the 184 inch Berkeley synchrocyclotron via proton bombardment of nuclei\cite{Bjorklund:1950}, as well as in the $\pi^{-} p \rightarrow \pi^{0}n$ reaction with stopped pions, and the dominant electromagnetic $\pi^{0} \rightarrow \gamma \gamma$ decay mode was detected\cite{Panofsky:1950,Panofsky:1950a} by observation of the approximately equal energy sharing of the two gamma rays.  The neutral pion was also detected in cosmic rays at 70,000 feet\cite{Carlson:1950}.
In the same year the $\pi^0$ was photoproduced at Berkeley and the coincidences between the two decay photons were observed for the first time\cite{Steinberger:1950}.  By the end of 1950  the following facts were established about the $\pi^{0}$ meson:
\begin{itemize}
\item{The value $m(\pi^{+}) -m(\pi^{0}) = 5.42 \pm  1.02\,\,{\rm MeV}
$\cite{Panofsky:1950,Panofsky:1950a}, consistent with the presently accepted number 4.59 MeV\cite{PDB}.
The dominant $\pi^{0} \rightarrow \gamma \gamma$ decay mode was observed\cite{Panofsky:1950,Panofsky:1950a,Steinberger:1950}}
\item{The cross sections for the $\gamma p \rightarrow \pi^{0} p$
\cite{Steinberger:1950,Panofsky:1952} and $\gamma p \rightarrow \pi^{+} n$\cite{Mozley:1950} reactions are roughly equal, indicating that the $\pi^{0}$ and $\pi^{+}$ mesons are "of the same type", indicating that the $\pi^{0}$ meson is a pseudoscalar.}
\item{The soft component of cosmic rays is due to the production and decay of $\pi^{0}$ mesons.}
\item{ An upper limit for the lifetime $\tau(\pi^{0}) < 5 \times 10^{-14}$ s was established by a measurement of the geometric size of the decay region\cite{Carlson:1950}.
}
\end{itemize}

It is impressive that, within a year of its discovery, so much was understood about the $\pi^0$, including an upper limit of $< 5 \times 10^{-14}$ s for the lifetime.  This value is far shorter than electronic detection resolution time and was obtained by setting an upper limit on the distance between the $\pi^0$ production and decay.  This upper limit\cite{Carlson:1950} utilized the best experimental method that was available for such short lifetimes---studying  $\pi^{0}$ production and decay in emulsions---since the resolutions are somewhat better than the grain size $ \approx 0.5\,\,\mu m$. As the mean decay distance is d($\pi^{0}$) = $\gamma \beta c \tau$, we find, using the  predicted lifetime $\tau = 0.80 \times 10^{-16}$ s\cite{bgh,kam,anm}, that $c \tau = 0.024\,\, \mu m\approx $ 5\% of a grain size!
With the benefit of hindsight then, it is not surprising that actual measurements (as opposed to upper limits) of the $\pi^0$ lifetime were much slower in coming.

In 1951 Dalitz proposed the existence of the $\pi^{0} \rightarrow \gamma e^{+} e^{-}$ decay mode and calculated a branching ratio of $\simeq$1.2\%\cite{Dalitz:1951}, in excellent agreement with the current experimental value of 1.174(0.035)\%\cite{PDB}.  Dalitz's primary point was that the observation of this decay mode would possibly enable a measurement of $\tau(\pi^{0})$ since the detection efficiency in emulsions for this decay mode would be much higher than for the two photon mode. In addition the opening angles of the electron-positron pair are on average larger then those due to pair production of one of the decay photons. In a cosmic ray interaction where a high energy particle creates many particles including a $\pi^{0}$ in an interaction ("star"), this leads to a radial distribution N(r) of $e^{+}e^{-}$ pairs, $ N(r) \simeq  const + (\delta/d)e^{-r/d}$ where $\delta$ is the relative probability to produce a $e^{+}e^{-}$ pair and d($\pi^{0}$)  is the mean $\pi^{0}$ decay distance.

In 1953 the first measurement of $N(r)$ in cosmic rays, as suggested by Dalitz, was carried out\cite{Anand:1953}.  The conclusion of this work was that "the most probable value of $\tau(\pi^{0})$ is $\approx 5 \times 10^{-15}$ s"\cite{Anand:1953}.  Perkins has pointed out that this value should be corrected downward due to the reduction in ionizing power when the $e^{+}e^{-}$ pair are very close together\cite{Perkins:1955} and this is the direction needed to bring this result into better agreement with the upper limits of  $2 \times 10^{-15}$ s and $ 1 \times 10^{-15}$ s found in cosmic ray emulsion experiments\cite{Lord:1950} as well as in low energy experiments on pion charge exchange at the Chicago cyclotron\cite{Lord:1952}.  It should also be noted that the long lifetime claimed by Anand\cite{Anand:1953} depends strongly on the $\pi^{0}$ momentum distribution in the cosmic rays, a quantity that is not well determined in the emulsion experiments.  In view of these issues the determination of Anand\cite{Anand:1953} must be considered as only a first tentative step in the road that lay ahead.

One of the main advances in this regard came through the development of higher energy particle accelerators, so that the intensity and control of the primary beam was greatly improved over the use of cosmic rays.  In 1957 an ingenious method was proposed to measure the $\pi^{0}$ lifetime from stopped $K^{+}$ mesons using the two-body decay mode $K^{+} \rightarrow \pi^{0} \pi^{+}$\cite{Harris:1957}.  The kaons were produced in the Berkeley Bevatron and were stopped in an emulsion.  The decay location was determined from the appearance of the $\pi^{+}$.  However,  the emulsion is insensitive to the gamma rays from the dominant $\pi^{0} \rightarrow \gamma \gamma$ decay.  Therefore the pair (Dalitz)  $\pi^{0} \rightarrow \gamma e^ {+} e^{-}$ decay mode, which occurs with a 1.2\% probability\cite{PDB}, was utilized.  For a
stopping kaon the pion momentum is 205 MeV/c and for an assumed $\pi^{0}$ lifetime of $10^{-15}$ s the mean decay distance is 0.3 $\mu m$.  The experiment indicated that the $\pi^{0}$ meson decayed in a significantly shorter distance so that $\tau(\pi^{0}) < 1 \times 10^{-15}$ s\cite{Harris:1957}.

\begin{figure}
\begin{center}
\epsfig{figure=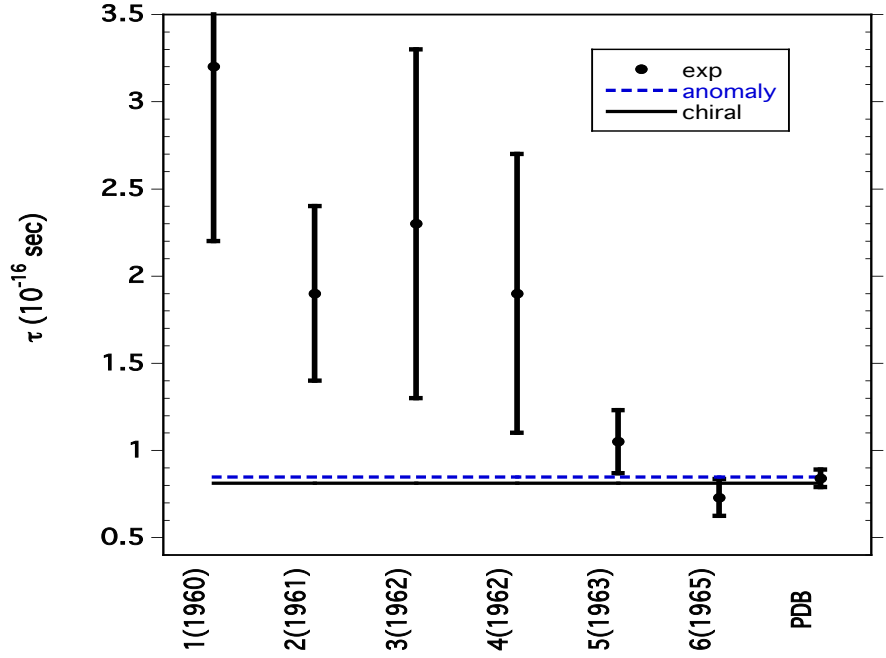,width=16cm,height=16cm}
\end{center}
\caption{The early measurements of the $\pi^{0}$ lifetime. From left to right in the same order as  in Table\ref{table:early-meas}. The first four measurements were performed with emulsions, 5) is the result of a direct measurement of the $\pi^{0}$ decay length induced by 5 GeV/c protons incident on a Pt foil  at CERN\cite{vonDardel:1963}($\tau(\pi^{0})$ is the corrected value presented in \cite{Atherton:1985}), and 6) the earliest Primakoff measurement at Frascati\cite{Bellettini:1965}. The last point is the 2011 particle data book average\cite{PDB-online}.
 \label{fig:tau-early}}
\end{figure}

Several years later, during the period from 1960 to 1963, the first definitive measurements of the $\pi^{0}$ lifetime were reported.  These experiments used the Berkeley Bevatron and the CERN PS (Proton Synchrotron), along with  emulsions with better spatial resolution by a factor of $\simeq$ 2, as well as having better statistics.  The results of these early measurements are summarized in Table \ref{table:early-meas} and Fig. \ref{fig:tau-early}.  The three earliest experiments that obtained results\cite{Blackie:1960,Glasser:1961,Tietge:1962} used the technique previously suggested\cite{Harris:1957} of using stopped kaons and observing the $K_{2\pi}$ decay mode.  To illustrate what a tour de force such experiments were, they observed a mean decay distance of $0.088 \pm 0.024\,\,\mu m$ (compared to a developed grain size of $\simeq 0.35\,\,\mu m$) which leads to a mean life $ \tau = (1.9 \pm 0.5 )\times 10^{-16}$ s\cite{Glasser:1961}.  This number is of the same order of magnitude as the current particle data book average\cite{PDB} and  predicted  value of $\tau(\pi^{0})=0.80 \times 10^{-16}$ s\cite{bgh,kam,anm} for which the mean decay distance is 0.037 $\mu$.  The fourth experiment at Berkeley utilized a 3.5 GeV $\pi^{-}$ beam to produce neutral pions in emulsion nuclei and then observed their Dalitz decay\cite{Shwe:1962}.  This method depended on understanding the $\pi^{0}$ momentum spectrum, and the assumption was made that this was identical to the measured $\pi^{+}$ spectrum.   All of these tour de force emulsion measurements obtained a lifetime $\simeq 2\times 10^{-16}$ s.   Nevertheless, as can be seen from Fig. \ref{fig:tau-early}, the values are higher than the presently accepted number and probably result from a systematic bias in the technique.  One possibility (mentioned above) pointed out by Perkins is that the lifetime value should be corrected downward due to the reduction in ionizing power when the $e^{+}e^{-}$ pair are very close together\cite{Perkins:1955}. Taking into account the experimental equipment of  the early 1960's this is a remarkable tour de force of experimental physics in which the emulsion technique was pushed to its limit.

The next step in performing a  measurement of the decay distance of the $\pi^{0}$ meson was to utilize higher energies so that there exists a large Lorentz boost.  In 1963, using an 18 GeV internal beam at the CERN proton synchrotron, the yield of 5 GeV/c positrons was measured from platinum targets of various thicknesses.  In this case the $\pi^{0}$ mesons, produced in the nuclear interactions of the protons, decay into two photons, some of which are converted in the target to $e^{+} e^{-}$ pairs.  The $\pi^{0}$ decay distance, inferred from the relative positron yield as a function of target thickness, was determined to be $ (1.5\pm 0.25)\,\,\mu m$ \cite{vonDardel:1963}.  To obtain $\tau(\pi^{0})$  the $\pi^{0}$ momentum spectrum must be known and, as in ref. \cite{Shwe:1962} mentioned above, this was taken to be the same as the measured $\pi^{+}$ spectrum.  With this assumption $<p_{\pi^{0}}>$, the average $\pi^{0}$ momentum that produced 5 GeV/c positrons in Pt, was 7.1 GeV and a lifetime $\tau(\pi^{0})= (0.95\pm 0.15)\times 10^{-16} $ s was obtained\cite{vonDardel:1963}\footnote{This is the corrected value presented in \cite{Atherton:1985}},  which is much closer to the  present values, both experimental\cite{PDB} and the theoretical, as can be seen from Fig. \ref{fig:tau-early}.

The first Primakoff measurement  was performed in Frascati in 1965\cite{Bellettini:1965}. This is the measurement of the cross section for the $\gamma + \gamma^{*} \rightarrow \pi^{0}$ reaction where one  photon is incident on the virtual photon $\gamma^{*}$ from the Coulomb field of a nucleus (the Primakoff effect is discussed in detail in section IV.A). In this case the incident photons were produced in an electron synchrotron with an endpoint energy of 1.0 GeV incident on a Pb target. The results of this experiment are in very good agreement with the present accepted and theoretical values as summarized in Table \ref{tab1} and Fig. \ref{fig:tau-early}.

 With these last two measurements  we have arrived at the beginning of the era on which the particle data book is based.  Before examining them, however, it is useful to review the corresponding theoretical studies of $\pi^0\rightarrow \gamma\gamma$ and their connection with QCD.

\begin{table}
\caption{First Measurements of the $\pi^{0}$ Lifetime.}
\label{tab1}
\begin{center}
{\small
\begin{tabular}{||c|c|c|c||}\hline\hline
 & & &  \\
Reference&  Reaction   & $\tau(\pi^{0})/10^{-16}$ s& number of events  \\
 & & &   \\ \hline
 & & &   \\
\cite{Blackie:1960} & a & 3.2$\pm$1.2 & 26 \\
 & & &  \\ \hline
\cite{Glasser:1961} & a  & 1.9$\pm$0.5 & 76\\
 & & &  \\ \hline
\cite{Shwe:1962} & b & $1.9_{-0.8}^{+1.3}$&44 \\
 & & &   \\ \hline
  \cite{Tietge:1962}& a & $2.3_{-1.0}^{+1.1}$& 61\\
  & & & \\ \hline
   \cite{vonDardel:1963}& c & 0.95 $\pm$ 0.15 &\\
    & & &\\ \hline
      \cite{Bellettini:1965}& d &  0.73$\pm$ 0.11 &\\
    & & &\\ \hline\hline
\end{tabular} }
\end{center}
Reaction footnotes: a) $K^{+} \rightarrow \pi^{+} \pi^{0}$ observing the Dalitz decay mode $\pi^{0} \rightarrow \gamma e^{+} e^{-}$ with stopped Kaons; b) $\pi^{-} \rightarrow \pi^{0}$ charge exchange reactions in emulsion nuclei with a 3.5 GeV pion beam; c) direct measurement of the $\pi^{0}$ decay length induced by 5 GeV/c protons incident on a Pt foil  at CERN($\tau(\pi^{0})$ is the corrected value presented in \cite{Atherton:1985}).; d) The earliest Primakoff measurement at Frascati. The first four experiments were performed at the Berkeley Bevatron, using emulsions as detectors.
\label{table:early-meas}
\end{table}

%% file: section_Barrymod12.tex

\section {$\pi^0\rightarrow \gamma\gamma$ Decay:  Theory\label{sec:Theory}}

\subsection{Early Theoretical History}
\begin{figure}
\begin{center}
\includegraphics
*[width=0.50\linewidth,clip=true]{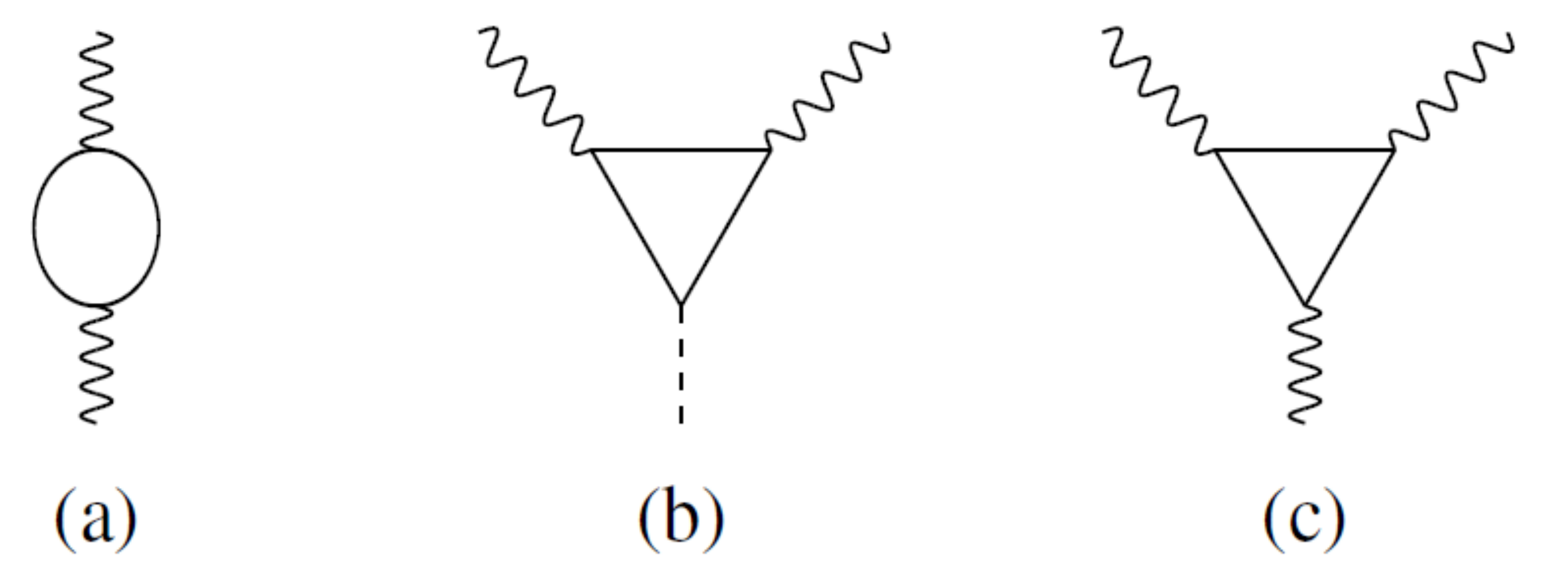}
\end{center}
\caption{Diagrams considered by early workers: a) vacuum polarization;
b) pseudoscalar to $\gamma\gamma$; c) axial current to $\gamma\gamma$.}
\label{fig:theory1}
\end{figure}

Just as the experimental study of $\pi^0\rightarrow\gamma\gamma$ took place over
many years, the corresponding theoretical understanding of neutral pion decay
evolved over several decades.  The theoretical examination of the decay amplitude for the mode
$\pi^0\rightarrow\gamma\gamma$ began contemporaneous with the work
on renormalization of quantum electrodynamics (QED).  In 1948
Sin-Itiro Tomonaga sent a letter to J. Robert Oppenheimer, who was
then director of the Institute for Advanced Study in Princeton, in
which he described some of the work that he and his group had been
doing in the area of QED.  This work was subsequently published
as a letter in Physical Review\cite{sit}, in which Tomonaga
described his successful work in dealing with divergences involving
the electron mass and charge. However, he was still having
difficulty dealing with the renormalization of the photon propagator,
in that the photon self-energy diagram shown in Fig. \ref{fig:theory1}a was not
only divergent but also violated gauge invariance, leading to a
nonzero value for the photon mass.  In order to study such issues
further, two of his associates----Hiroshi Fukuda and Yoneji
Miyamoto---undertook a calculation of the process $\pi^0\rightarrow
\gamma\gamma$, which involves the pseudoscalar meson-vector current-vector
current (PVV) triangle diagram shown in Fig. \ref{fig:theory1}b\cite{fum}.
They also examined the axial current-vector current-vector current (AVV)
triangle diagram in Fig. \ref{fig:theory1}c, which is a three-point function
connecting the axial vector current to two photons.  Such triangle diagrams are linearly
divergent, and this problem was dealt with by the use of a
Pauli-Villars regulator\cite{pvi},\cite{pvii}. The calculation raised
interesting problems in that the PVV amplitude was found to be gauge
invariant, while the AVV amplitude was not.

A parallel calculation was undertaken by Jack Steinberger at
Princeton (then a theorist), who was aware of the Fukuda-Miyamoto work and, also using
the Pauli-Villars method, obtained similar results\cite{jst}.
Defining
\begin{equation}
{\cal
L}_{\pi\gamma\gamma}=A_{\pi\gamma\gamma}\pi^0F^{\mu\nu}\tilde{F}_{\mu\nu}
\end{equation}
where
$$F_{\mu\nu}=\partial_\mu A_\nu-\partial_\nu A_\mu$$ is the
electromagnetic field tensor and
$$\tilde{F}_{\mu\nu}={1\over 2}\epsilon_{\mu\nu\alpha\beta}F^{\alpha\beta}$$ is the
dual tensor, Steinberger determined, using a proton loop, that
\begin{equation}
A_{\pi\gamma\gamma}={e^2g_{\pi NN}\over 16\pi^2m_N}\label{eq:kh}
\end{equation}
where $g_{\pi NN}$ is the strong pseudoscalar $\pi NN$ coupling
constant.  Using the Goldberger-Treiman relation\cite{gtr}
\begin{equation}
g_{\pi NN}={m_Ng_A\over F_\pi}
\end{equation}
where $g_A\simeq 1.27$ is the neutron axial decay amplitude and
$F_\pi\simeq 92.2$ MeV is the pion decay constant, Eq. \ref{eq:kh}
can be rewritten as
\begin{equation}
A_{\pi \gamma\gamma}={e^2g_A\over 16\pi^2 F_\pi}
\end{equation}
which is remarkably similar to the value
\begin{equation}
A_{\pi \gamma\gamma}^{anom}={e^2\over 16\pi^2F_\pi}
\end{equation}
predicted by the chiral anomaly, as shown below.  The corresponding decay
rate predicted by Steinberger\cite{jst}
\begin{equation}
\Gamma_{\pi\gamma\gamma}={|A_{\pi\gamma\gamma}|^2 m_\pi^3\over 4\pi}=g_A^2\Gamma_{\pi\gamma\gamma}^{anom}
=g_A^2\times 7.76\,\,{\rm eV}
\end{equation}
is about 60\% larger than the later prediction of the chiral anomaly (and the PDG experimental value).

Julian Schwinger also visited these problems in 1951\cite{jsc}. He
showed how to handle the issues with the photon self-energy, and
confirmed that there were difficulties with the triangle diagrams, but
he did not succeed in resolving them.

\subsection{The Adler-Bell-Jackiw Anomaly}

The resolution of this problem and its connection with symmetry was
not understood until the late 1960's, when at CERN John Bell and
Roman Jackiw examined the problem of $\pi^0$ decay within the
$\sigma$ model, which is known to obey what we now call chiral
symmetry\cite{bej}. At the time, chiral symmetry was manifested
in the validity of PCAC, which asserts that the divergence of the axial current
\begin{equation}
J^a_{5\mu}(x)=\bar{\psi}(x){1\over 2}\tau_a\gamma_\mu\gamma_5\psi(x)
\end{equation}
can be used as an interpolating field for the pion
\begin{equation}
\partial^\mu J_{5\mu}^a(x)=F_\pi m_\pi^2 \phi_\pi^a(x)\label{eq:kn}
\end{equation}
By this logic the divergence of the axial triangle diagram should be
related to the pion decay triangle diagram, but this was clearly not
the case since the PVV amplitude is gauge invariant while its AVV
analog is not. By a careful analysis within the $\sigma$ model, Bell
and Jackiw were able to show that the origin of the problem is the
breaking of chiral symmetry associated with quantizing the theory.  That is, the
chiral symmetry is valid classically but is destroyed via
quantization---a situation which is called anomalous symmetry
breaking or simply the anomaly.  (At the same time as Bell and
Jackiw were resolving the problem, Steve Adler at the Institute for
Advanced Study came to the same conclusion in his study of spinor
field theory\cite{sdl}, and for this reason the phenomenon is often
called the ABJ anomaly.)

The basic reason underlying this behavior is that, because quantum
field theory involves an infinite number of degrees of freedom, the
short distance properties of the theory do not coincide with what is suggested by
naive manipulations.  That one must be very careful in this region is suggested by
the feature that  a spinor field theory obeys the anticommutation relation
\begin{equation}
\{\psi_a(t,\vec{x}),\psi_b^\dagger
(t,\vec{y})\}=\delta^3(\vec{x}-\vec{y})\delta_{ab}.
\end{equation}
Thus one must deal very carefully with a current such as $J^a_{5\mu}(x)$, which
involves both the field and its conjugate defined at the same point and
can be handled in a variety of ways, but in the end all such methods lead to identical results.

Since the result of all techniques is the same, we shall detail here only the most
intuitive of these procedures---perturbation theory---which was the method employed
by the early investigators. Before examining this calculation, however, we review the
simple symmetry aspects that one might expect.  We consider a simple massless spinor
field carrying charge $e$ coupled to the electromagnetic field, for which the Lagrangian
density is
\begin{equation}
{\cal L}=\bar{\psi}(i\not\!{\partial}-e\not\!\!{A})\psi-{1\over
4}F_{\mu\nu}F^{\mu\nu}
\label{eq:mn}
\end{equation}
We note that this Lagrangian is invariant under a global phase
transformation of the spinor field
\begin{equation}
\psi\rightarrow \exp(i\beta)\psi\equiv {\psi_V}'\label{eq:ku}
\end{equation}
which, by Noether's theorem, leads to a conserved polar-vector
current
\begin{equation}
J_\mu=\bar{\psi}\gamma_\mu\psi\quad{\rm with}\quad \partial^\mu J_\mu=0.
\end{equation}
Alternatively, the Lagrangian of Eq. \ref{eq:mn} is also invariant under a global axial
phase transformation
\begin{equation}
\psi\rightarrow\exp(i\zeta\gamma_5)\psi\equiv {\psi_A}'\label{eq:jn}
\end{equation}
which, by Noether's theorem, leads to a conserved axial-vector
current
\begin{equation}
J_{5\mu}=\bar{\psi}\gamma_\mu\gamma_5\psi\quad{\rm with}\quad \partial^\mu J_{5\mu}=0
\end{equation}

Consider now the three point AVV amplitude, which we designate by
\begin{equation}
T_{\mu\nu\gamma}(q_1,q_2)=-ie^2\int d^4x d^4ye^{-iq_1\cdot
x-iq_2\cdot y}<0|T(J_\mu^{em}(x)J_\nu^{em}(y)J_{5\gamma}(0))|0>
\end{equation}
where it is understood that $q_1^2=q_2^2=0$.  Current conservation
for the vector and axial-vector currents  yields the conditions
\begin{equation}
q_1^\mu T_{\mu\nu\gamma}(q_1,q_2)=q_2^\nu
T_{\mu\nu\gamma}(q_1,q_2)=(q_1+q_2)^\gamma
T_{\mu\nu\gamma}(q_1,q_2)=0\label{eq:hf}
\end{equation}
The requirement that all three conditions in Eq. \ref{eq:hf} be
satisfied then leads to the vanishing of the
$\pi^0\rightarrow\gamma\gamma$ decay amplitude, a result which is
called the Sutherland-Veltman theorem\cite{sut},\cite{vel}. This
result can be demonstrated by writing the most general form for
$T_{\mu\nu\gamma}(q_1,q_2)$ which satisfies the strictures of Bose
symmetry, parity conservation, and gauge invariance
\begin{eqnarray}
T_{\mu\nu\gamma}(q_1,q_2)&=&\epsilon_{\lambda\sigma\alpha\beta}\left[
p_\gamma {g^\lambda}_\mu {g^\sigma}_\nu q_1^\alpha q_2^\beta
G_1(p^2)+\left({g^\sigma}_\mu q_{2\nu}-{g^\sigma}_\nu
q_{1\mu}\right)q_1^\alpha q_2^\beta {g^\lambda}_\gamma
G_2(p^2)\right.\nonumber\\
&+&\left.\left(({g^\sigma}_\mu q_{1\nu}-{g^\sigma}_\nu
q_{2\mu})q_1^\alpha q_2^\beta -{1\over 2}p^2{g^\sigma}_\mu
{g^\alpha}_\nu (q_1-q_2)^\beta\right){g^\lambda}_\gamma
G_3(p^2)\right]\nonumber\\
\quad
\end{eqnarray}
where $p=q_1+q_2$ is the momentum carried by the axial current.
Imposing the condition for axial current conservation yields
the constraint
\begin{equation}
0=p^\gamma
T_{\mu\nu\gamma}(q_1,q_2)=\epsilon_{\mu\nu\alpha\beta}q_1^\alpha
q_2^\beta p^2(G_1(p^2)+G_3(p^2))
\end{equation}
Defining the off-shell $\pi^0\rightarrow\gamma\gamma$ amplitude as
\begin{equation}
<\gamma\gamma|\pi^0>=\epsilon_1^{\mu *} \epsilon_2^{\nu *}
A_{\mu\nu}(q_1,q_2)
\end{equation}
where
\begin{equation}
A_{\mu\nu}(q_1,q_2)= A(p^2)\epsilon_{\mu\nu\alpha\beta}q_1^\alpha
q_2^\beta
\end{equation}
we have, using the Lehmann-Symanzik-Zimmermann reduction\cite{dbj} and Eq. \ref{eq:kn},
\begin{equation}
A(p^2)={(m_\pi^2-p^2)\over F_\pi m_\pi^2}p^2(G_1(p^2)+G_3(p^2)).
\end{equation}
Unless  $G_1(p^2)$ and $G_2(p^2)$ develop poles at $p^2=0$, which is excluded on
physical grounds, we conclude that $A(0)=0$, which is the content of the Sutherland-Veltman
theorem, and asserts that in the chiral symmetric limit, where $m_\pi^2=0$,
the $\pi^0\rightarrow\gamma\gamma$ decay amplitude vanishes.  Of course,
in the real world $m_\pi^2\neq 0$ and we must extrapolate from the
chiral limit. However, this scenario would suggest a decay amplitude
of size
\begin{equation}
A(m_\pi^2)\sim {e^2\over 16\pi^2 F_\pi}\times {m_\pi^2\over
\Lambda_\chi^2}\label{eq:bh}
\end{equation}
where $\Lambda_\chi\sim 4\pi F_\pi\sim 1$ GeV is the chiral
scale\cite{chs},\cite{chsi}.  Here the factor $m_\pi^2/\Lambda_\chi^2$ represents the
feature that this amplitude is {\it two} chiral orders higher than the vanishing lowest
order term, the factor $e^2/4\pi$ is needed because we have a two photon amplitude with a loop
diagram, and the "extra" $4\pi F_\pi$ is required for dimensional purposes.  In any case, Eq.
\ref{eq:bh} would lead to a $\pi^0$ lifetime
\begin{equation}
\tau_{\pi^0\rightarrow\gamma\gamma}=1/\Gamma_{\pi^0\rightarrow\gamma\gamma}\sim
10^{-13}\,{\rm s},
\end{equation}
{\it three} orders of magnitude longer than observed.

\begin{figure}
\begin{center}
\includegraphics*[width=0.50\linewidth,clip=true]{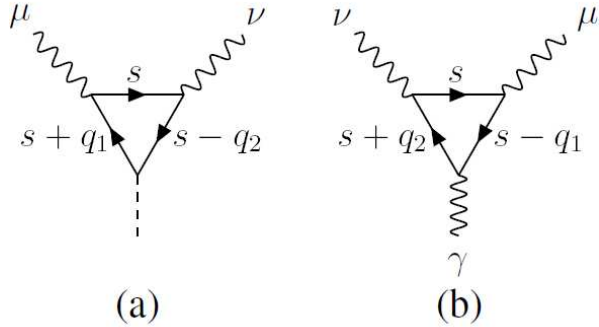}
\end{center}
\caption{Perturbation theory diagrams for PVV and AVV processes.  Here the indices
$\mu,\nu$ are Lorentz indices of vector currents while $\gamma$ is the Lorentz index of
the axial-vector current.
\label{fig:theory2}}
\end{figure}

In order to study this phenomenon further we shall examine the
$\pi^0$ decay process in perturbation theory, wherein the
three point function is described by the Feynman diagram in Fig.
\ref{fig:theory2}b
\begin{equation}
T_{\mu\nu\gamma}(q_1,q_2)=U_{\mu\nu\gamma}(q_1,q_2)+U_{\nu\mu\gamma}(q_2,q_1)
\end{equation}
where
\begin{equation}
U_{\mu\nu\gamma}(q_1,q_2)=-i{e^2K_F\over 2}\int{d^4s\over
(2\pi)^4}{\rm Tr}\left({1\over
\not\!{s}+\not\!{q}_1}\gamma_\mu{1\over \not\!{s}}\gamma_\nu{1\over
\not\!{s}-\not\!{q}_2}\gamma_\gamma\gamma_5\right)
\end{equation}
Note that $U_{\mu\nu\gamma}(q_1,q_2)$ arises from
colored quark loops and includes a factor
\begin{equation}
K_F=N_c\sum_{u,d}Q_q^2\tau_{3q}=3\left[({2\over 3})^2-(-{1\over
3})^2\right]=1
\end{equation}
This proportionality of the decay amplitude to $N_c$ has led many
authors to assert that the agreement between experimental and
theoretical values of the $\pi^0\rightarrow\gamma\gamma$ decay rates
offers "proof" that $N_c=3$.  However, Baer and Wiese\cite{bae} and earlier Gerard and
Lahna\cite{jgm} have noted
that anomaly cancelation {\it in a gauge theory} requires, in a two-flavor picture with
$N_c$ colors, that the $u,d$ quark charges must have the values
\begin{equation}
Q_u={1\over 2}[{1\over N_c}+1]\quad{\rm and}\quad Q_d={1\over
2}[{1\over N_c}-1]
\end{equation}
so that the factor $K_F$ has the value
\begin{equation}
K_F={N_c\over 4}[({1\over N_c}+1)^2-({1\over N_c}-1)^2]=1
\end{equation}
in {\it any} consistent theory.  Note that the Steinberger
calculation\cite{jst} is then a special case wherein $N_c=1$ and this feature
is the reason that his calculation agrees with the usual quark model
result.  (Of course, if one also examines other anomalous processes
such as $\eta^0\rightarrow\pi^+\pi^-\gamma$ agreement between
experiment and theory {\it does} require $N_c=3$, and this provides
the real proof.)

We can now check the validity of the various conservation
conditions---Eq. \ref{eq:hf}, the first of which reads
\begin{eqnarray}
q_1^\mu T_{\mu\nu\gamma}(q_1,q_2)&=&-{ie^2\over 2}\int{d^4s\over
(2\pi)^4}{\rm Tr}\left[({1\over \not\!{s}}-{1\over
\not\!{s}+\not\!{q}_1})\gamma_\nu{1\over
\not\!{s}-\not\!{q}_2}\gamma_\gamma\gamma_5\right.\nonumber\\
&+&\left.{1\over \not\!{s}+\not\!{q}_2}\gamma_\nu({1\over
\not\!{s}}-{1\over
\not\!{s}-\not\!{q}_1})\gamma_\gamma\gamma_5\right]
\end{eqnarray}
However, the integrals which involve a {\it single} factor of photon
momentum $q_1$ or $q_2$ vanish, since the epsilon tensor associated
with the trace
$${\rm Tr}\gamma_\mu\gamma_\nu\gamma_\alpha\gamma_\beta\gamma_5=4i\epsilon_{\mu\nu\alpha\beta},$$
requires contraction with {\it two} independent momenta in order
to be nonvanishing.  Thus, defining
\begin{equation}
W_{\nu\gamma}(s)={\rm Tr}\left({1\over \not\!{s}}\gamma_\nu{1\over
\not\!{s}-\not\!{q}_1-\not\!{q}_2}\gamma_\gamma\gamma_5\right)
\end{equation}
we have
\begin{equation}
q_1^\mu T_{\mu\nu\gamma}(q_1,q_2)=-{ie^2\over 2}\int{d^4s\over
(2\pi)^4}[W_{\nu\gamma}(s+q_1)-W_{\nu\gamma}(s+q_2)]\label{eq:dg}
\end{equation}
If the integrals in Eq. \ref{eq:dg} were convergent, or diverged no
worse than logarithmically, then we could shift the integration
variables freely, thereby obtaining zero and verifying gauge invariance.  However,
because there exists a {\it linear} divergence at large $s$ we must be
more careful.  Using Taylor's theorem
\begin{equation}
\int{d^4s\over (2\pi)^4}F(s+a)=\int{d^4s\over
(2\pi)^4}\left[F(s)+a^\alpha\partial_\alpha F(s)+\ldots\right]
\end{equation}
we can write
\begin{equation}
q_1^\mu T_{\mu\nu\gamma}(q_1,q_2)=-{ie^2\over
2}(q_2-q_1)^\alpha\int{d^4s\over (2\pi)^4}[\partial_\alpha
W_{\nu\gamma}(s)+\ldots]
\end{equation}
where the higher order terms, denoted by the ellipses, vanish, while
the piece we have retained can be evaluated via Gauss' theorem, yielding
\begin{equation}
q_1^\mu T_{\mu\nu\gamma}(q_1,q_2)={-ie^2\over
8\pi^2}\epsilon_{\mu\nu\gamma\delta}q_1^\mu q_2^\delta\label{eq:fds}
\end{equation}
Similarly we find
\begin{equation}
q_2^\nu T_{\mu\nu\gamma}(q_1,q_2)={ie^2\over
8\pi^2}\epsilon_{\mu\nu\gamma\delta}q_1^\mu q_2^\delta\label{eq:fdt}
\end{equation}
and
\begin{equation}
(q_1+q_2)^\gamma T_{\mu\nu\gamma}(q_1,q_2)=0
\end{equation}
From Eqs. \ref{eq:fds} and \ref{eq:fdt} then we observe that electromagnetic gauge invariance
is violated, which would have serious
consequences for photon interactions.  This problem can be solved by
appending a polynomial in the external momenta, which can be done
without affecting the absorptive component of the amplitude.  Thus
defining the physical decay amplitude via
\begin{equation}
t^{phys}_{\mu\nu\gamma}(q_1,q_2)=T_{\mu\nu\gamma}(q_1,q_2)-{ie^2\over
8\pi^2}\epsilon_{\mu\nu\gamma\delta}(q_1-q_2)^\delta
\end{equation}
we see that gauge invariance is restored
\begin{equation}
q_1^\mu t^{phys}_{\mu\nu\gamma}(q_1,q_2)=q_2^\nu
t^{phys}_{\mu\nu\gamma}(q_1,q_2)=0
\end{equation}
However, taking the axial current divergence now yields
\begin{equation}
(q_1+q_2)^\gamma t^{phys}_{\mu\nu\gamma}(q_1,q_2)={ie^2\over
4\pi^2}\epsilon_{\mu\nu\gamma\delta}q_1^\gamma q_2^\delta
\end{equation}
so that the axial current is no longer conserved.  Thus, the axial
symmetry has been {\it broken} via a proper quantization of the
theory---there exists an anomaly.  We have then
\begin{equation}
t^{phys}_{\mu\nu\gamma}(q_1,q_2)={q_{1\gamma}+q_{2\gamma}\over
(q_1+q_2)^2+i\epsilon}{ie^2\over 4\pi^2}\epsilon_
{\mu\nu\alpha\beta}q_1^\alpha q_2^\beta
\end{equation}
which corresponds to the operator condition
\begin{equation}
\partial^\mu J^3_{5\mu}={e^2\over
16\pi^2}F_{\mu\nu}\tilde{F}^{\mu\nu}
\end{equation}
Using the PCAC condition, we have then
\begin{equation}
<2\gamma|\partial^\mu J^3_{5\mu}|0>=F_\pi m_\pi^2 {1\over
m_\pi^2}<2\gamma|\pi^0>={e^2\over
4\pi^2}\epsilon_{\mu\nu\alpha\beta}q_1^\mu \epsilon_1^\nu q_2^\alpha
\epsilon_2^\beta
\end{equation}

Since this evaluation represents a simple (LO) perturbative
calculation, one should ask about the influence of
interactions, whether these are from higher order electromagnetic or
strong interactions. The answer, as shown by Adler and
Bardeen\cite{adb}, is that such effects do {\it not} modify the
chiral anomaly. The basic argument is that any such interactions
will result in changes to the triangle diagrams in which the linear
divergence is removed---the diagrams become more convergent. Because
of this modification the vanishing of Eq. \ref {eq:dg} from such
diagrams is assured and we conclude that the lowest order
calculation given above must be true to {\it all} orders.

\subsection{Alternative Approaches:  All Roads Lead to Rome}

What are some of the alternative methods that can be used in order
to deal with the short distance behavior?  Some of the possibilities
include:

\begin{itemize}

\item [i)] Pauli-Villars Regularization:  As has already been mentioned, a Pauli-Villars regulator
can be used in order to make the results finite.  In this procedure
one defines the physical amplitude as the difference of the
amplitude calculated as above and the same amplitude calculated with
quarks having large mass $M$, {\it i.e.},
\begin{equation}
T^{physical}_{\mu\nu\lambda}(q_1,q_2)=\lim_{M\rightarrow\infty}
\left[T_{\mu\nu\lambda}(q_1,q_2)-T^M_{\mu\nu\lambda}(q_1,q_2)\right]
\end{equation}
where $T^M_{\mu\nu\lambda}(q_1,q_2)$ is identical to
$T_{\mu\nu\lambda}(q_1,q_2)$ but with the massless fermion
propagators replaced by propagators having mass $M$, {\it i.e.},
\begin{equation}
T^M_{\mu\nu\gamma}(q_1,q_2)=U^M_{\mu\nu\gamma}(q_1,q_2)+U^M_{\nu\mu\gamma}(q_2,q_1)
\end{equation}
where
\begin{eqnarray}
U^M_{\mu\nu\gamma}(q_1,q_2)&=&-{ie^2\over 2}K_F\int{d^4s\over
(2\pi)^4}\nonumber\\
&\times&{\rm Tr}\left({1\over
\not\!{s}+\not\!{q}_1-M}\gamma_\mu{1\over
\not\!{s}-M}\gamma_\nu{1\over
\not\!{s}-\not\!{q}_2-M}\gamma_\gamma\gamma_5\right)
\end{eqnarray}
Taking the axial divergence we find
\begin{equation}
(q_1+q_2)^\lambda U^M_{\mu\nu\lambda}(q_1,q_2)={e^2\over
16\pi^2}K_F\epsilon_{\mu\nu\alpha\beta}q_1^\alpha q_2^\beta
\end{equation}
so that
\begin{equation}
(q_1+q_2)^\lambda T^{physical}_{\mu\nu\lambda}(q_1,q_2)={e^2\over
4\pi^2}K_F\epsilon_{\mu\nu\alpha\beta}q_1^\alpha q_2^\beta
\end{equation}
We reproduce in this way then the axial anomaly
\begin{equation}
\partial^\lambda J^3_{5\lambda}={e^2\over 16\pi^2}F_{\mu\nu}\tilde{F}^{\mu\nu}
\end{equation}

\item [ii)] Path Integration: Path integral methods can also be
used.  In this case one quantizes by using the generating
functional, which sums over all field configurations
\begin{equation}
W=\int[d\psi][d\bar{\psi}][dA_\mu]\exp iS[\psi,\bar{\psi}, A_\mu]
\end{equation}
Now if we alter the field transformation Eq. \ref{eq:ku} or
Eq. \ref{eq:jn} from a {\it global} to a {\it local one} the action becomes
\begin{equation}
S[\psi',\bar{\psi}', A_\mu]=S[\psi,\bar{\psi}, A_\mu]+\int
d^4x\beta(x)\partial^\mu J_\mu(x)
\end{equation}
(Note that by using a local rather than a global field transformation
we are required to include a gauge boson field $A_\mu$ in addition to the
Dirac fields $\psi$ and $\bar{\psi}$.)
Kazuo Fujikawa showed that under a vector transformation---Eq.
\ref{eq:ku}---the integration measure is unchanged\cite{kfj}
\begin{equation}
 [d{\psi'}_V][d{\bar{\psi}'}_V][dA_\mu]=[d\psi][d\bar{\psi}][dA_\mu]
\end{equation}
so that, requiring equality of the two representations for arbitrary
$\beta(x)$ yields
\begin{equation}
\partial^\mu J_\mu(x)=0
\end{equation}
{\it i.e.}, the classical field symmetry is also a quantum field
symmetry. However, this is {\it not} the case for a local axial
transformation---Eq. \ref{eq:jn}. In this case we have a non-unit
Jacobian
\begin{equation}
[d\psi_A'][d\bar{\psi}_A'][dA_\mu]=[d\psi][d\bar{\psi}][dA_\mu]J
\end{equation}
where
\begin{equation}
J=\exp(-2i{\rm Tr}\zeta(x) \gamma_5)
\end{equation}
Here the trace is over not only the Dirac indices but also over
spacetime and must be regulated in order not to diverge.  Employing
a covariant regulator of the form $$\exp(-\not\!\!{D}^2/M^2)$$ which
cuts off the high energy (short distance) modes, and the result
\begin{equation}
\not\!\!{D}^2=D^2+{eQ\over 2}\sigma_{\mu\nu}F^{\mu\nu}
\end{equation}
Fujikawa demonstrated that\cite{kfj}
\begin{equation}
J=\exp(-i\int d^4x \zeta(x){e^2\over
16\pi^2}F_{\mu\nu}\tilde{F}^{\mu\nu})
\end{equation}
The connection with the chiral anomaly can be made by noting that
the generating functional
\begin{equation}
W=\int[d\psi][d\bar{\psi}][dA_\mu]\exp iS[\psi,\bar{\psi}, A_\mu]
\end{equation}
after a local axial transformation Eq. \ref{eq:jn}, assumes the form
\begin{eqnarray}
W&=&\int[d\psi_{A}'][d\bar{\psi}_{A}'][dA_\mu]\nonumber\\
&\times&\exp i\left[S[\psi,\bar{\psi},
A_\mu]+\int d^4x \zeta(x)(\partial^\lambda
J^3_{5\lambda}(x)-{e^2\over
16\pi^2}F_{\mu\nu}\tilde{F}^{\mu\nu})\right]\nonumber\\
\quad
\end{eqnarray}
so that invariance of the functional integration for arbitrary
$\zeta(x)$ yields the anomaly condition
\begin{equation}
\partial^\lambda J^3_{5\lambda}={e^2\over 16\pi^2}F_{\mu\nu}\tilde{F}^{\mu\nu}
\end{equation}

\item[iii)] Point splitting:  Since the problems arise when the
field and its conjugate are at the same spacetime point, one can
define the axial current via the definition\cite{rjr}
\begin{equation}
J^3_{5\mu}(x)\equiv\lim_{\epsilon\rightarrow 0}\bar{\psi}(x+{1\over
2}\epsilon){1\over 2}\tau_3\gamma_\mu\gamma_5\psi(x-{1\over
2}\epsilon)\exp(ie\int_{x-{1\over 2}\epsilon}^{x+{1\over
2}\epsilon}dy_\beta A^\beta(y))
\end{equation}
If we now take the divergence we find
\begin{eqnarray}
i\partial^\mu J^3_{5\mu}(x)&=&\lim_{\epsilon\rightarrow
0}\left\{\bar{\psi}(x+{1\over 2}\epsilon){1\over
2}\tau_3\gamma_5\left[e\not\!\!{A}(x+{1\over
2}\epsilon)-e\not\!\!{A}(x-{1\over 2}\epsilon)\right]\psi(x-{1\over
2}\epsilon)\right.\nonumber\\
&-&\left.e\bar{\psi}(x+{1\over 2}\epsilon){1\over
2}\tau_3\gamma_\mu\gamma_5\psi(x-{1\over
2}\epsilon)\epsilon_\nu\partial^\mu
A^\nu\right\}\exp(ie\int_{x-{1\over 2}\epsilon}^{x+{1\over
2}\epsilon}dy_\beta A^\beta(y))\nonumber\\
&=&\lim_{\epsilon\rightarrow
0}e\epsilon_\nu\left\{\bar{\psi}(x+{1\over 2}\epsilon){1\over 2
}\tau_3\gamma_\mu\gamma_5\psi(x-{1\over
2}\epsilon)F^{\mu\nu}\exp(ie\int_{x-{1\over 2}\epsilon}^{x+{1\over
2}\epsilon}dy_\beta A^\beta(y))\right\}\nonumber\\
\quad
\end{eqnarray}
In taking the limit $\epsilon\rightarrow 0$ we use the short
distance behavior of the Dirac field\cite{rjr}
\begin{equation}
\lim_{\epsilon\rightarrow 0}{\rm Tr}\left[\epsilon_\nu{1\over
2}\tau_3\gamma_\mu\gamma_5<0|T\left(\psi(x-{1\over 2}\epsilon)\bar{\psi}(x+{1\over 2}
\epsilon)\right)|0>\right]={e\over 16\pi^2}\tilde{F}_{\mu\nu}
\end{equation}
to once again obtain the axial anomaly
\begin{equation}
\partial^\lambda J^3_{5\lambda}={e^2\over 16\pi^2}F^{\mu\nu}\tilde{F}_{\mu\nu}
\end{equation}

\item[iv)] Geometric Approach: Using geometric methods, Bardeen was
able to identify the full form of the anomaly\cite{wba}, which was soon thereafter
written in terms of an effective action by Wess and Zumino\cite{wzx}.  In the
case of SU(2) and electromagnetism the anomalous Lagrangian density
is
\begin{equation}
{\cal L}_A=-{N_c\over
48\pi^2}\epsilon^{\mu\nu\alpha\beta}\left[eA_\mu{\rm Tr}(QL_\nu
L_\alpha L_\beta-QR_\nu R_\alpha R_\beta)+ie^2F_{\mu\nu}A_\alpha
T_\beta\right]\label{eq:zx}
\end{equation}
where, defining
$$U=\exp(i\vec{\tau}\cdot\vec{\phi}_\pi/F_\pi),$$
\begin{eqnarray}
L_\mu&=&\partial_\mu UU^\dagger,\qquad R_\mu=\partial_\mu U^\dagger
U\nonumber\\
T_\beta&=&{\rm Tr}\left(Q^2L_\beta-Q^2R_\beta+{1\over 2}QUQU^\dagger
L_\beta-{1\over 2}QU^\dagger QUR_\beta\right)\label{eq:gx}
\end{eqnarray}
the piece of Eq. \ref{eq:zx} responsible for
$\pi^0\rightarrow\gamma\gamma$ can be found by expanding to first
order in the pion field
\begin{eqnarray}
{\cal L}_A&=&{e^2N_c\over 16\pi^2F_\pi}{\rm
Tr}(Q^2\tau_3)\epsilon^{\mu\nu\alpha\beta}F_{\mu\nu}A_\alpha\partial_\beta\pi^0\nonumber\\
&=&{e^2N_c\over 24\pi^2F_\pi}F_{\mu\nu}\tilde{F}^{\mu\nu}\pi^0
\end{eqnarray}
which once again reproduces the anomaly prediction.
\end{itemize}

In each case then, one is forced to modify short distance properties
in order to produce a consistent quantum field theory, and it is
this feature which breaks the classical symmetry and produces the
anomaly.  No matter how it is obtained, the result in the case of
the two photon decay amplitude of the neutral pion is that the decay
amplitude is precisely predicted to be
\begin{equation}
T_{\pi^0\gamma\gamma}={e^2\over 4\pi^2
F_\pi}\epsilon_{\mu\nu\alpha\beta}\epsilon_1^\mu\epsilon_2^\nu
q_1^\alpha q_2^\beta ,
\end{equation}
leading to a decay rate
\begin{equation}
\Gamma^{anom}_{\pi^0\gamma\gamma}={\alpha^2m_\pi^3\over
64\pi^3F_\pi^2}=7.76\,\,{\rm eV}\label{eq:lk}
\end{equation}
where $\alpha=e^2/4\pi$ is the fine structure constant, in good
agreement with the experimental value---{\it cf.} Fig. \ref{fig:width}.  However,
before a careful comparison of theory and experiment can be made, one must confront
the fact that the prediction Eq. \ref{eq:lk} is made in the limit of
chiral symmetry, rather than the real world.

\subsection{Real World Corrections}\label{sec:HO}

As mentioned above, the prediction Eq. \ref{eq:lk} is unsatisfactory
in that the chiral limit, in which both the $u,d$ quarks and the
pion are massless does not represent the real world. The $u,d$
quarks are light but not massless, and in turn the pion is the
lightest hadron but certainly does not possess zero mass. More
importantly, because the light quarks are nondegenerate, the
physical $\pi^0$ meson is {\it not} a pure U(3) state $|P_3>$ but is
instead a mixture of $|P_3>,\,|P_8>,\,$ and $|P_0>$ states.  This
mixing is important since a simple U(3) picture of the decay
predicts the bare amplitudes for the two photon decay to be
\begin{equation}
A_{P_3\gamma\gamma}:A_{P_8\gamma\gamma}:A_{P_0\gamma\gamma}=1:\sqrt{1\over
3}:2\sqrt{2\over 3}\label{eq:vx}
\end{equation}
so that, even if the mixture of $|P_8>,\,|P_0>$ states in the
neutral pion is small, they can play an important role in the
$\pi^0\rightarrow\gamma\gamma$ decay amplitude.  One should
note here that the predictions given in Eq. \ref{eq:vx} are given
via the use of U(3) symmetry, wherein the wavefunctions of the
$\eta_8$ and $\eta_0$ are taken to be identical.  This procedure is perhaps
somewhat surprising, since the $\eta_8$ is a Goldstone boson and is massless in the chiral
symmetric limit, while $\eta_0$ is not. The $\eta_0$ mass is known to arise
due to the axial singlet anomaly, but does vanish in the
$N_c\rightarrow\infty$ limit\cite{dgh}. Nevertheless, here and in
other circumstances the use of U(3) symmetry is known to give good
results\cite{nmc}. In terms of chiral perturbation theory, the
prediction of the chiral anomaly for the
$\pi^0\rightarrow\gamma\gamma$ decay amplitude, since it involves
the four-dimensional Levi-Civita tensor, is already
four-derivative--${\cal O}(q^4)$---and is two orders higher than the
leading order strong interaction chiral effects, which are ${\cal
O}(q^2)$\cite{chl}. (Note that we follow ref. \cite{uch} here in using the counting
${\cal O}(e)={\cal O}(q)$.) Modifications of the pion decay amplitude due to
particle mixing are also ${\cal O}(q^4)$. Since the pion is an isovector
while the $\eta$ and $\eta^{'}$ are isoscalars the associated mixing matrix elements must be proportional to
$${m_u-m_d}.$$  (Of course, there exist in addition corrections of ${\cal
O}(q^6)$ which involve factors such as $m_\pi^2/\Lambda_\chi^2$, but
these are presumably higher order and somewhat smaller than those which come
from mixing.)

It is useful to make a simple back of the envelope calculation of
the modifications due to mixing effects by use of the representation
of the physical states $|\pi^{0},\eta,\eta^{'}>$ in terms of the bare U(3) states
P = $|\pi_{3},\eta_{8}, \eta_{0}>$. In order to do this it is useful to first review
the $P \rightarrow \gamma \gamma$ decay rates predicted by the anomaly
\begin{eqnarray}
A_{P \gamma\gamma}&=&{e^2\over 16\pi^2F_P} \\ \nonumber
\Gamma_{\pi\gamma\gamma}&=&{|A_{\pi\gamma\gamma}|^2 m_P^3\over 4\pi}
\label{eq:Gamma-P}
\end{eqnarray}
where $m_P, F_P$ represents the mass, decay constants of the respective pseudoscalars.
In Eq. \ref{eq:vx} the relative amplitudes were given in lowest order, where all of
the decay constants are equal. In the next chiral order there are corrections
to this equality\cite{Donoghue:1985} which lead to $F_{8}/F_{3} = F_{\eta}/F_{\pi} \simeq$ 1.25 and
$F_{0}/F_{3} = F_{\eta^{'}}/F_{\pi} \simeq$ 1.  We shall use these modified values to amend
Eq. \ref{eq:vx}.  With this adjustment, the two-gamma widths are calculated to be
$\Gamma (\eta \rightarrow \gamma \gamma)$ = 0.11 keV and $\Gamma (\eta^{'} \rightarrow \gamma \gamma)$
= 7.40 keV and are {\it not} in agreement with the experimental results
$\Gamma (\eta \rightarrow \gamma \gamma) \simeq ( 0.51 \pm 0.05)$ keV and
$\Gamma (\eta^{'} \rightarrow \gamma \gamma) \simeq 4.28 \pm 0.38$ keV\cite{PDB}.
The fact that the calculated width of the $\eta$ is too low, while that of the $\eta^{'}$ is too high,
is strong evidence for $\eta, \eta^{'}$ mixing, which one can introduce by writing\cite{Donoghue:1985}
\begin{eqnarray}
|\eta>  &=& \cos\theta|\eta_{8}> - \sin\theta |\eta_{0}> \\ \nonumber
|\eta^{'}>  &=& \sin\theta |\eta_{8}> + \cos\theta|\eta_{0}>
\label{eq:theta}
\end{eqnarray}
Using this representation one obtains for the decay amplitudes
\begin{eqnarray}
A_{\eta \gamma\gamma}  &=& \frac{\alpha}{4 \pi \sqrt{3}} ~\Bigg{[ }\frac{ \cos\theta}{ F_{\eta} }- \frac{\sin\theta \sqrt{8} }{ F_{\eta^{'}}}~\Bigg]\\ \nonumber
A_{\eta^{'} \gamma\gamma}  &=& \frac{\alpha}{4 \pi \sqrt{3}} ~\Bigg{[ }\frac{ \sin\theta}{ F_{\eta} } +\frac{\cos\theta \sqrt{8} }{ F_{\eta^{'}}}~\Bigg]
\label{eq:mixing}
\end{eqnarray}
and, treating $\theta$ as a free parameter, one can obtain agreement with the experimental values\cite{PDB} by using $-25\, {\rm deg}\leq \theta\leq -20 \,{\rm deg}$\cite{Donoghue:1985}.

There exists another method to estimate this mixing, which is instructive and will be helpful in estimating the chiral corrections to the $\pi^0$ decay rate.  One simply diagonalizes the mass matrix to obtain the physical $\eta, \eta^{'}$ states and the mixing angle
\begin{equation}
\left(
\begin{array} {cc}
m_{8}^{2} -M^{2} & m_{08}^{2} \\
m_{08}^{2} & m_{0}^{2} -M^{2}
\end{array}
\right )
 \left(
\begin{array} {c}
|\eta>\\
|\eta^{'}>
\end{array}
\right)
~ = ~0
\label{eq:matrix}
\end{equation}
where $m_{8}, m_{0}, m_{08}$, represent the masses of the unperturbed $\eta_{8},\eta_{0}$ states and the off-diagonal mass mixing matrix element, which we take as a parameter.  (The eigenvalue equation involves the square of the masses since $\eta ,\eta'$ are bosons.) There exist two eigenvalues for ${\cal M}$ which are $m^2_{\eta}, m^2_{\eta^{'}}$.  To solve this eigenvalue equation we need the value of $m_{8}$ which can be obtained from the relationship
\begin{eqnarray}
m_{8}^{2}  &=& \frac{1}{3} (4 m_{K}^{2} -m_{\pi}^2) (1+ \delta) \\ \nonumber
m_{K}^{2} &=& \frac{1}{2} (m_{K^0}^{2} + m_{K^{+}}^{2})
\label{eq:GMO}
\end{eqnarray}
For $\delta$ = 0 Eq. 77 is simply the Gell-Mann-Okubu mass formula
\begin{equation}
4m_K^2=3m_\eta^2+m_\pi^2,\label{eq:bz}
\end{equation}
which is derived via the use of SU(3) symmetry.  Since SU(3) is not exact, there exist chiral corrections, for which a leading-log estimate gives $\delta \simeq$ 0.16\cite{Donoghue:1985}. The value of $m_{0}$ can be obtained by observing that, from the trace of the mass matrix in  Eq. \ref{eq:matrix} we have $m_{8}^{2} + m_{0}^{2} = m_{\eta}^{2} + m_{\eta^{'}}^{2}$. Taking $\delta$ as a free parameter, the eigenvectors (specified by the angle $\theta$ of Eq. \ref{eq:theta})  can be obtained and from this the decay rates $\Gamma(\eta, \eta^{'} \rightarrow \gamma, \gamma)$, yielding $0.16\leq\delta\leq 0.22$ in approximate agreement with the leading-log estimate\cite{Donoghue:1985}.

Now we are ready to estimate the magnitude of $\eta, \eta^{'}$ mixing in the $\pi^{0}$ amplitude, which can be approximately written as
\begin{eqnarray}
|\pi^{0} > \simeq | \pi_{3}> + \theta_{\eta} |\eta> + \theta_{\eta^{'}} |\eta^{'}>
\label{eq:pi0-mixing}
\end{eqnarray}
since the mixing angles are small.  The mixing amplitudes $\theta_{\eta}, \theta_{\eta^{'}}$ can be obtained using perturbation theory. The off diagonal matrix elements of the mass matrix $M$ (see, {\it e.g.}, ref. \cite{dgh}) and the mass matrix mixing amplitudes in terms of the quark masses are
\begin{eqnarray}
m_{38}^{2}  &=& < \eta_{8} | M | \pi_{3}> = \frac{B_{0}}{\sqrt{3}} (m_{d} - m_{u})\nonumber \\
m_{30}^{2} &=&  < \eta_{0} | M | \pi_{3}> = \frac{\sqrt{2} B_{0}}{\sqrt{3}} (m_{d} - m_{u})
\nonumber \\
\theta_{\eta} &=& \frac{\cos\theta m_{38}^{2} - \sin\theta m_{30}^{2}}{m_{\eta}^{2} - m_{\pi^{0}}^{2}} \nonumber \\
\theta_{\eta^{'}} &=& \frac{\sin\theta m_{38}^{2} + \cos\theta m_{30}^{2}}
{m_{\eta^{'}}^{2} - m_{\pi^{0}}^{2}}
\label{eq: pi-mixing}
\end{eqnarray}
where $B_0$ is related to the vacuum expectation value of the
quark scalar density via
\begin{equation}
B_0=-{1\over F_0^2}<0|\bar{q}q|0>\simeq {m_\pi^2\over 2\hat{m}}
\label{eq:B0}
\end{equation}
Using the value of $B_{0}$ in terms of $m_{\pi^{0}}$ and the quark mass ratio $m_{u}/m_{d} \simeq 0.56$\cite{Leutwyler:2009} numerical calculations can be performed.  A number of interesting features emerge. First the decay rate is increased  by $\simeq$ 4\%. This value is only weakly dependent on the parameter $\delta$ which corrects the Gell-Mann-Okubo mass formula Eq. \ref{eq:bz}. For $\delta \simeq$ 0.18, for which the $\eta,\eta^{'}$ rates are in agreement with experiment, the values of
$\theta_{\eta} \simeq$ 0.015 rad and $\theta_{\eta^{'}} \simeq$ 0.0032 rad are obtained. The resultant value of $\Gamma(\pi^{0} \rightarrow \gamma \gamma) \simeq $8.1 eV represents a $\simeq$ 4.5\% increase over the value predicted by the chiral anomaly. The contribution of the $\eta$ is
$\simeq$ 3 \% while from the $\eta^{'} \simeq$ 1\%. As we shall see, it is remarkable that this back of the envelope estimate is
in good agreement with various careful chiral perturbation theory calculations.

Stimulated by the PrimEx experiment, QCD corrections to the chiral anomaly prediction
for the $\pi^0\rightarrow\gamma\gamma$ decay amplitude were estimated by a number of groups
with remarkably similar results.  As mentioned above, the leading order (LO) anomaly amplitude already involves four derivations and
is ${\cal O}(q^4)$, while higher order (HO) corrections have been estimated within various approximation schemes.  (It is because
"higher order" means different things depending on the calculation that we have chosen to use this notation, rather than the usual NLO.)
\begin{itemize}

\item [a)] A sum rule estimate by Ioffe and Oganesian\cite{ioo}
including {\it only} $\pi^0,\,\eta^0$ mixing yielded a 3\%
enhancement---$\Gamma_{\pi^0\rightarrow\gamma\gamma}=7.93\pm 0.12$
eV.  The fact that this result is nearly 2{\%} lower than the other three studies can be understood by the omission of
the $\eta -\eta'$ mixing effect given above.
\end{itemize}

\noindent Other authors {\it have} included mixing with the $\eta'$ within the
context of various approximations to QCD:

\begin{itemize}

\item [b)] The work of Goity, Bernstein, and Holstein\cite{bgh} involves
the use of chiral $U(3)\times U(3)$ symmetry and the large $N_c$
limit in order to include the $\eta'$ as a Goldstone boson and
includes modifications to the anomaly prediction of ${\cal
O}(q^6)$ and ${\cal O}(q^4\times {1\over N_c})$.  Such corrections
predict a 4.5\% enhancement to the decay
rate---$\Gamma_{\pi^0\rightarrow\gamma\gamma}=8.13\pm 0.08$ eV.

\item [c)] An alternate approach was taken by Ananthanarayan and
Moussallam\cite{anm}, who employed chiral perturbation theory in the
anomaly sector with the inclusion of dynamical photons.  In this way
they looked both at quark mass effects and at electromagnetic
corrections of ${\cal O}(q^6)$.  The result was a predicted decay
rate---$\Gamma_{\pi^0\rightarrow\gamma\gamma}=8.06\pm 0.02\pm 0.06$
eV, in excellent agreement with the Goity et al. calculation.

\item [d)] The work of Kampf and Moussallam\cite{kam}
involved an NNLO calculation within two flavor chiral perturbation theory.  However,
there exist a number of undetermined counterterms which were estimated by use of a modified
counting scheme wherein $m_s$ is taken to be of ${\cal O}(q)$. The
result was a prediction---$\Gamma_{\pi^0\rightarrow\gamma\gamma}=8.09\pm 0.11$
eV.
\end{itemize}

\begin{figure}
\begin{center}
\includegraphics*[width=1\linewidth,clip=true]{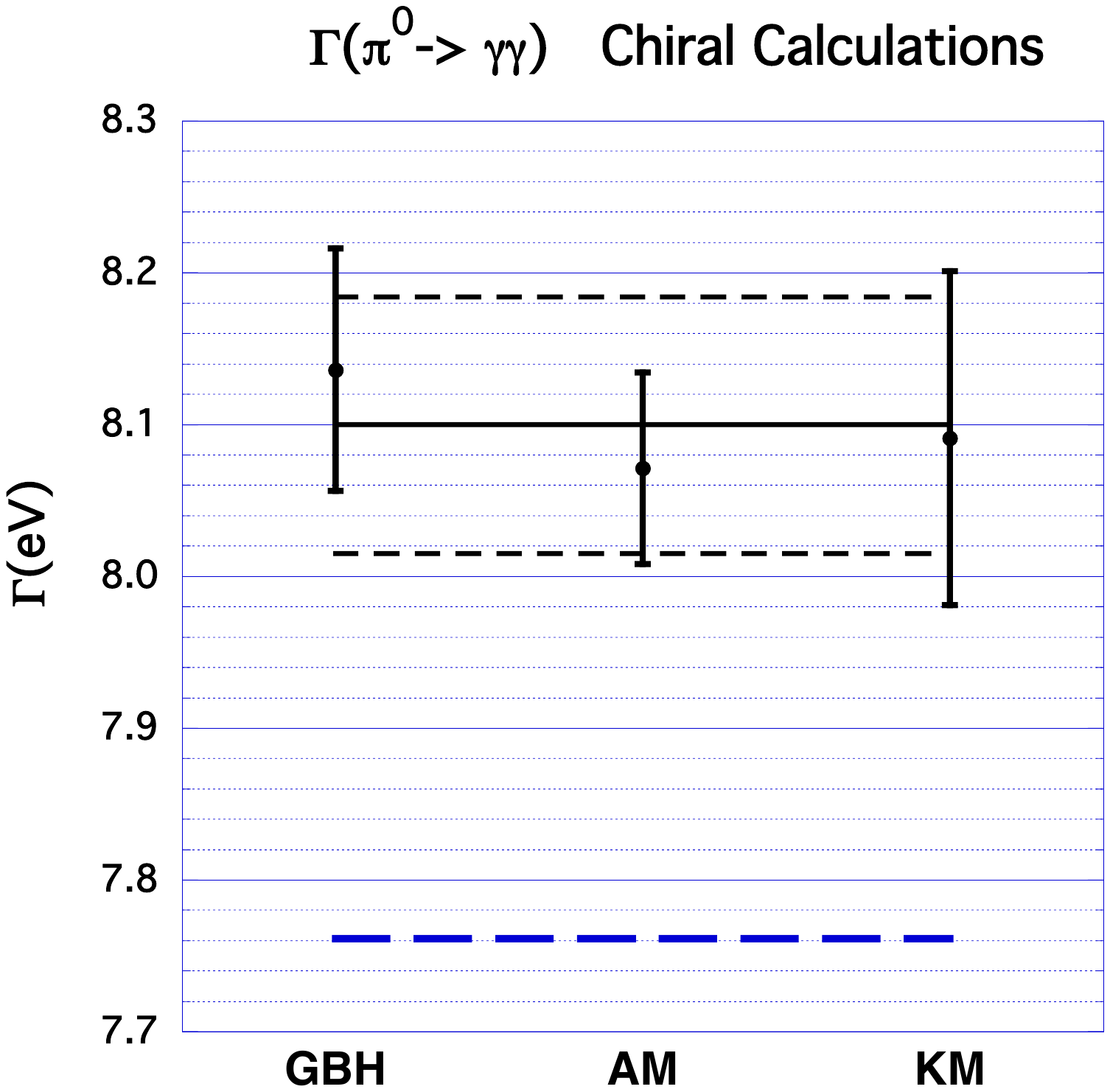}
\end{center}
\vspace{-4cm}
\caption{The $\Gamma(\pi^{0} \rightarrow \gamma \gamma)$ width in eV predicted by the NLO chiral calculations, GBH\cite{bgh}, AM\cite{anm}, and KM\cite{kam}. The lower dashed line is  the predictions of the LO chiral anomaly\cite{bej,sdl} ($\Gamma(\pi^{0} \rightarrow \gamma \gamma) = 7.760\,\,{\rm eV}, \tau(\pi^{0}) = 0.838\times 10^{-16}$ s.). The upper solid line is  the average value of the NLO chiral prediction and the dotted lines show the estimated 1\% error ($\Gamma(\pi^{0} \rightarrow \gamma \gamma)= 8.10\,\,{\rm eV}, \tau(\pi^{0})= 0.80\times 10^{-16}$ s.).
\label{fig:Gamma-ChPT2}}
\end{figure}

These values are plotted in Fig. \ref{fig:Gamma-ChPT2}.
There is obviously little scatter among these theoretical
calculations which assert that chiral symmetry breaking quark mass effects
increase the decay rate of the neutral pion from its $7.76$ eV value
predicted by the anomaly to  8.10 eV, the average of these results.  The basic
reason responsible for this 4.5\% enhancement can be found in the
pseudoscalar mixing estimate given above.   On the theoretical side, the possible
$\sim 1$\% errors in the $\pi^0\gamma\gamma$ decay rate estimates arise not from
convergence issues in the chiral expansion but rather from uncertainty in some of the calculational
input parameters as well as the mixing estimates due to isospin breaking.
Since these predictions are already at the one loop level in the chiral 
expansion, it is unlikely that they will be significantly improved in the near future.
However, in order to confirm this enhancement it is clearly necessary to perform an
experiment looking at $\pi^0\rightarrow\gamma\gamma$ decay at the percent level.

\subsection{Other Anomalous Processes}\label{sec:Other}

From the form of the full SU(2) anomalous effective Lagrangian Eq.
\ref{eq:zx}, it is clear that, in addition to the triangle diagrams
which we have considered above, there also exist anomalous box and
pentagon diagrams.  The presence of the Levi-Civita tensor means
that the processes described by Eq. \ref{eq:zx} are of a different
character than those described by the conventional chiral
Lagrangian\cite{chl}.  Witten\cite{wit} has pointed out that under
what he calls an "intrinsic parity transformation" under wherein pseudoscalar fields undergo a change of sign
but spacetime coordinates remain unchanged
$$\phi^P\rightarrow -\phi^P\quad {\rm so}\quad U=\exp({i\over F_P}\sum_{j=1}^8\lambda_j\phi_j^P)\rightarrow
\exp({-i\over F_P}\sum_{j=1}^8\lambda_j\phi_j^P)=U^\dagger ,$$
the conventional chiral Lagrangian ala Gasser and Leutwyler\cite{chl},\cite{chl1}, which contains terms
such as
$${\rm Tr}\partial_\mu U^\dagger\partial^\mu U,\,\,[{\rm Tr}\partial_\mu U^\dagger\partial^\mu U]^2,\,\,
{\rm Tr}(\partial_\mu U^\dagger\partial_\nu U){\rm Tr}(\partial^\mu U^\dagger\partial^\nu U),\,\, etc.,
$$
includes only the interactions of an {\it even} number of pseudoscalars (or
axial currents) or the coupling of one or two photons to two, four,
six, {\it etc.} pseudoscalars or axial currents, while the SU(2) anomalous Lagrangian
of Eq. \ref{eq:zx} (and its SU(3) generalization) is antisymmetric under $U\rightarrow U^\dagger$ and $L_\mu\rightarrow R_\mu$
and so describes transitions between processes involving an {\it odd}
number of pseudoscalars (or axial currents) or of the coupling of
one or two photons to one, three, five, {\it etc.} pseudoscalars or axial
currents\cite{wit}.  The existence of the chiral anomaly then, in addition to predicting the decay
amplitude for $\pi^0\rightarrow\gamma\gamma$, also makes {\it parameter-free} predictions for processes such as
\begin{itemize}
\item[a)] $\gamma\pi^0\rightarrow\pi^+\pi^-$
\item[b)] $\pi^+\rightarrow e^+\nu_e\gamma$
\item[c)] {\it etc}.
\end{itemize}

or, by extending our analysis to the case of SU(3), for reactions
such as
\begin{itemize}
\item[d)] $K^+K^-\rightarrow\pi^+\pi^-\pi^0$
\item[e)] $\eta\rightarrow\pi^+\pi^-\gamma$
\item[f)] $K^+\rightarrow\pi^+\pi^-e^+\nu_e$
\item[g)] $K^+\rightarrow e^+\nu_e\gamma$
\item[h)] {\it etc.}
\end{itemize}

Note here that when a weak interaction is involved, the coupling to the lepton pair $e^+\nu_e$ is associated
with both a polar vector as well as an axial vector current.  The axial current behaves like a pseudoscalar
under an intrinsic parity transformation so that, for example, the process $\pi^+\rightarrow e^+\nu_e\gamma$
has even intrinsic parity when mediated by the axial current and is not anomalous.  However, if the leptons
couple via a polar vector current, the same process has odd intrinsic parity and occurs via the anomaly.

While it is possible in principle to probe the validity of the chiral anomaly by measurements of any
these reactions, analysis of the $K_{\ell 4}$ decay (f) generally {\it assumes} the value for the appropriate vector form factor predicted by the anomaly.  In the radiative $\eta$ decay reaction (e), significant mixing
with the $\eta'$ obscures a precise test of the chiral anomaly\cite{vgp}.  In the case of the $\gamma 3\pi$
vertex (a) a test was performed many years ago by Antipov et al. with the result\cite{atv},\cite{atw}
\begin{equation}
{\rm Amp}_{\gamma 3\pi}^{exp}=(12.9\pm 0.9\pm 0.5)\,{\rm GeV}^3\quad{\rm vs.}\quad
 {\rm Amp}_{\gamma 3\pi}^{anomaly}={eN_c\over 12\pi^2 F_\pi^3}=9.7\,{\rm GeV}^3
\end{equation}
However, comparison of theoretical and experimental results is not as straightforward as in the case of neutral pion decay.  Of course, "real world" chiral symmetry breaking and electromagnetic corrections similar to those calculated in the case of the $\pi^0\rightarrow\gamma\gamma$ decay amplitude affect ${\rm Amp}_{\gamma 3\pi}^{anomaly}$ and must be included\cite{rhc}.  But also, the reaction  $\gamma\pi^0\rightarrow \pi^+\pi^-$ has an energy dependence that must be accounted for in comparing with the anomaly prediction\cite{khb}, which is calculated in the chiral limit.  Certainly, an up to date modern measurement of the $\gamma 3\pi$ amplitude would be of interest.

Another reaction that has been utilized as a test of the chiral anomaly is radiative pion decay (b).  Since this process is used by the PDG in their determination of the neutral pion decay lifetime, it will be discussed next in more detail.  However, while examination of any of the reactions listed above and their connection to the chiral anomaly is of interest, in the present paper, we shall focus on the $\pi^0\rightarrow\gamma\gamma$ process.

\subsection{Radiative Pion Beta Decay}\label{sec:pi-beta}
The PIBETA experiment mentioned in the introduction uses a {\it measured} weak polar vector form factor to predict the neutral pion decay amplitude by isospin invariance and is used as an input to the PDG average\cite{Bychkov:2009}. The purpose of the PIBETA experiment was to measure the rate of the pion beta decay reaction $\pi^+\rightarrow \pi^0e^+\nu_e$ as a test of the conserved vector current (CVC) relation or, by using CVC, to measure the CKM parameter $V_{ud}$ in a reaction where strong interaction uncertainties are not important.Because this is such a small ($\sim 10^{-8}$) branching ratio process, the experiment also detected other higher branching ratio reactions such as radiative pion decay $\pi^+\rightarrow e^++\nu_e+\gamma$.
Here we discuss this process in a bit more detail. The transition amplitude can be written in general as
\begin{equation}
{\cal M}_{\pi^+\rightarrow e^+\nu_e\gamma}=-{eG_F\over \sqrt{2}}V_{ud}M_{\mu\nu}(p,q)
\epsilon^{\mu *}\bar{u}(p_\nu)\gamma^\nu(1+\gamma_5)v(p_e)
\end{equation}
where $G_F$ is the weak decay constant, $V_{ud}$ is the CKM parameter connecting the $u$ and $d$ quarks, and\cite{dgh}
\begin{eqnarray}
M_{\mu\nu}(p,q)&=&\int d^4xe^{iq\cdot x}<0|T[J_\mu^{em}(x)J_\nu^{1-i2}(0)]|\pi^+(p)>\nonumber\\
&=&-\sqrt{2}F_\pi{(p-q)_\nu\over (p-q)^2-m_\pi^2}<\pi^+(p-q)|J_\mu^{em}|\pi^+(p)>+\sqrt{2}F_\pi g_{\mu\nu}\nonumber\\
&-&{1\over m_\pi}F_A((p-q)_\mu q_\nu-g_{\mu\nu}q\cdot(p-q)) + i {1\over m_\pi}F_V\epsilon_{\mu\nu\alpha\beta}q^\alpha p^\beta\label{eq:bg}
\end{eqnarray}
Here the first two terms on the right hand side represent the Born diagram together with a contact term required for gauge invariance and the subscripts $V$ and $A$ in the remaining terms indicate whether the weak vector or axial-vector of the weak currents are involved. Ordinarily a radiative decay is primarily sensitive to the Born amplitude, which simply generates a correction to the non-radiative decay process via
\begin{equation}
{d\Gamma\over d\Omega dk}\sim {\alpha\over k}{d\Gamma_0\over d\Omega}+\ldots
\end{equation}
so that direct decay amplitudes are hidden under this huge bremsstrahlung background.  However, because the nonradiative decay process $\pi^+\rightarrow e^+\nu_e$ is helicity-suppressed the direct decay amplitudes $F_V,\,F_A$
above can both be measured and this was done by the PIBETA experimenters.  The direct axial-vector decay amplitude $F_A$ determines, via PCAC, the charged pion polarizability\cite{gxr}, while the direct polar-vector amplitude $F_V$ is related to the pion decay amplitude via\cite{dgh}
\begin{equation}
A_{\pi^0\gamma\gamma}={e^2\over 2\sqrt{2}m_\pi}F_V\label{eq:lz}
\end{equation}
That this should be the case is clear from the Wess-Zumino anomaly Lagrangian Eq. \ref{eq:zx} or simply from isotopic spin invariance.  Indeed since the pion as well as the isoscalar component of the electromagnetic current have negative G-parity while the isovector piece of the electromagnetic current has positive G-parity, the two-photon decay of the $\pi^0$ must involve both isoscalar and isovector pieces of $J_\mu^{em}$, so
\begin{equation}
2A_{\pi\gamma\gamma}\epsilon_{\mu\nu\alpha\beta}q^\alpha(p-q)^\beta=e^2\int d^4xe^{iq_1\cdot x}<0|T(^{I=0}J_\mu^{em}(x)^{I=1}J_\nu^{em}(0))|\pi^0(\vec{p})>
\end{equation}
G-invariance also requires the direct polar vector radiative pion decay amplitude $F_V$ to involve only the isoscalar component of the electromagnetic current
\begin{equation}
{1\over m_\pi}F_V\epsilon_{\mu\nu\alpha\beta}q^\mu(p-q)^\nu=\int d^4xe^{iq_1\cdot x}<0|T(^{I=0}J_\mu^{em}(x)^{1-i2}J_\nu(0))|\pi^0(\vec{p})>
\end{equation}
By CVC these two transition amplitudes are related via
\begin{eqnarray}
\int d^4xe^{iq_1\cdot x}<0|T(^{I=0}J_\mu^{em}(x)^{I=1}J_\nu^{em}(0))|\pi^0(\vec{p})>\nonumber\\
=\sqrt{1\over 2}\int d^4xe^{iq_1\cdot x}<0|T(^{I=0}J_\mu^{em}(x)^{1-i2}J_\nu(0))|\pi^0(\vec{p})>
\end{eqnarray}
from which Eq. \ref{eq:lz} follows.  Thus a measurement of $F_V$ can be used to yield an experimental value for the $\pi^0\rightarrow\gamma\gamma$ decay rate
\begin{equation}
\Gamma_{\pi\gamma\gamma}={1\over 2}\pi m_\pi\alpha^2|F_V|^2
\end{equation}
The result of the PIBETA experiment gives\cite{Bychkov:2009}
\begin{equation}
\Gamma_{\pi^0\gamma\gamma}^{\rm pi-beta}=7.7\pm 1.0 \,{\rm eV},\label{eq:zj}
\end{equation}
and this number is included in the PDG average.

Note, however, that since isospin invariance is broken at the $\sim$1\% level, the present 11\% precision of this method is not an issue, but the uncertainty associated with isospin violation ultimately limits its use at the $\sim$1\% level unless this breaking is included.  The breaking associated with the neutral pion decay amplitude was discussed in the previous section and amounts to a $\simeq$ 2.3\% increase in the decay amplitude.  However, we also need to know the isospin violation---both from electromagnetic corrections as well as from the u-d quark mass difference---in the radiative pion decay amplitude.  These breaking effects have been calculated by Unterdorfer and Pichel\cite{unp} and were included in the PIBETA\cite{Bychkov:2009} analysis except for a $\simeq$ 0.9\% increase to the radiative pion decay vector form factor $F_V$.  Thus the proper way in which to determine the neutral pion decay amplitude from the experimental value of $F_V$ is to increase the CVC-predicted value of the neutral pion decay amplitude---Eq. \ref{eq:lz}---by 2.3-0.9\%=1.4\%, so that the pion decay rate predicted in Eq. \ref{eq:zj} becomes $7.9\pm 1.0$ eV.  Since the uncertainty in the PIBETA\cite{Bychkov:2009} value of $F_V$ is at the 11\% level, this modification is not important at the present time, but could be a factor in a future precision determination.

\subsection{Physics of the Anomaly}

Above we have shown how the chiral anomaly leads to a remarkably
successful agreement between experiment and theory for the decay
rate of the neutral pion.  However, this derivation is somewhat
formal and it remains to show what the "physics" of the anomaly
is---that is, {\it why} must quantization destroy the classical
axial symmetry.  In order to present an answer to this question it
is useful to first examine the Schwinger model, which is the name
generally given to massless electrodynamics in one plus one
dimensions\cite{jsc}.  Here the Lagrange density is given by
\begin{equation}
{\cal L}=\bar{\psi}i\not\!\!{D}\psi-{1\over 4}F_{\mu\nu}F^{\mu\nu}
\end{equation}
where
\begin{equation}
D_\mu=\partial_\mu+ieA_\mu
\end{equation}
is the covariant derivative and the $2\times 2$ Dirac matrices are
given in terms of the Pauli matrices via
\begin{equation}
\gamma^0=\sigma_1\quad{\rm and}\quad\gamma^1=i\sigma_2
\end{equation}
It is then easy to see at the classical level that we have equations
of motion
\begin{equation}
i\not\!\!{D}\psi=0\quad{\rm and}\quad\Box A_\mu=ej_\mu\label{eq:vc}
\end{equation}
where
\begin{equation}
j_\mu=\bar{\psi}\gamma_\mu\psi
\end{equation}
is the vector current and is conserved---$\partial^\mu j_\mu=0$. There also exists an axial current
\begin{equation}
j_\mu^5=\bar{\psi}\gamma_\mu\gamma_5\psi\quad{\rm with}\quad
\gamma_5=-\gamma^0\gamma^1=\sigma_3
\end{equation}
which is conserved---$\partial^\mu j_\mu^5=0$.  For later use, we
note also that
\begin{equation}
j_\mu^5=\epsilon_{\mu\nu}j^\nu\label{eq:hb}
\end{equation}
where $\epsilon_{\mu\nu}$ is the two-dimensional Levi-Civita tensor.
So far then, this looks simply like a two-dimensional version of
massless QED.

However, upon quantization it can be shown that the Lagrangian can
be written as
\begin{equation}
{\cal L}=\bar{\psi}'i\not\!{\partial}\psi'-{1\over
4}F_{\mu\nu}F^{\mu\nu}-{e^2\over 2\pi}A_\mu A^\mu
\end{equation}
that is, in terms of a noninteracting system of massless spin 1/2
particles and free "photons" having mass $m^2_\gamma=e^2/\pi$.  Also
in the quantized theory the axial current is no longer conserved.
Rather we have
\begin{equation}
\partial^\mu j_\mu^5=-{e\over 2\pi}\epsilon^{\mu\nu}F_{\mu\nu}
\end{equation}
That is, axial current conservation is broken by
quantization---there exists an anomaly.

The physical origin of the anomaly can be seen via an argument due
to Widom and Srivastava\cite{srw} by considering the vacuum state in
the quantized theory, which, according to Dirac, can be considered
as a filled set of negative energy states.  In the absence of an
external electric field there exists a density of electron states
$dp/2\pi$ with momenta evenly distributed between $p=-\infty$ and
$p=\infty$, and so there is no net current.  Now consider what
happens in the presence of a constant electric field $E$---a net
current flow develops, which increases with time. In terms of the
current density $j$ we have
\begin{equation}
{dj\over dt}=e\int_{-\infty}^\infty {dp\over 2\pi}{dv\over dt}
\end{equation}
where, using the Lorentz force law---$dp/dt=eE$---we have
\begin{equation}
{dv\over dt}={d\over dt}{d\over dp}\sqrt{m^2+p^2}={eEm^2\over
(m^2+p^2)^{5\over 2}}
\end{equation}
Performing the integration we find a result
\begin{equation}
{dj\over dt}={e^2E\over \pi}
\end{equation}
which is {\it independent} of the mass.  Since the vacuum charge
density $\lambda$ is independent of position---$d\lambda/dx=0$---we
have, defining $j^\mu=(\lambda,j)$ and using Eq. \ref{eq:hb}
\begin{equation}
e\partial^\mu j_\mu^5=-{e^2\over
2\pi}\epsilon^{\mu\nu}F_{\mu\nu}\label{eq:jg}
\end{equation}
which is the chiral anomaly.  Note that Eq. \ref{eq:jg} can also be
written as
\begin{equation}
0=\epsilon_{\mu\nu}\partial^\mu\left(ej^\nu+{e^2\over
\pi}A^\nu\right)
\end{equation}
In Lorentz gauge---$\partial^\mu j_\mu=0$---we have
\begin{equation}
j_\mu=-{e^2\over \pi}A_\mu
\end{equation}
so that the equation of motion in Eq. \ref{eq:vc} becomes
\begin{equation}
(\Box+{e^2\over \pi})A_\mu=0
\end{equation}
which indicates that the "photon" has developed a mass
$m^2_\gamma=e^2/\pi$.  We see then that in this picture the origin
of the anomaly is clear and arises from the feature that the vacuum
state of the quantized system is altered in the presence of an
applied electric field.

\begin{figure}
\begin{center}
\includegraphics*[width=0.50\linewidth,clip=true]{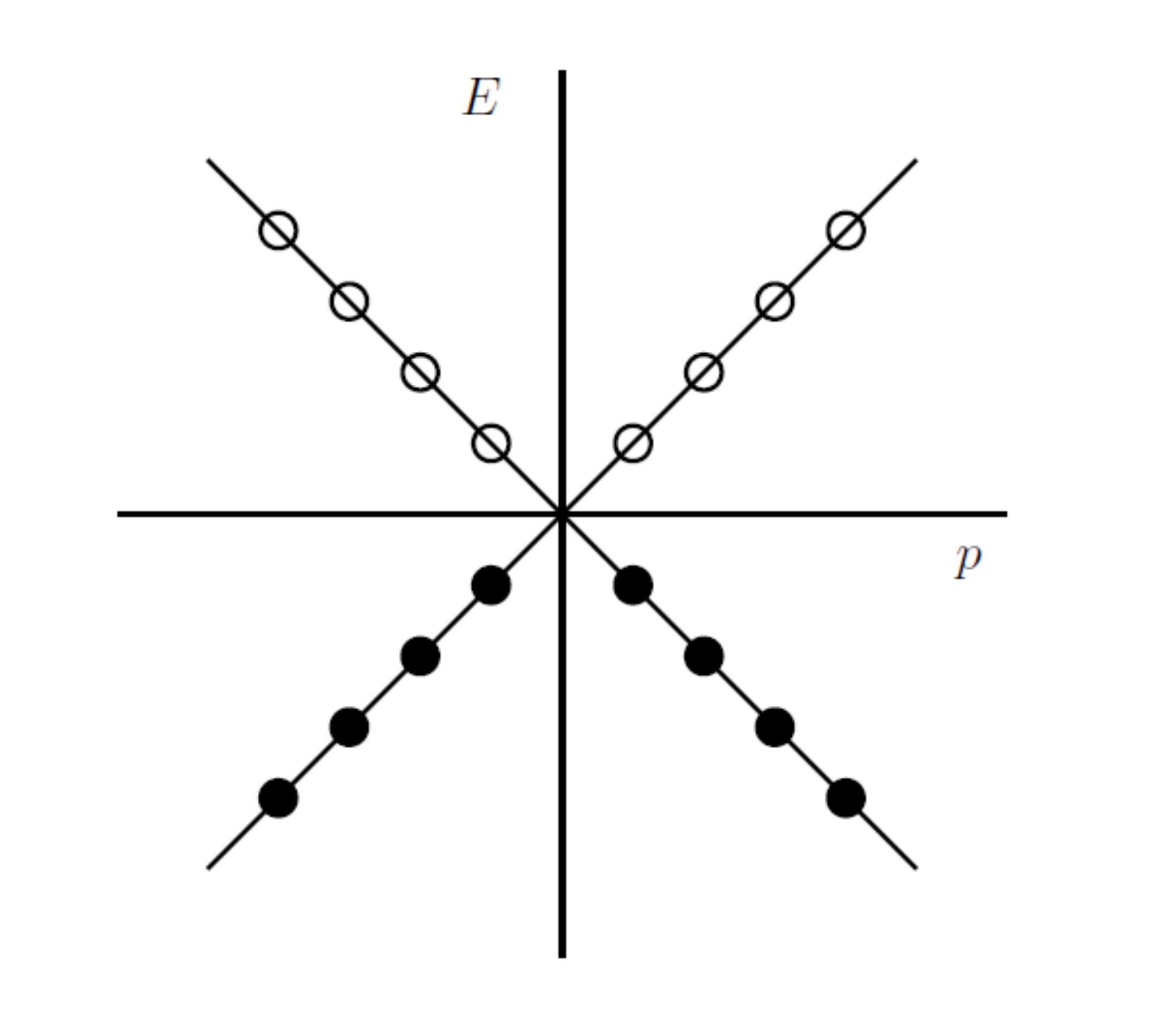}
\end{center}
\vspace{-5mm} \caption{The Dirac Sea in the case of a vanishing vector
potential.  Here a solid dot represents a filled state while an
empty dot signifies an empty state.
\label{fig:theory3}}
\end{figure}

An alternative way to understand the same result has been given by
Jackiw\cite{rjk}, wherein one looks at solutions of the
time-independent Dirac equation
\begin{equation}
E\psi=\gamma_0\gamma_1\left(-i{\partial\over \partial
x}-eA\right)\psi
\end{equation}
In the case of a constant vector potential then there are two
classes of solutions
\begin{eqnarray}
\psi_+(x)&=&\left(\begin{array}{c} e^{ipx}\\0
\end{array}\right)\quad{\rm with}\quad E=p-eA\nonumber\\
\psi_-(x)&=&\left(\begin{array}{c}0\\e^{ipx}\end{array}\right)\quad{\rm
with}\quad E=-p+eA
\end{eqnarray}
where the subscript $\pm$ specifies the chirality of the
solution---{\it i.e.}, the eigenvalues of the operators ${1\over
2}(1\pm \gamma_5)$.  If $A=0$ we see then that the vacuum consists
of (negative energy) states with $p<0$ for positive chirality and
$p>0$ for negative chirality---{\it cf.} Fig. \ref{fig:theory3}.

Now suppose we make an adiabatic change from $A=0$ to a nonzero
field with $A=\epsilon$.  In the presence of the field, the vacuum
states become those with $p<e\epsilon$ for positive chirality and
$p>e\epsilon$ for negative chirality, meaning that there is a net
chirality production
\begin{equation}
\Delta\chi=2\int_0^{e\epsilon}{dp\over 2\pi}={e\epsilon\over
\pi}\label{eq:mz}
\end{equation}
This result should be expected from the chiral anomaly, which
requires a time-varying axial charge
\begin{equation}
{d\over dt}Q_5={e\over \pi}E={e\over \pi}{dA\over dt}\label{eq:bv}
\end{equation}
Since
\begin{equation}
Q_5=\int dx\psi^\dagger\sigma_3\psi
\end{equation}
corresponds to chiral charge, we find, integrating both sides of Eq.
\ref{eq:bv}
\begin{equation}
\Delta\chi={e\over \pi}\Delta A={e\epsilon\over \pi}
\end{equation}
in agreement with Eq. \ref{eq:mz}.  Once again we see that the
origin of the anomaly is the modification of the vacuum in the
presence of an applied electric field.

A simple handwaving argument can be used to generalize the latter
argument to four spacetime dimensions\cite{amu}.  A constant
magnetic field in the $z$-direction can be represented by the vector
potential
\begin{equation}
\vec{A}={1\over 2}\vec{B}\times\vec{r}
\end{equation}
The energies in the presence of the field are given by the Landau
levels
\begin{equation}
E_k=\pm\sqrt{p_z^2+2eBk}\quad{\rm with}\quad k=0,1,2,...
\end{equation}
If we now turn on an electric field, also in the
z-direction, then the energy levels change in accord with the
Lorentz force law
\begin{equation}
{d\vec{p}_z\over dt}=e\vec{E}
\end{equation}
As $p_z$ changes, the energy levels change, but negative energy
levels remain negative so that no particle creation occurs. However,
if $k=0$ then the structure of the levels is as in the case of the
Schwinger model, so that helicity is produced at the rate
\begin{equation}
{dh\over dt}=\rho{e\over 2\pi}E
\end{equation}
where $\rho$ is magnetic flux density.  Since magnetic flux density
is quantized in terms of $2\pi/e$, we have
\begin{equation}
\rho={eB\over 2\pi}
\end{equation}
and thus
\begin{equation}
{dh\over dt}={e^2\over 4\pi^2}EB
\end{equation}
Since helicity and chirality are the same for a massless system we
can write this as the covariant equation
\begin{equation}
{dQ_5\over dt}={e^2\over 4\pi^2}F_{\mu\nu}\tilde{F}^{\mu\nu}
\end{equation}
which is the standard chiral anomaly.

\subsection{Theoretical Summary}

We have seen above that a consistent quantum field theory requires
modification of the naive short distance behavior and leads to the
breaking of axial symmetry---
\begin{equation}
\partial^\mu J^3_{5\mu}=F_\pi m_{\pi^0}^2\phi_{\pi}^3+{e^2\over 16\pi^2}F_{\mu\nu}\tilde{F}^{\mu\nu}
\end{equation}
This anomalous symmetry breaking leads to a decay rate
\begin{equation}
\Gamma^{anom}_{\pi\rightarrow\gamma\gamma}=7.76\,\,{\rm eV}
\end{equation}
When real world corrections such as chiral symmetry breaking and
mixing are included this prediction is raised by about 4.5\% to
\begin{equation}
\Gamma^{theo}_{\pi\rightarrow\gamma\gamma}=8.10\,\,{\rm eV},
\end{equation}
with remarkably little theoretical uncertainty.  An ${\cal
O}(1\%)$ experimental verification of this prediction would then constitute a
validation of QCD.

Although rigorous quantum field theoretical arguments lead unambiguously to this
violation of axial symmetry, it is also useful to understand the
"physics" of this phenomenon and we presented arguments which show
that origin of the anomaly is the modification of the vacuum state of the field theory
in the presence of an electromagnetic field.  Having then understood the connection of
the decay rate of the $\pi^0$ with QCD, we move now to examine the experiments.


%% file: section_PDB.tex
\section{Measurement of the $\pi^{0}$ Lifetime: The Particle Data Book }\label{sec:PDB}

As discussed in section \ref{early-exp}, by the end of 1965 it was known that $\tau(\pi^{0}) \sim 10^{-16}$ s,
based on measurement of the $\pi^{0}$ decay distance\cite{vonDardel:1963} and determination of the Primakoff cross section\cite{Bellettini:1965}. This section will cover experiments performed during the period from 1970 through 1988, on which the  value in the 2011 particle data book is primarily based\cite{PDB,PDB-online}. As an overview  we gave in Fig. \ref{fig:width} the results which the 2011 particle data average is based, which will be presented in this section, as well as the recent Primakoff measurement\cite{Larin:2010} and the current theoretical predictions.
The first three measurements  shown in Fig. \ref{fig:width} were performed using the Primakoff effect\cite{Bellettini:1970,Kryshkin:1970,Browman:1974}  at  DESY, Tomsk, and Cornell respectively.  The fourth result is a direct measurement of the $\pi^{0}$ decay distance\cite{Atherton:1985} performed at CERN at high energies, and the fifth result was obtained via a $\pi^{0}$ production cross section measurement in  $e^{+}e^{-}$ collisions performed at DESY\cite{Williams:1988}. The sixth result is due to  a measurement of radiative pion decay($\pi^{+} \rightarrow e^{+}\nu \gamma$)\cite{Bychkov:2009}.  Using isospin invariance, the weak polar-vector form factor contributing to this decay channel is related by a simple isospin rotation to the amplitude for $\pi^0\rightarrow\gamma\gamma$ (see Sec.\ref{sec:Other} for further discussion). The last point is a recent Primakoff effect measurement\cite{Larin:2010} (discussed in Sec. V). We now discuss each measurement that is used in the 2010 version of the Particle Data Book\cite{PDB} and the 2011 online update\cite{PDB-online}. As a result of the PrimEx experiment\cite{Larin:2010}, this review,  and private communications with the particle data group the 2012 version will be changed\cite{PDG:private}.


\input{subsection-Primakoff}
\input{subsection_direct}

\input{subsection-ee}

%% file: subsection-Primakoff.tex
\subsection{Primakoff Effect Measurements}\label{subsection:Primakoff}

In the decade between 1965 and 1974 there were four experiments performed \cite{Bellettini:1965,Bellettini:1970,Kryshkin:1970,Browman:1974} which used the Primakoff effect\cite{Primakoff:1951}.  All of these early experiments utilized bremsstrahlung photons, which have a continuous  energy spectrum up to $E_{0}$, the energy  of the electrons which produced them.  For two of these experiments\cite{Kryshkin:1970,Browman:1974} an approximate measure of the photon energy was obtained from the opening angle distribution from the two photon decay(see the appendix at the end of this section).
The Primakoff experiments determine the cross sections for the $\gamma + A \rightarrow \pi^{0} + A$ reaction. At small angles this reaction is dominated by the $\gamma + \gamma^{*} \rightarrow \pi^{0}$ process, in which one of the gamma rays is due to the Coulomb field of the nucleus, which remains in its ground state.  The neutral pions are also produced by the photons interacting with the nucleons leaving the nucleus in its ground state (nuclear coherent production) or excited or continuum states with the production of pions (nuclear incoherent).  The Primakoff effect dominates at very small angles, but the small contributions of the nuclear effects must be subtracted from this small angle signal.  In practice this is accomplished by fitting the parameters of model calculations for the nuclear effects using the larger angle data.  We shall now sketch the basics of the Primakoff and nuclear effects before examining the experiments.

\begin{figure}
\begin{center}
\epsfig{figure=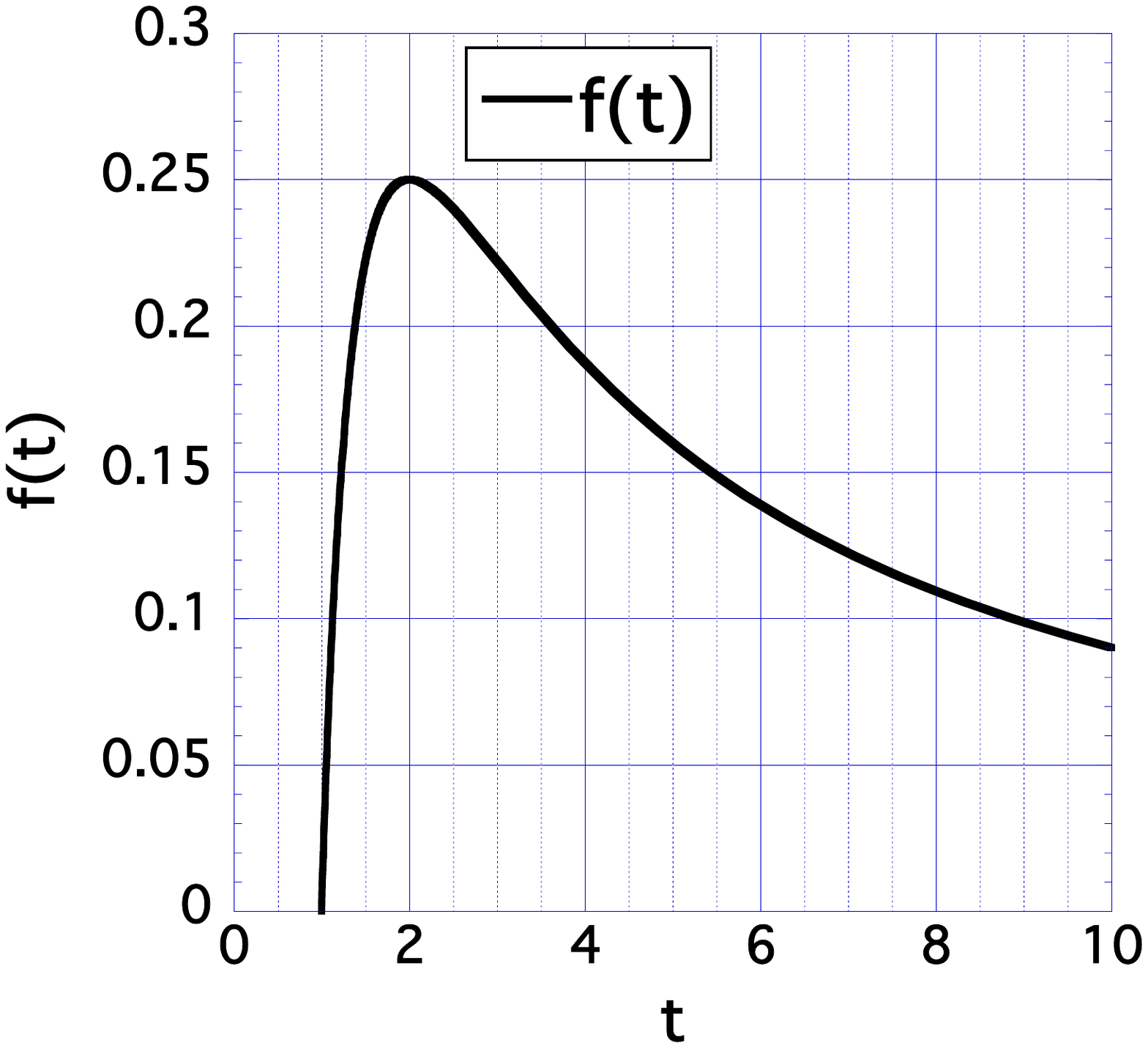,width=6.75 cm,height=9cm}
\epsfig{figure=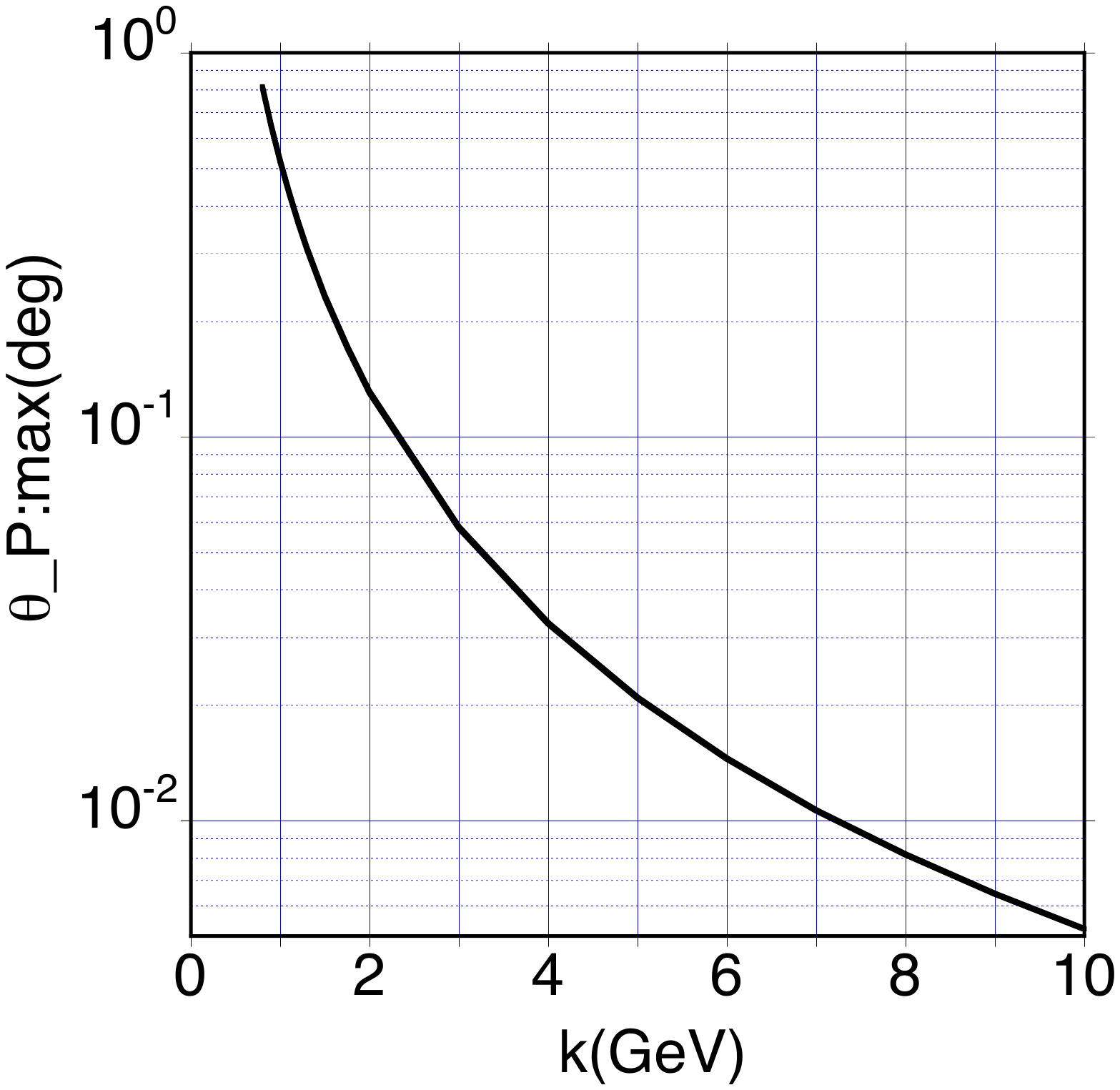,width=6.75 cm,height=9cm}
\end{center}
\caption{(Left) The energy independent Primakoff function f(t) (see Eq.\, \ref{eq:Gourdin} and text). The four momentum transfer t is units of $Q^{2}_{min}$ and is dimensionless. (Right) The center of mass angle in degrees  for which the Primakoff cross section is 
maximal versus photon energy k. }\label{fig:diff-cross}
\end{figure}

The Primakoff cross section peaks at small angles and is quite narrow. The general features of the cross section can be seen by following the treatment of Gourdin\cite{Gourdin:1971} in the high energy limit
\begin{eqnarray}
\frac{d\sigma_P}{d\Omega}&=&\Gamma_{\gamma
\gamma}\frac{8{\alpha}Z^2}{m^3}\frac{{k^2}}{Q_{min}^{2}}|F_{e.m.}(Q)|^2 f(Q^{2}/Q_{min}^{2})\nonumber \\
f(t&=&Q^{2}/Q_{min}^{2}) = (t-1)/t^{2} \nonumber \\
Q^{2}_{min} &\simeq& (m^{2}/2k)^{2} \quad
\theta_{P:max} \simeq m^{2}/(2k^{2})
\label{eq:Gourdin}
\end{eqnarray}
where  $\Gamma_{\gamma \gamma}$ is the pion decay width (the primary objective of the Primakoff experiments), $Z$ is the atomic number of the target nucleus, $m$ is the mass of the produced meson, $k$ is the energy of incoming photon, $Q^{2}_{min}$ is the minimum value for the square of the momentum transfer, and $\theta_{P:max}$ is  the angle for which the Primakoff cross section reaches its maximum  value. One advantage of this formulation is that  the four momentum transfer t is in units of $Q^{2}_{min}$ and is dimensionless. The energy-independent function $f(t)$ is shown in Fig. \ref{fig:diff-cross} and can be seen to rise rapidly from forward angles ($\theta$ =0, t= 1) and its peak at t = 2.  The angle at  which the Primakoff  cross section is a maximum $\theta_{P:max}$ decreases rapidly with photon energy as shown in Fig. \ref{fig:diff-cross}.  The small value of $\theta_{P:max}$ means that the experiments that can measure the Primakoff effect must detect $\pi^{0}$ production at small angles.  In addition, from the shape of $f(t)$ it can be inferred that the width of the peak is $\simeq 2 \theta_{P:max}$ so that the detector needs to have excellent angular resolution. Furthermore, in order to suppress background, it is very useful for the detector to have good energy resolution as well.

In the Primakoff cross section the target photon is virtual  since the square of its four momentum transfer is not zero as it is for a real photon. It is easy to show that this is a very small effect.  If we consider the four momenta of the process
\begin{eqnarray}
p_{\gamma} + p_{A} = p_{\pi^{0}} + p_{A^{'}} \nonumber \\
 p_{A^{'}} - p_{A}  = p_{\gamma^{*}} \nonumber \\
F_{off-shell} -1 \rightarrow | p_{\gamma^{*}}^2| R_{A}^2/6
\label{eq:q}
\end{eqnarray}
where the four momenta $p_{\gamma},p_{A}, p_{\pi^{0}}, p_{A^{'}},  p_{\gamma^{*}}$  represent the incident photon, the target nucleus, the outgoing $\pi^{0}$, the recoil nucleus, and the virtual photon. The estimate of the off shell effect is given by the low momentum transfer estimate of the the nuclear form factor where $R_{A}$ is the nuclear radius which is $\simeq 1.2 A^{1/3}$ fm where A is the nuclear mass number.To estimate the magnitude of the off shell effect the value of the momentum transfer at the Primakoff peak $|q^2| = 2 Q^{2}_{min}$ from Eq.\,\ref{eq:Gourdin} is used. For 5 GeV incident photons this leads to a negligible off shell correction of $\simeq 8\times 10^{-6} (5\times 10^{-5}) $ for C(Pb).

The  photoproduction of pions from a complex nucleus, $\gamma+A\to \pi^0+A$, can be described by the sum of Coulomb  $T_C$ and Strong $T_S$ amplitudes. Including incoherent production, the differential cross section is\cite{Browman:1974}:

\begin{eqnarray}
\frac{d\sigma}{d\Omega} &=&
\mid{T_C+e^{i \phi} T_S}\mid^2 +\frac{d\sigma_{inc}}{d\Omega} \nonumber  \\
 &=& \frac{d \sigma_P}{d\Omega} +
\frac{d\sigma_S}{d \Omega} + \frac{d \sigma_{inter}}{d\Omega} +\frac{d\sigma_{inc}}{d\Omega} \nonumber \\
\frac{d\sigma_P}{d\Omega}&=&\mid{T_C}\mid^2=\Gamma_{\gamma
\gamma}\frac{8{\alpha}Z^2}{m^3}\frac{\beta^3{k^4}}{Q^4}|F_{e.m.}(Q)|^2 \sin^{2}\theta_{\pi}\nonumber \\
\frac{d\sigma_S}{d \Omega} &=& \mid{T_S}\mid^2=C_{S} \sigma_N A^2 |F_S(Q)|^2 \sin^2\theta_{\pi} \nonumber \\
Q^{2} &\simeq& (m^{2}/2k)^{2} + k^{2} \sin^2\theta_{\pi}
  \label{eq:diff_cross}
\end{eqnarray}

\noindent
where $T_C, T_S$ are the Coulomb (Primakoff) and strong amplitudes.  The phase $\phi$  originates from the $\gamma p \rightarrow \pi^{0}p$ amplitude and is fitted to the data. The first two terms in the first line represent the coherent cross section for which the  nucleus is left in its ground state. $d\sigma_{i}/d\Omega (i = P,S,inter,inc)$
are the cross sections for Primakoff, strong, interference, and incoherent processes (the latter involving target nucleus excitation or break up).  $\beta$, $\theta_{\pi}$ are the velocity and production angle of the pion.  For the spin 0 targets employed in these experiments the coherent cross sections are non-spin flip and as a consequence have a $\sin^2\theta_{\pi}$ dependence, ensuring that no scattering occurs at forward or backward angles.  $Q$ is the momentum transfer to the nucleus, and $F_{e.m.}(Q), F_S(Q)$ are the nuclear electromagnetic and strong form factors, corrected for final state interactions of the outgoing pion\cite{Morpurgo:1964,Faldt:1972,Gevorkyan:2009}. (Note that  due to the absorption of the outgoing pions these form factors are complex.)  The shape of the strong cross section $ d\sigma_S/d \Omega$ is determined by the dependence of the absolute value of the strong form factor $\mid{F_S(Q)}\mid$ and the $\sin(\theta_{\pi})^2$ factor. $\sigma_N$ is the the non-spin flip part of the neutral pion photoproduction  cross section on the nucleon. The order of magnitude of this term is  $\sigma_N \simeq 100 k^{2} \mu b$ where the photon energy k is in GeV\cite{Browman:1974}. The Primakoff cross section can be easily shown to be equal to Eq.\ref{eq:Gourdin} with $\beta$ = 1 which is appropriate for the high energy limit.
The experimental results are fit with the theoretical cross sections with four free parameters $\Gamma_{\gamma \gamma}, C_{S}, C_{inc},\phi$. The fitting parameter $C_{inc}$, which is not shown in Eq. \ref{eq:diff_cross}, is introduced to vary the magnitude of the theoretical incoherent cross sections. This latter contribution is small.

\begin{figure}
\begin{center}
\epsfig{figure=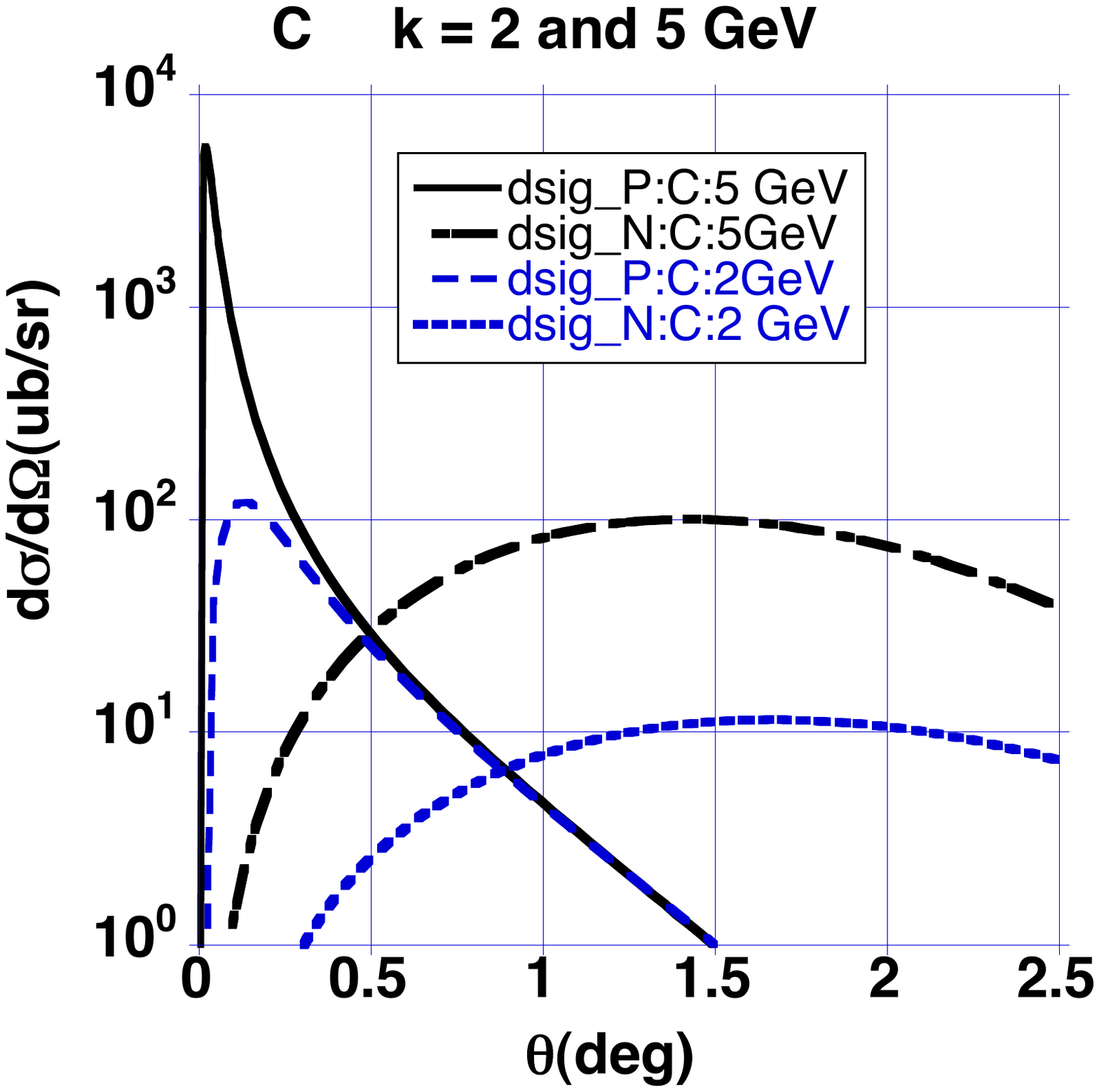,width=6.75cm,height=11cm}
\epsfig{figure=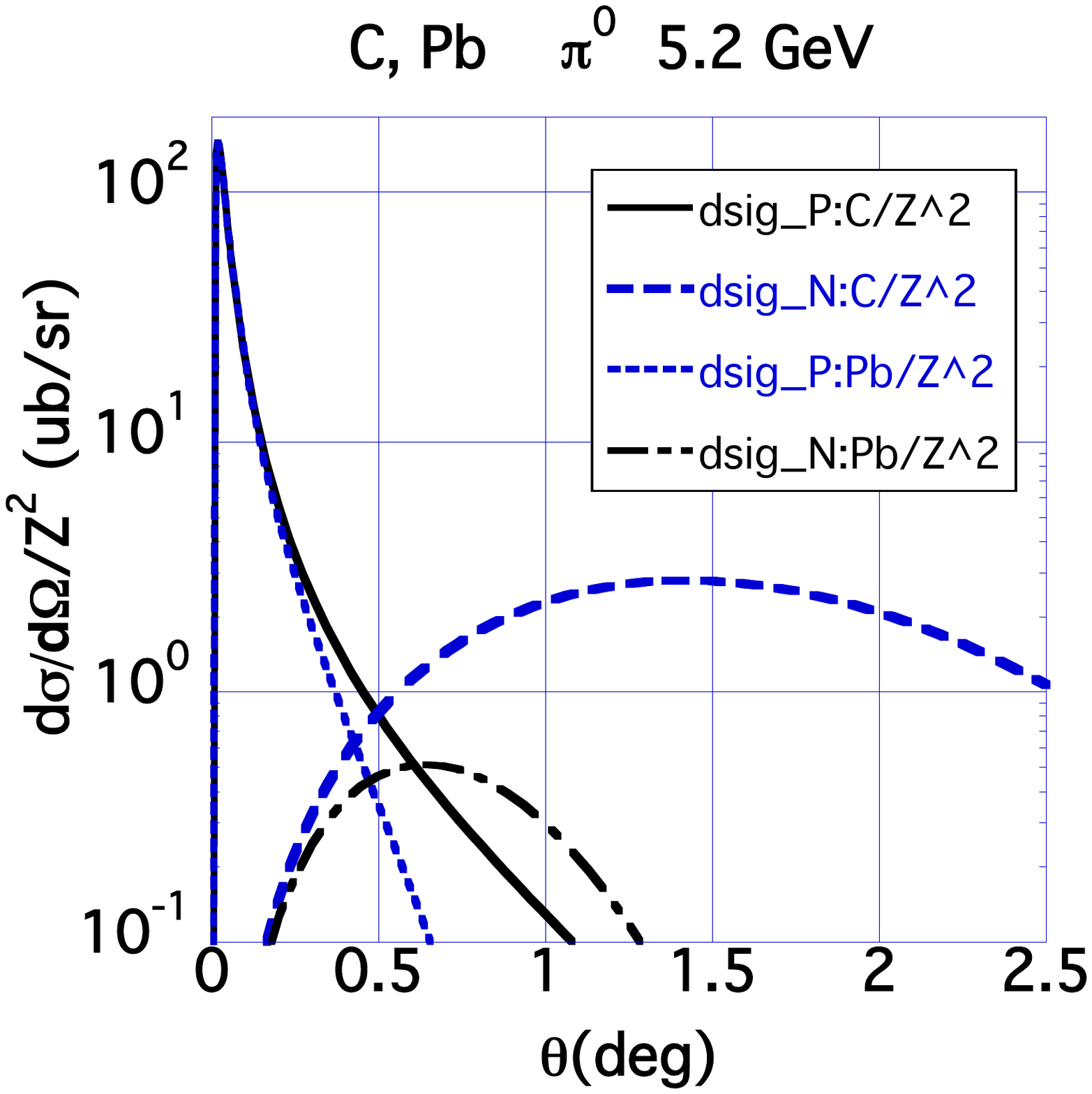,width=6.75cm,height=11cm}
\end{center}
\caption{Left figure: The Primakoff and nuclear coherent cross sections for C for $\pi^{0}$ production for photon energies of 2 and 5 GeV. Right figure: The Primakoff and  nuclear coherent cross sections for $\pi^{0}$ production divided by $Z^{2}$ for C and Pb at a photon energy of 5.2 GeV.   For these figures the parameters for nuclear coherent production are taken from the Cornell experiment\cite{Browman:1974}.}\label{fig:P-sig}
\end{figure}

The general features of the cross section for $\pi^{0}$ production are shown in Fig. \ref{fig:P-sig}. Here the contribution of the Primakoff, nuclear coherent, and their interference cross sections are shown (the small contribution of the nuclear incoherent cross section is not given). The figure demonstrates the dramatic increase in the cross section as the photon energy is increased and also indicates how the Primakoff cross section increases relative to the nuclear coherent cross section for a heavy nucleus relative to a lighter one.  The reason for this increase is nuclear absorption of the outgoing neutral pions.  If this effect were absent the nuclear coherent cross section would scale as $A^{2}$, where A is the nuclear mass number.  However, for strong nuclear absorption of the outgoing pions the cross section only increases as $A^{2/3}$, which is close to the actual physical case.

The first Primakoff experiments was performed at Frascati in 1965\cite{Bellettini:1965} using bremsstrahlung photon beams with end point energy $E_{0} \simeq$ 1 GeV and a Pb target.  The experiment was then repeated at DESY in 1970 by a similar group with $E_{0} =1.5$ on 2.0 GeV and C, Zn, Ag, and Pb targets\cite{Bellettini:1970}. Another experiment was performed at Tomsk in 1970 with $E_{0} =1.1$ GeV and a Pb target\cite{Kryshkin:1970}.  These experiments were analyzed with the early nuclear coherent calculations of Morpurgo\cite{Morpurgo:1964}.  This calculation was a significant advance over previous ones which neglected final state interactions of the outgoing pions and explained the main features of the nuclear coherent photoproduction, which included absorption of the outgoing pions in the nuclear medium. However, the further  approximation of a uniform nuclear density was made, which limited the accuracy of these calculations. In addition it was later shown that they did not include the effect of rescattering of the outgoing pions back to small angles\cite{Faldt:1972}.  In addition, the phase of the interference amplitude was assumed to be angle-independent.  This phase is important since this amplitude is the only correction that must be applied to the small angle cross section in an accurate determination of the $\Gamma( \pi^{0} \rightarrow \gamma \gamma)$ width.  We therefore conclude that, due to the low energy of the first three Primakoff experiments and the use of the Morpurgo calculation\cite{Morpurgo:1964}, these experiments are not sufficiently precise to include in a modern average of the $\pi^{0}$ lifetime.  This conclusion differs from the 2010 particle data book average\cite{PDB} which includes both the DESY\cite{Bellettini:1970} and Tomsk\cite{Kryshkin:1970} measurements. Following our recommendation we expect that the 2012 particle data book average will not include these measurements\cite{PDG:private}.

The last of the early Primakoff experiments was performed at Cornell using bremsstrahlung beams of endpoint energy $E_{0} =4.4$ and 6.6 GeV with Be, Al, Cu, Ag, and U targets\cite{Browman:1974}.  This experiment had the advantages of higher beam energies and also utilized the improved calculations of F\"aldt\cite{Faldt:1972}.  This  was the first to include the effect of rescattering of the outgoing pions back to small angles and also included the real part of pion rescattering in the final state. These effects lead to complex  nuclear form factors $F_{e.m.}(Q), F_{S}(Q)$ of Eq. \ref{eq:diff_cross} which makes the relative phase of the Primakoff-nuclear interference cross section angle-dependent.  F\"aldt also utilized a more modern non-uniform nuclear matter distribution.  All of these theoretical advances were utilized by the Cornell group and their
result for the $\pi^{0}\rightarrow \gamma \gamma$ decay width is $\Gamma= 7.92
\pm0.42$ eV(5.4\%)\cite{Browman:1974}. Taking the branching ratio of 1.2\% of the Dalitz decay $\pi^{0} \rightarrow e^{+}e^{-}\gamma$ into account yields $\tau(\pi^{0}) = 0.821(0.043) \times 10^{-16}$ s.  The error in the Cornell experiment results from combining the spread in values obtained using several different kinematic conditions with  the
systematic uncertainty. Contributions to the latter are estimated for the uncertainty in the nuclear shape parameters, the outgoing pion-nucleon cross sections, accelerator energy, beam luminosity and for the maximum opening
angle cut. At this level of accuracy, it is worth noting that it was assumed that the incoherent cross section is independent of $\theta(\pi^{0})$.  This is contrary to the previous assumption, which reduced this contribution at small angles due to Pauli blocking.  Since its magnitude is determined at larger angles where it becomes relatively more important, this might lead to a larger contribution under the small angle Primakoff peak and could in turn lead to a small reduction of the reported value of $\Gamma(\pi^{0} \rightarrow \gamma \gamma)$.  It should also be pointed out, however, that the companion $\eta$  measurement\cite{Browman:1974a} deviates strongly from the average value of other experiments.
Taking these last factors into account suggests that the quoted error of 5.4\%  is possibly underestimated. However, in our opinion, this is the first modern measurement of the $\pi^{0}$ lifetime and should be included in a updated PDG average.


\subsection{Appendix: $\pi^{0} \rightarrow \gamma \gamma$ Opening Angle Distributions}

\begin{figure}
\begin{center}
\epsfig{figure=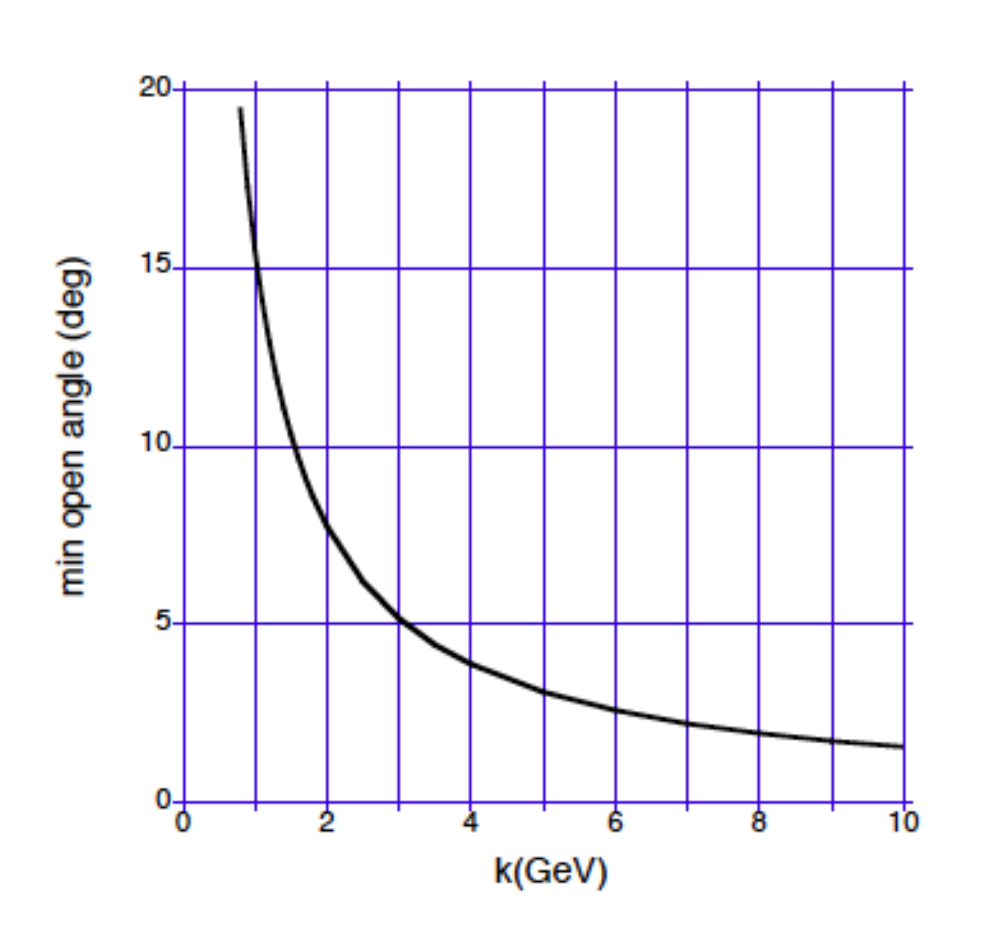,width=6.75cm,height=7.3cm}
\epsfig{figure=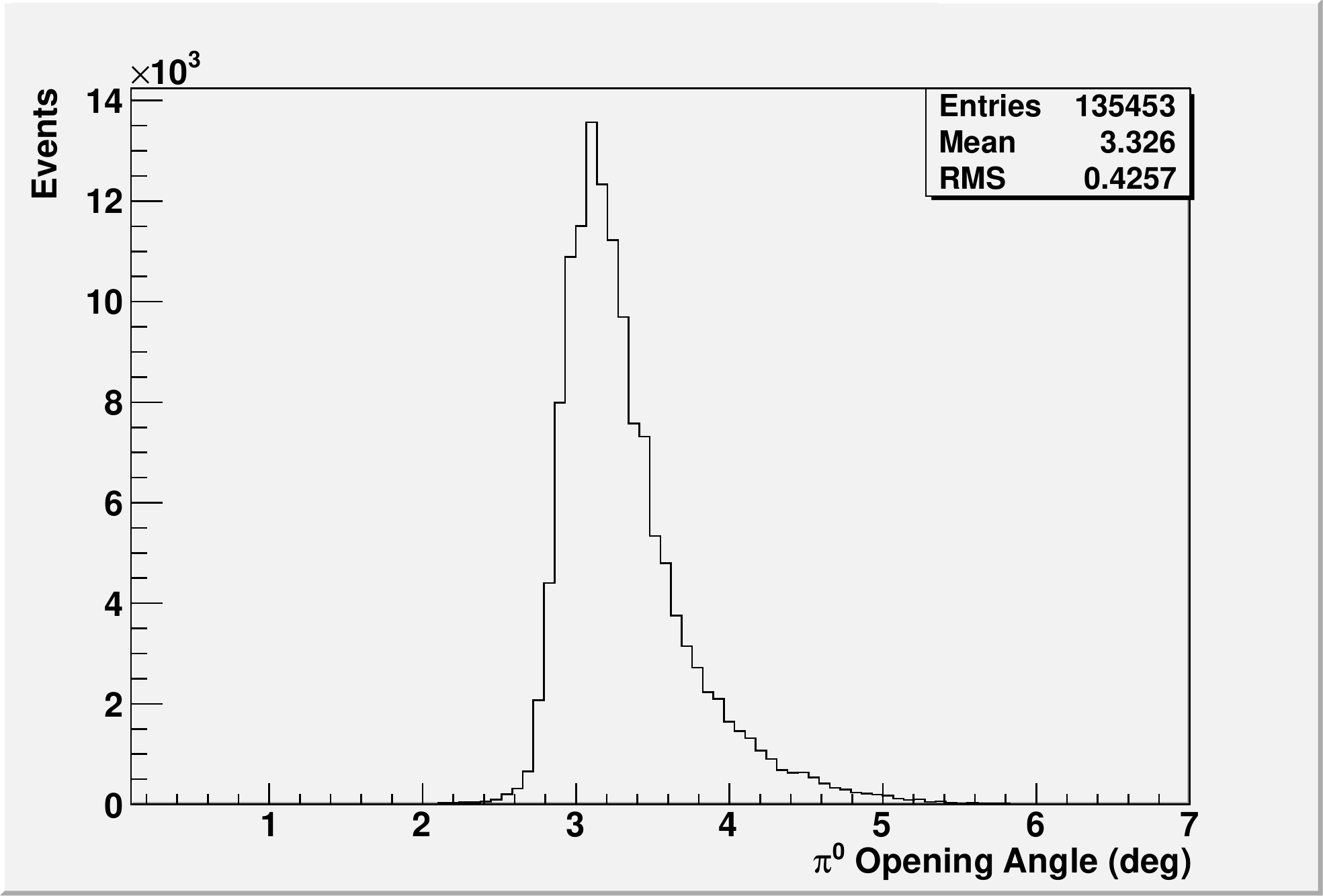,width=6.75cm,height=7cm}
\end{center}
\caption{Left figure: the laboratory opening angle for the $\pi^{0} \rightarrow \gamma \gamma$ versus the photon energy k. Right figure: Recent results for the $\pi^{0}$ opening angle\cite{McNulty:2011} from the PrimEx experiment\cite{Larin:2010} . }\label{fig:th-opening}
\end{figure}

In its rest frame, the $\pi^{0}$  decays into two gamma rays of energy $m_{\pi^{0}}/2$ with an isotropic angular distribution. In the laboratory frame, this distribution is boosted forward towards $\vec{p_{\pi^{0}}}$, with the angular distribution
\begin{eqnarray}
\frac{dN}{d\theta_{12}}  &\propto&
\frac{  \cos(\theta_{12}/2)}{\sin^{2}(\theta_{12}/2)[\sin^{2}(\theta_{12}/2)-\sin^{2}(\theta_{12-min}/2)}\nonumber \\
\sin(\theta_{12-min}/2) &=& m_{\pi^{0}}/E_{\pi^{0}}
  \label{eq:th-opening}
\end{eqnarray}
where $\theta_{12}$ is  the opening angle between the two gamma rays and $\theta_{12-min}$ is its minimum value. It can be seen that this results in a very sharply peaked angular distribution starting at $\theta_{12-min}$.  The singularity at that value is removed by the angular resolution of the detector. The values of $\theta_{12-min}$ versus photon energy are shown in Fig. \ref{fig:th-opening}, and it can be seen that a measurement of the opening angle  provided the early Primakoff method experimenters with a method to approximately determine the photon energy which was useful since they were employing bremsstrahlung beams.  The opening angle distribution\cite{McNulty:2011} from the recent PrimEx experiment\cite{Larin:2010} is also shown in Fig. \ref{fig:th-opening}. This was initiated by tagged photons between 4.9 and 5.5 GeV, so that $\theta_{12-min}$ was between 2.8 and 3.2 degrees.
The sharply peaked nature of the opening angle distribution is apparent from this figure.

%% file: subsection_direct.tex
\subsection{The Direct Measurement} \label{subsection:direct}

The most precisely determined lifetime measurement reported in the 2011 particle data book was performed at CERN in 1985\cite{Atherton:1985}.  This was a direct measurement of the $\pi^{0}$ decay distance at higher energies than the original 1963 direct measurement discussed in Sec.\ref{early-exp}\cite{vonDardel:1963}. In the 1985 experiment neutral
pions were produced by 450 GeV/c protons from the CERN SPS(Super Proton Synchrotron) incident on a tungsten foil\cite{Atherton:1985}.  The measurement consisted of two parts: the first was a precise determination of the mean decay length; the second  was an estimate of the $\pi^{0}$ momentum spectrum.

\begin{figure}
\epsfig{figure=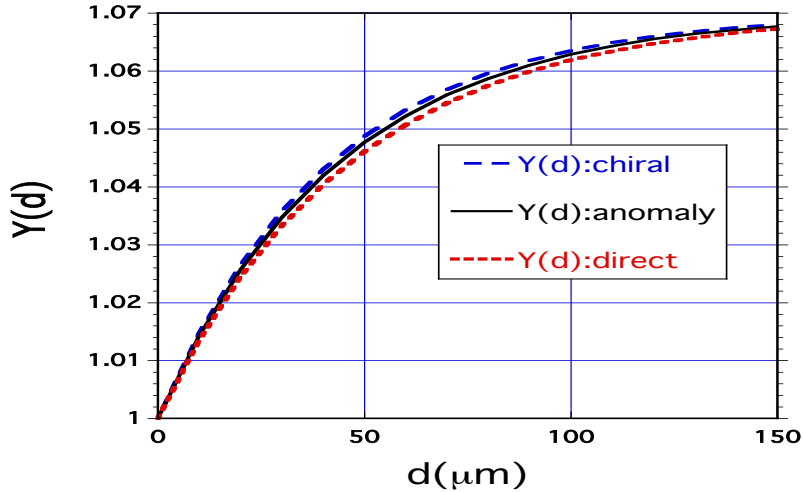,width=14cm,height=12cm}
\caption{Yield versus the separation of the two plates in the direct experiment at CERN\cite{Atherton:1985}. The three curves are for the values of $\Gamma(\pi^{0} \rightarrow \gamma \gamma)$ which correspond to the result of the CERN experiment\cite{Atherton:1985} and the predictions of the LO axial anomaly and HO chiral calculations as shown in Fig. \ref{fig:width}. The yield is relative to the direct production of the 150 GeV positrons (which is independent of the plate separation d).  \label{fig:Y-d}}
\end{figure}

To measure the mean decay length, a production target of  70 $\mu$m W  was placed in the 450 GeV/c proton beam. A second foil, placed at a variable distance d from the first, was used to convert some of the  gamma rays from the $\pi^{0}$ decay into electron, positron pairs. Downstream a magnetic spectrometer system measured the yield Y(d) of positrons of momentum 150 GeV/c. To illustrate the sensitivity of this experiment to the $\pi^{0}$ lifetime we calculated Y(d) for mean momentum $ <p_{\pi}> \simeq$ 235 GeV, the mean $\pi^{0}$ momentum of the CERN experiment\cite{Atherton:1985}.  At this momentum the relativistic boost factor $\gamma \simeq 1700$ and the mean decay distance predicted by the LO axial anomaly\cite{bej,sdl} is $43.8\,\,\mu$m.  The result is shown in Fig. \ref{fig:Y-d} to show its general shape. To illustrate the sensitivity curves are also shown for the chiral prediction\cite{bgh,kam,anm} and for the result of the direct experiment, a spread of approximately 10\% in lifetime. It can be seen that the sensitivity is greatest in the region of $\simeq$ 50$ \mu$m. These curves show that the experiment had to be performed with excellent accuracy, which it was. The reported experimental result result was the ratio
\begin{eqnarray}
R= [Y(250\,\,\mu) -Y(45 \mu)]/[Y(250\,\,\mu) - Y(0)] = 0.3787 \pm 0.0078(2.1\%).
\end{eqnarray}
A mean decay length $\lambda = 46.5 \pm 1.0\,\, \mu$m(2.1\%) was obtained\cite{Atherton:1985}.

\begin{figure}
\epsfig{figure=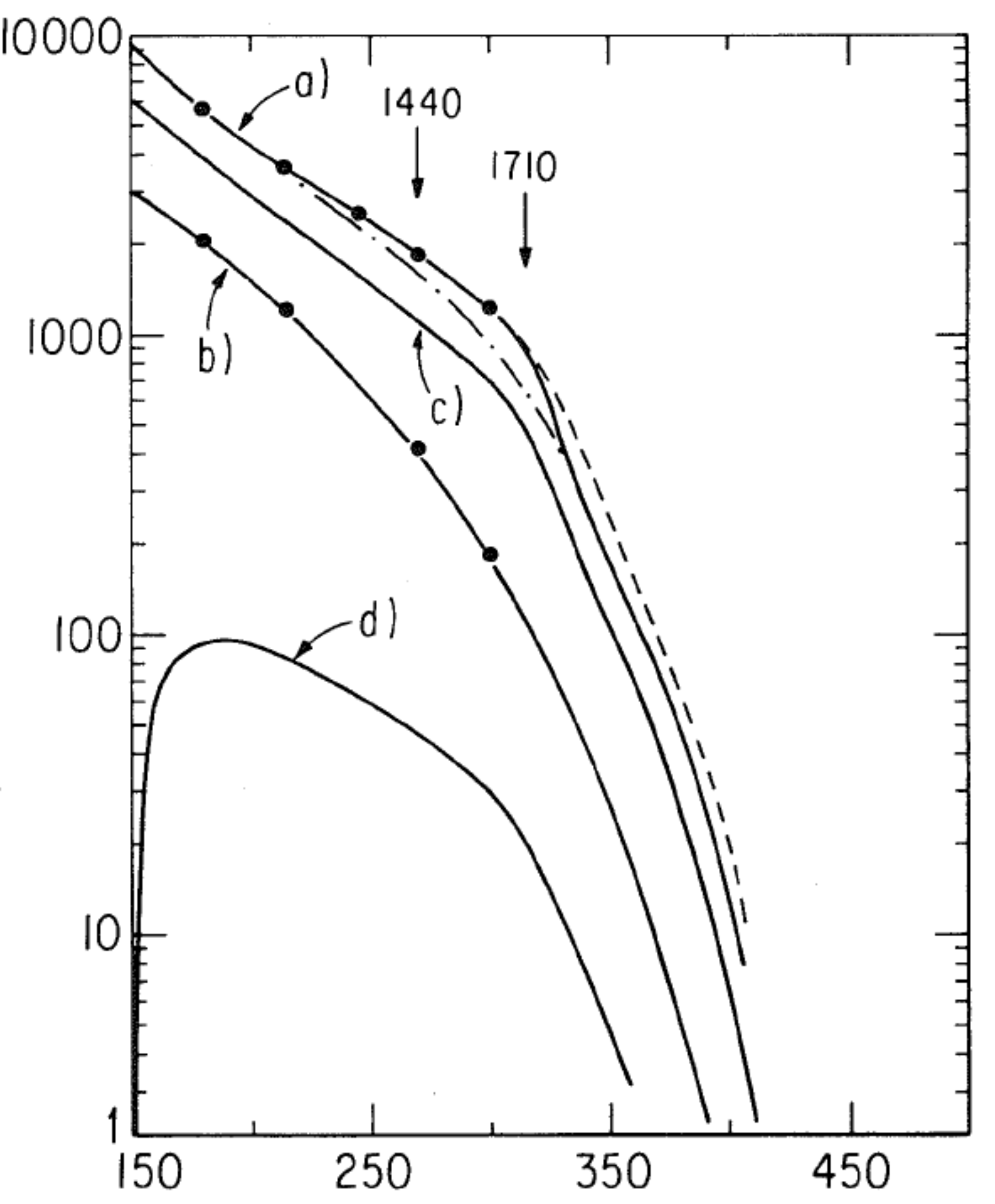,width=8cm,height=8cm}
\caption{The momentum distributions of the direct experiment at CERN as a function of pion momentum in GeV/c: from \cite{Atherton:1985}. These are the spectra produced at $0^{0}$ by 450 GeV protons on Ta. The solid curves are the spectra used in the analysis.  (a) $\pi^{+}$ production spectrum: solid circles indicate measured points.  (b)$\pi^{-}$ production spectra; solid circles indicate measured points.  (c) $\pi^{0}$ production spectra for $\kappa$ = 0.50(see Eq. \ref{eq:sig_pi0}).  (d) $\pi^{0}$ spectrum which gives 150 GeV positrons. The dot-dashed curve and the dashed curves represent variations used of the $\pi^{+}$ spectrum used to estimate the systematic error. Nucleon isobars give peaks at the indicated momenta.
 \label{fig:N_p}}
\end{figure}

To infer the $\pi^{0}$ momentum spectrum $N(p_{\pi})$, measurements were made of the charged pion momentum distributions $N(p_{\pi^{\pm}})$.  In terms of these results the $\pi^{0}$ momentum distribution was assumed to be
\begin{eqnarray}
 N(p_{\pi}) = \kappa N(p_{\pi^{+}}) +(1-\kappa)N(p_{\pi^{-}}).
 \label{eq:sig_pi0}
 \end{eqnarray}
 Note that only the relative magnitudes of the momentum distributions are relevant for the $\pi^{0}$ lifetime determination.  For the lifetime and systematic error determination $\kappa$ was taken to lie in the range  $0.50\pm 0.25$ and to be momentum-independent. The momentum distributions from that experiment \cite{Atherton:1985} are shown in Fig. \ref{fig:N_p}.  It can be seen that the measurements that were made at CERN were all below $p_{\pi^{\pm}}$ = 300 GeV/c, the  upper limit of the  magnetic spectrometer\cite{Millikan:1985}.  For higher momenta other experiments and estimates were used\cite{Atherton:1985,Millikan:1985}.  The CERN direct experiment utilized forward-produced neutral pions (transverse momentum $p_{t} \simeq 0$). Due to a paucity of pion production data, the extrapolations utilized experiments performed at larger values of the transverse momenta for which the contributions of excited nucleon resonances did not contribute. These resonances could contribute in the vicinity of the arrows shown in Fig. \ref{fig:N_p}.  The extrapolations also utilized experiments performed on different targets and energies using Feynman scaling; they were checked as carefully as possible by comparison with the CERN data\cite{Millikan:1985}.

From the $\pi^{0}$ momentum distribution $N(p_{\pi})$ the production probability of 150 GeV/c positrons $N_{e}(p_{\pi})$ must be obtained.
\begin{eqnarray}
N_{e}(p_{\pi}) = f(p_{e}, p_{\pi}) N(p_{\pi})
\label{eq:N(e)}
\end{eqnarray}
where $ f(p_{e}, p_{\pi})$ is the probability that the gamma rays from the $\pi^{0} \rightarrow \gamma \gamma$ decay of a pion with momentum $p_{\pi}$ produce a positron of momentum $p_{e}$=150 GeV/c. This function is evaluated from an integration over intermediate photon momenta and is given in Ref. \cite{Atherton:1985}.
To a good approximation the results for the $\pi^{0}$ lifetime depend only on the average $\pi^{0}$ momentum
\begin{eqnarray}
<p_{\pi}> =\frac{ \int d p_{\pi} ~p_{\pi}~ N_{e}(p_{\pi})}
{ \int d p_{\pi}~ N_{e}(p_{\pi})}
\label{eq:p_pi_av}
\end{eqnarray}
With this assumption the mean decay length of the $\pi^{0}$ is
\begin{eqnarray}
\lambda = \frac{<p_{\pi}> c \tau(\pi^{0})}{m_{\pi^{0}}}
\label{eq:lambda}
\end{eqnarray}
In the evaluation of the experimental results the full momentum distribution was used\cite{Atherton:1985}.  The use of the average $\pi^{0}$  momentum $<p_{\pi}> \simeq$ 235 GeV/c  reduced the final lifetime result by 0.8\% compared to the use the full momentum spectrum\cite{Atherton:1985,Millikan:1985}.

\begin{figure}
\epsfig{figure=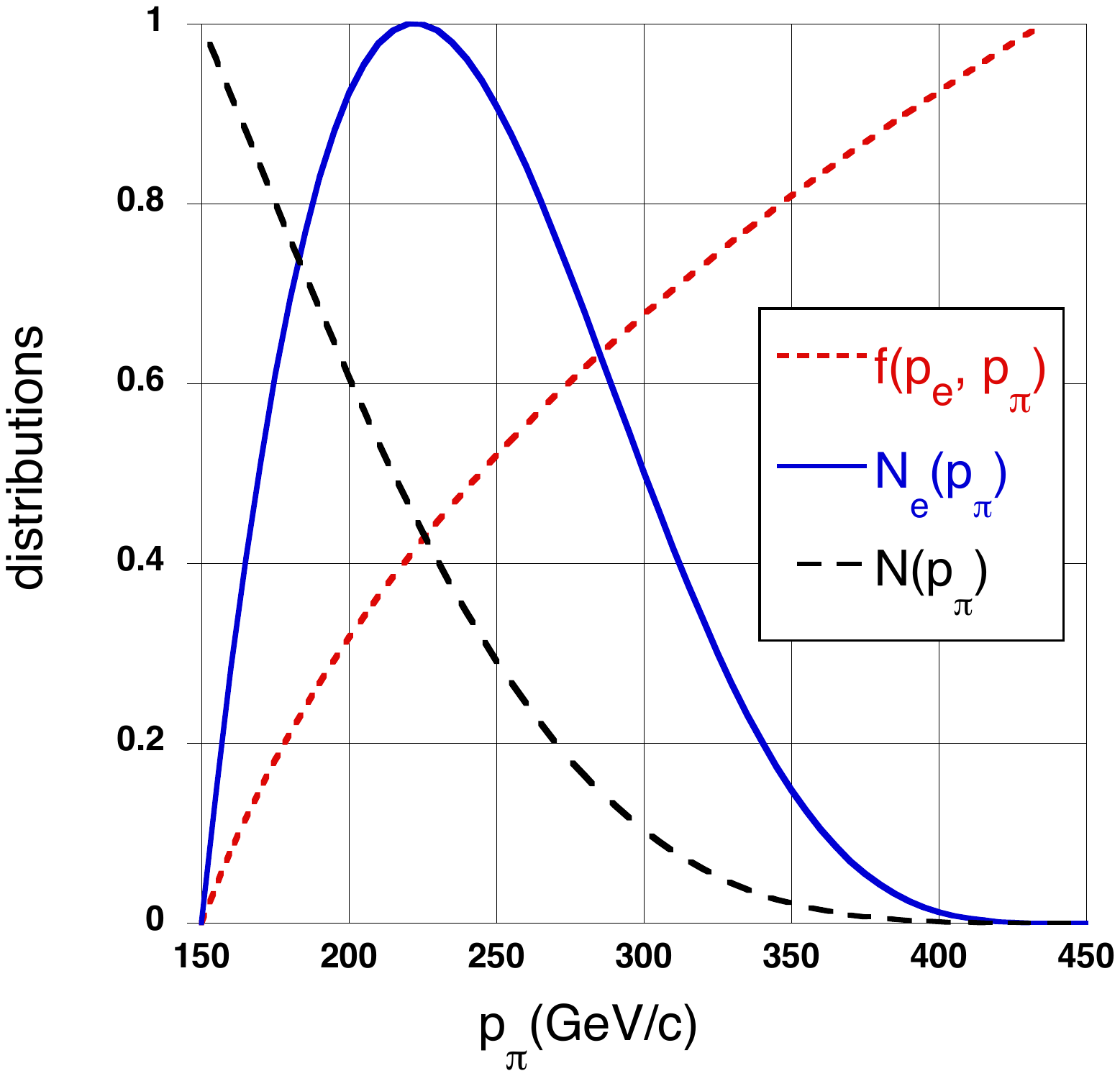, width=16cm,height=16cm}
\caption{The $\pi^{0}$ and $e^{+}$ probability distributions for the direct experiment at CERN\cite{Atherton:1985}.
The (black)dashed curve is the $\pi^{0}$ distribution(normalized to unity at  $p_{\pi} = $150 GeV/c),
the (red) dashed  curve is $ f(p_{e},p_{\pi})$
(which tends to unity as $p_{\pi} \rightarrow$ 450 GeV),
the(blue) solid curve is $N_{e}(p_{\pi}) =f(p_{e},p_{\pi})~N(p_{\pi})$(normalized to unity at its maximum).
See Eq. \ref{eq:N(e)} for the definitions and the text for discussion.
 \label{fig:dist}}
\end{figure}

To obtain a physical understanding of the pion momentum-dependent quantities, we made a graph of the individual ingredients in Eqs. \ref{eq:N(e)} and \ref{eq:p_pi_av}. The results are plotted in
Fig. \ref{fig:dist}, which illustrates the sensitivity to the high momentum components of the $\pi^{0}$ momentum spectrum $N(p_{\pi})$. The reason is that $ f(p_{e}, p_{\pi})$  increases with  $p_{\pi}$ even though $N(p_{\pi})$ is falling. The product $N_{e}(p_{\pi})$ peaks near $<p_{\pi}> \simeq$ 235 GeV/c.

The quoted experimental result is $\tau(\pi^{0}) = (0.892 \pm 0.022 \pm 0.017) \times 10^{-16}$ sec, where the first error(2.5\%) is statistical and the second(1.9\%) is systematic \cite{Atherton:1985}.  Adding them in quadrature gives a total uncertainty of 3.1\%.  Knowing the lifetime, the width $ \Gamma(\pi^{0})  = \hbar / \tau(\pi^{0})$ can be obtained, yielding $\Gamma(\pi^{0} \rightarrow \gamma \gamma ) = (7.25 \pm 0.22 \pm 0.18)$ eV \footnote{For this article we shall quote the value of $\Gamma(\pi^{0} \rightarrow \gamma \gamma ) = BR(\pi^{0} \rightarrow \gamma \gamma) \Gamma(\pi^{0})$, where  $BR(\pi^{0} \rightarrow \gamma \gamma)$ = 0.98798(0.032)\cite{PDB}.}. This result is  10.3\% ( 3$\sigma$) below the HO prediction of the axial anomaly with chiral corrections.
In this experiment the momentum distribution of the neutral pions was estimated to be the average of the $\pi^{+}$ and $\pi^{-}$ momentum distributions as indicated by lower energy experiments.  As discussed above, the quoted systematic error takes the variation in the relative weighting of these two cross sections (see Eq. \ref{eq:sig_pi0}), and their variation in shape, into account\cite{Atherton:1985}.  However, one cannot be sure of the accuracy of these assumptions until the $\pi^{0}$ momentum spectrum is explicitly measured.

It is of interest to discuss how seriously to take the discrepancy between the direct measurement and the theoretical predictions, particularly those involving the HO chiral corrections to the axial anomaly.  The measurement of the decay distance is very strong, so any problems most likely are in the $\pi^{0}$ momentum distribution which was not directly measured.  Based on Eq. \ref{eq:lambda}, agreement with the HO predictions of the axial anomaly  plus chiral corrections would require an increase of the average $\pi^{0}$ momentum $<p_{\pi}>$ of 10.3\%.  This seems to be a rather large effect in view of the effort put into their determination\cite{Atherton:1985, Millikan:1985}.  Nevertheless, considering how fundamental the prediction of $\Gamma(\pi^{0} \rightarrow \gamma \gamma)$ is, a measurement of the $\pi^{0}$ momentum spectrum $N(p_{\pi})$ would be valuable. Fortunately the Compass experiment at CERN is looking into this possibility\cite{Friedrich:2011}.

%% file: subsection-ee.tex
\subsection{$e^{+}e^{-}$ Colliding Beam Measurements}\label{subsection:ee}

Another technique uses the $e^{+}e^{-} \rightarrow e^{+}e^{-}\gamma^{*}\gamma^{*} \rightarrow e^{+}e^{-}P \rightarrow
e^{+}e^{-}\gamma\gamma$ reaction with colliding beams where $P=
\pi,\eta,\eta^{'}$, and $\gamma^{*}$ are almost real photons because the final state leptons scatter at small angles and are not detected.  An experiment was performed at the DORIS II ring at DESY\cite{Williams:1988} which detected the $P \rightarrow \gamma \gamma$ decays in a crystal ball gamma ray detector which covered 93\% of $4\pi$ solid angle. Cuts were made on the invariant mass of the two gamma rays at the $ \pi,\eta,\eta^{'}$ masses, and the luminosity was measured via elastic $e^{+}e^{-}$ scattering.  The efficiency was determined primarily by Monte Carlo calculations. The backgrounds due to beam, residual gas interactions and cosmic ray events were eliminated by separate measurements and also by
stringent cuts on the transverse momenta of the produced meson, $\Sigma | p_{t} | \leq M_{\gamma,\gamma}/10$.

This method is a generalization of the Primakoff method in which the cross section for the $\gamma + \gamma^{*} \rightarrow P$ transition is measured. It is unique in that it is based on purely electromagnetic physics, whereas the Primakoff effect involves a target nucleus.  Furthermore it is the only experiment which, when carried out at sufficiently high energies, can measure the widths of all pseudoscalar mesons simultaneously.  In this case both of the incident photons are slightly off-shell ({\it i.e.}, the four momentum transfer $q^{2} < 0$), whereas for the Primakoff effect this is true for only one.

For $e^{+}e^{-}$ collisions the production cross section calculated from QED is
\begin{eqnarray}
\sigma_{\gamma \gamma \rightarrow P}(q_{1}^{2},q_{2}^{2})  = \Gamma(P \rightarrow \gamma \gamma)~ 16 \pi^{2} \delta((q_{1}+ q_{2})^{2} - m_{P}^2) \frac{ | \vec{q}|}{m_{P}^{2}}
F^{2}( q_{1}^{2},q_{2}^{2})\label{eq:sig_ee}
\end{eqnarray}
where $m_{P}$ is the mass of the produced pseudoscalar meson, ~$ \vec{q}$ is the three momentum transfer of either of the two virtual photons, and $ F( q_{1}^{2},q_{2}^{2})$ is the form factor for the $\gamma^{*}(q_{1}^{2}) + \gamma^{*}(q_{2}^{2}) \rightarrow P$ vertex which is not specified by QED. To estimate how much $F$ deviates from unity the vector dominance form $F( q_{1}^{2},q_{2}^{2}) = (1-q_{1}^{2}/m_{\rho}^{2})^{-1}(1-q_{2}^{2}/m_{\rho}^{2})^{-1}$, where $m_{\rho}$ is the mass of the $\rho$ meson, was used.  It was estimated by Monte Carlo calculations that the stringent cut on $\Sigma | p_{t} |$ restricts $< -q^{2}> = 10\,\,{\rm MeV}^{2}$ for $\pi^{0}$ production (it is larger for $\eta, \eta^{'}$ production) so that the form factor deviation from unity is negligible, as is the case for the Primakoff effect.

The result for the $\pi^{0} \rightarrow \gamma \gamma$ decay width is $\Gamma = (7.7 \pm0.5 \pm0.5)$ eV\cite{Williams:1988}. It is important to have a modern version of this purely electromagnetic determination of  $\Gamma(\pi^{0} \rightarrow \gamma \gamma)$ with $\simeq$ 1\% errors. In this connection it is also of great interest to perform measurements of the $\eta$ and $\eta^{'}$ lifetimes with comparable accuracy.  Fortunately there are groups at the Beijing Electron Synchrotron (BES)\cite{Denig:2011} and Frascati\cite{Babusci:2011} and  looking into this possibility. At BES the beam energy is higher so that all of the $\pi,\eta,\eta^{'}$ mesons can be studied at the same time. At Frascati the energy is more limited so that the only the $\pi$ and $\eta$ can be studied (the latter with reduced cross section). Monte Carlo simulations have  been prepared\cite{Czyz:2012} and plans have been made for both single and double electron tagging. An estimated statistical accuracy $\sim$ 1\% seems feasable for the $\pi^{0}$ width\cite{Babusci:2011}. 

%% file: section_PrimEx.tex
\section{PrimEx Experiment \label{sec:exp}}

After a gap of three decades an accurate measurement of the $\pi^{0}$ lifetime, performed using the Primakoff effect, was recently published\cite{Larin:2010}. This experiment benefited from the huge improvement in accelerator and detector technology during this period.  The CW structure of the electron beam allowed for the first time the use of tagged photons, through which the photon energy was determined to $\sim$ 0.1\% for each event.  The small beam emittance allowed placing detectors very close to the beam, while the huge improvement in detector technology resulted in a far superior energy and angular resolution of the decay photons.  The experiment was performed by the PrimEx collaboration in Hall B of Jefferson Lab. The Primakoff effect itself and the previous experiments have been described in subsections \ref{subsection:Primakoff} so that only the specific improvements for the PrimEx experiment will be presented below.

In addition to experimental advances, the improvements over previous Primakoff measurements also include advances in the theoretical tools used to extract the results.  Specifically, the PrimEx experiment utilized recent calculations of the nuclear coherent amplitude\cite{Gevorkyan:2009} which represent the strong amplitude $T_{s}$ of Eq. \ref{eq:diff_cross}  as
\begin{eqnarray}
T_{s} = ( \vec{h}\cdot \vec{q_{t}}) \Phi(0) (F_{s}(Q) -w F_{I}(Q) ) \nonumber \\
{\rm with}\quad\vec{h} = \vec{k} \times \vec{\epsilon}/k,\quad
Q^{2} = q_{t}^{2} + Q_{min}^{2}  \nonumber \\
 q_{t} = p_{\pi^{0}} \sin (\theta_{\pi^{0}}), \quad
 Q_{min} \simeq m_{\pi ^{0}}^{2}/ k
\label{eq: Ts}
\end{eqnarray}
where $\vec{k}$ is the incident photon momentum, $\vec{\epsilon}$ its polarization vector, and $\Phi(0) \neq 0$ the forward scattering non-spin-flip amplitude for photo-pion production on the nucleon. The transverse momentum transfer $q_{t}$ insures that the coherent cross section will have the required $\sin^2\theta$ dependence.  $F_{s}(Q)$ is the strong form factor, which takes the final state pion interaction into account using the Glauber approach and includes the F\"aldt correction\cite{Faldt:1972}, which describes the rescattering of the outgoing pions as well as their absorption. For light nuclei such as $^{12}$C the next order in the multiple scattering series was also included.  Such effects were {\it not} taken into account in the first Primakoff experiments\cite{Bellettini:1965, Bellettini:1970,Kryshkin:1970}.  The second term in parentheses in Eq. \ref{eq: Ts} accounts for photon shadowing in the initial state.  This is a two-step process in which the incoming photon produces a vector meson (primarily the $\rho$ ) which in turn produces the emerging $\pi^{0}$.  $F_{I}$ is the associated form factor taking the final state interaction into account and  the photon shadowing parameter $w$ lies between 0 and 1 (none to full shadowing).  Following the empirical evidence the PrimEx analysis assumed $w$ = 0.25 [see \cite{Gevorkyan:2009} for references and details]. As part of the systematic error estimate a range of $w$ from 0 to 0.5 was assumed\cite{Larin:2010}.  For the incoherent cross section two recent calculations\cite{Gevorkyan:2009a,Rodrigues:2010} were utilized. One used a multiple scattering approach using Glauber theory\cite{Gevorkyan:2009a}. The second used a cascade model approach\cite{Rodrigues:2010}. The nuclear incoherent cross section is small at small angles (the Primakoff region) and only plays a minor role in the extraction of $\Gamma(\pi^{0} \rightarrow \gamma \gamma)$ from the data.

In a recent theoretical development, photoproduction of $\pi^{0},\eta,\eta^{'}$ mesons has been developed in a unified field theoretical basis using photon and Regge exchange\cite{Mosel:2011}. Full vector dominance has been included. This has been done for the Primakoff {\it and} nuclear coherent (but not the incoherent) cross sections.  It is impressive that there is good agreement between the results of this calculation, without any empirical parameters, and the PrimEx data. In addition, there have been two other calculations of the coherent production of the pseudoscalars $\pi^{0},\eta,\eta^{'}$ from the proton\cite{Laget:2005, Sibirtsev:2009}. The second article\cite{Sibirtsev:2009} shows an extensive fit of the Regge parameters to the existing data base, which is primarily nuclear production. Both articles point out the possibility of performing future Primakoff effect measurements with a proton target. This is attractive for the $\eta$ and $\eta^{'}$ because the Primakoff cross section decreases rapidly with increasing mass of the produced meson, reducing the signal to background ratio (for more detail see sec. \ref{Section: Related-Measurements}). The proton target also has the advantage that the nuclear physics complications are not present. On the other hand one loses the factor of $Z^{2}$ advantage that the Primakoff reaction has. For each case a detailed analysis of the optimum targets and photon energies must made to determine the optimum choice of target.

\begin{figure}
\begin{center}
\includegraphics[height=0.8\textwidth,width=0.9\textwidth] {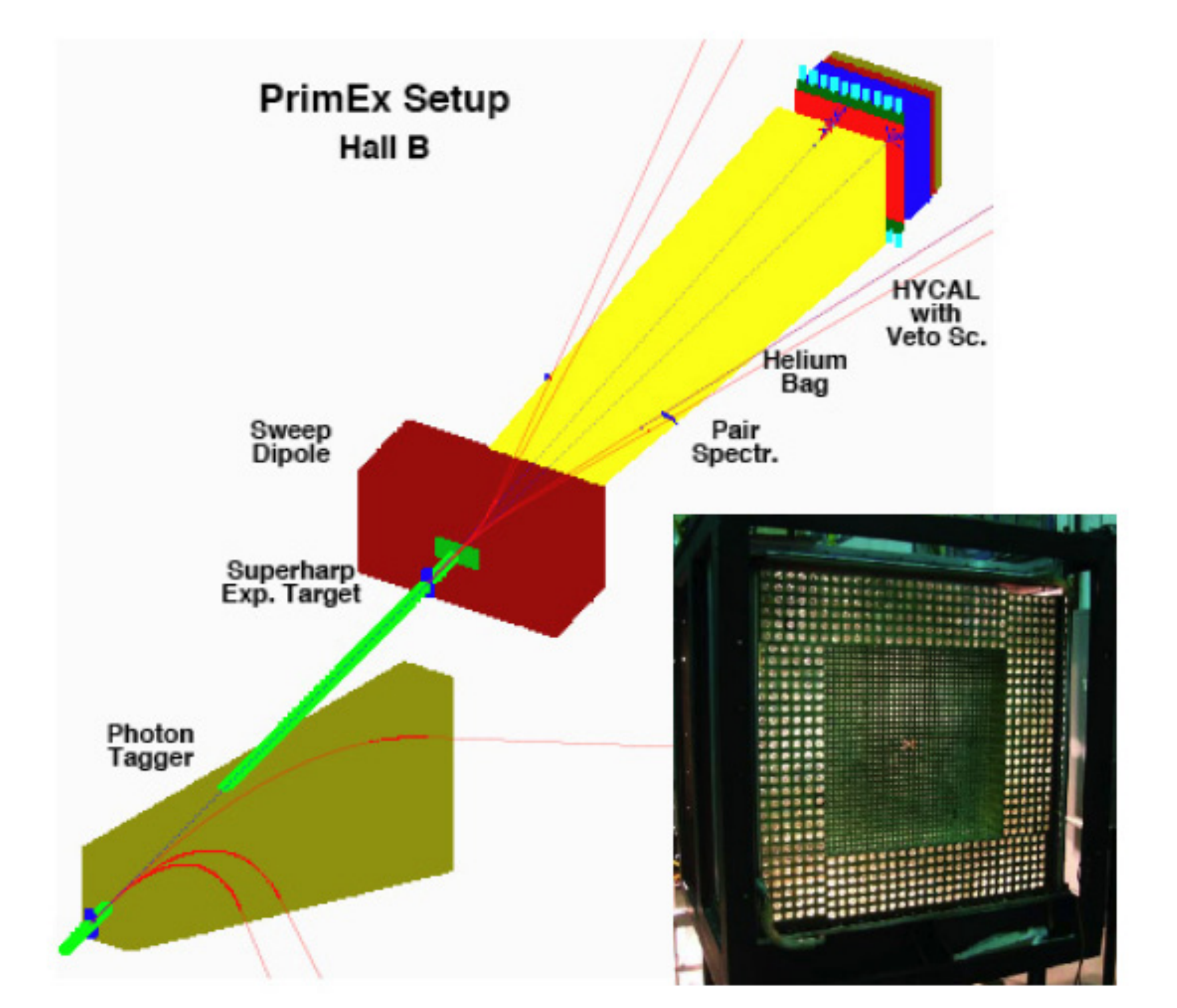}
\end{center}
\caption{ Schematic layout of the {\em PrimEx} experimental setup showing the incident electron beam, photon energy tagging system,  target, sweeping magnet, He bag, electron pair spectrometer, veto counter, and HYCAL detector  shown in more detail in the insert. It consists of an inner section of PbWO$_4$ crystals and an outer section of Pb-glass detectors.}
\label{fig:Layout}
\end{figure}

In the PrimEx experiment incident photons interact with the Coulomb field of a nucleus to produce $\pi^{0}$ mesons, which quickly decay into two photons and are detected in a forward calorimeter as shown schematically in Fig. \ref{fig:Layout}.
Tagged photons were used to measure the absolute cross section of small angle $\pi^{0}$ photoproduction from
C and Pb nuclei which make up the target.  The invariant mass, energy, and angle of the pions were  reconstructed by detecting the decay photons from the $\pi^{0} \rightarrow \gamma \gamma$ reaction in a forward calorimeter (HYCAL), which was constructed for this experiment.  A photograph of this  detector is shown in Fig. \ref{fig:Layout}.  The 1152 PbWO$_4$ crystals which formed the heart of the detector are 2 cm by 2 cm by 20 radiation lengths at a distance of 7.5 m from the target; the published results primarily came from these detectors. An energy resolution of $\simeq$ 1.8\% and angular resolution of $\simeq 0.02^{0}$ were achieved. There is an aperture of 2 by 2 crystals for the highly collimated  photon beam. The outer crystals in HYCAL are Pb-glass. The schematic diagram (Fig. \ref{fig:Layout}) does not show the shielding and the beam dump.  These and the clean electron beam were sufficient to allow for a low background experiment with a sensitive forward calorimeter.  The function of the He bag was to reduce the number of photons interacting between the target and the detector.  A plastic scintillator was placed in front of the HYCAL colorimeter to veto charged particles.  The magnet, which was placed directly behind the target, swept the produced electrons away from the detector. Detectors were placed behind the magnet to monitor the luminosity using pair production.  By turning the magnet to lower fields, measurements of $e^{+}e^{-}$ pair production were made using  plastic scintillators in coincidence mode.  By turning the sweeping magnet off, and again using the plastic scintillators in coincidence mode, Compton scattering cross sections were also measured.  To measure the tagging efficiency, HYCAL was moved out of the beam, and was replaced by a total absorption counter.  This replacement must be performed at very low currents.  The pair production monitors are linear in both the higher flux production runs as well as the low flux tagging efficiency runs and could be used to interpolate between them.  In order to measure the $\pi^{0}$ production cross sections, precisely measured\cite{Martel:2008} targets of C and Pb, approximately 5\% of a radiation length, were used.
Further details of the experiment can be found in the PrimEx publication\cite{Larin:2010}, on the PrimEx web site\cite{PrimEx} and in a recent review article by Miskimen\cite{Miskimen:2011}.

\begin{figure}
\begin{center}
\includegraphics[height=0.7\textwidth,width=0.8\textwidth] {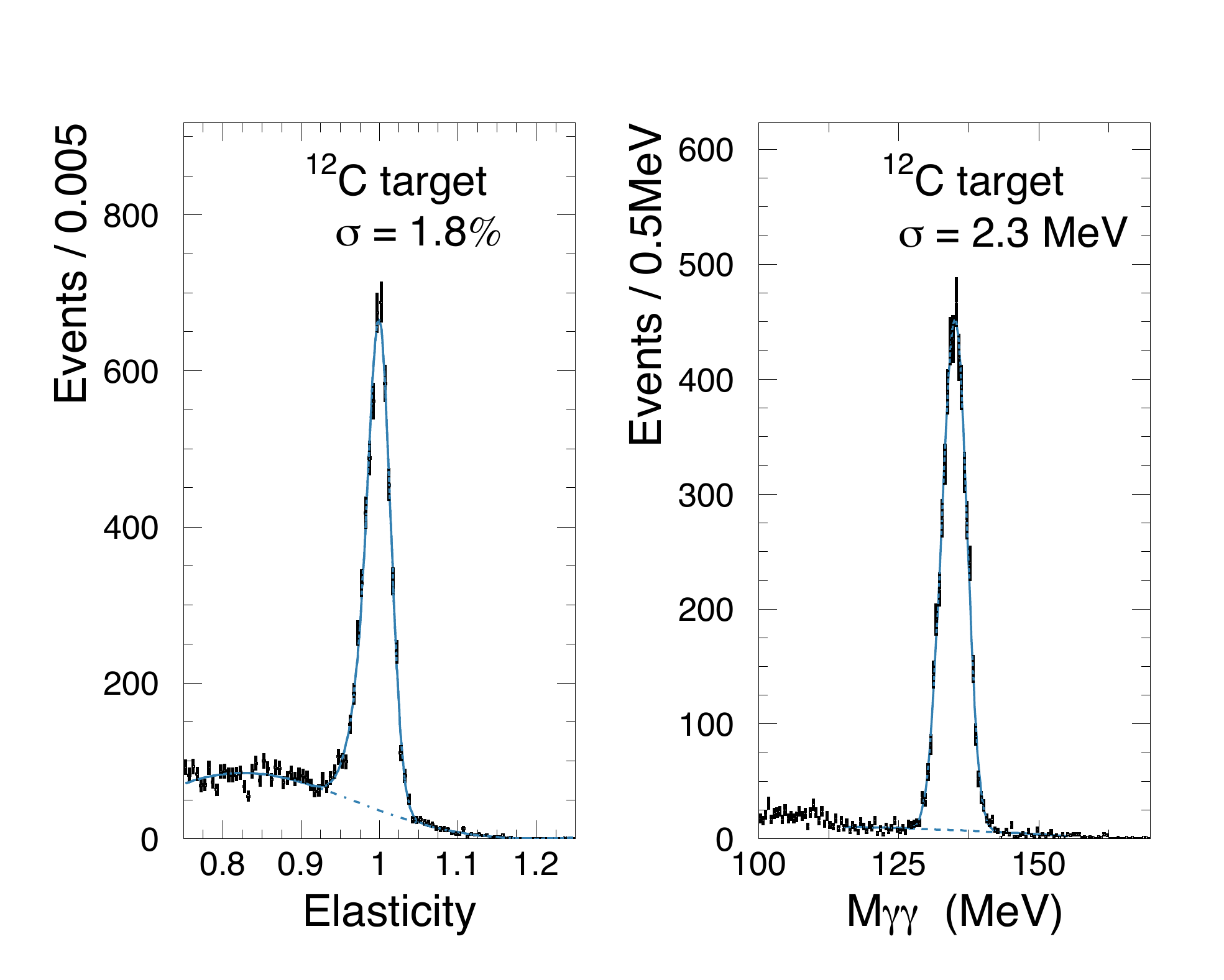}
\end{center}
\caption{Observed inelasticity (left) invariant mass(right) distributions:from \cite{Larin:2010}.The background fits are shown as dashed lines and the total fits as solid lines. See text for discussion.}
\label{fig:obs}
\end{figure}

To measure the $\gamma + A \rightarrow \pi^{0} + A $ reaction followed by the $\pi^{0} \rightarrow \gamma(p_{1}) \gamma (p_{2})$ decay, where $p_{1},p_{2}$ are the four-vectors of the decay photons, two cluster events are identified in the HYCAL detector.  The center of the hit distribution is obtained from the distribution of energies in the individual detector crystals, and the energy of each photon is determined from the sum of these energies. The invariant mass distribution $m_{\gamma \gamma}^{2} = (p_{1} + p_{2})^2 = 4  k_{1} k_{2} \sin^{2}(\theta_{12}/2)$ was obtained. It is important to determine whether the target nucleus is left in its ground state. This is measured by the elasticity = $ (k_{1}+k_{2})/k $, where k is the incident photon energy.  Typical results for these quantities in the Primakoff peak region are shown in Fig. \ref{fig:obs}.  At these small angles the two-photon final states are totally dominated by $\pi^{0}$ production as can be seen from the sharp peak for $m_{\gamma \gamma} = m_{\pi^{0}} \simeq $ 135 MeV.  The elasticity distribution shows that higher mass states are produced along with coherent $\pi^{0}$ production.  One important example of such a final state is $\omega$ meson photoproduction followed by the $\omega \rightarrow \pi^{0} \gamma$ decay. Another possible inelastic mechanism is nuclear excitation or quasi-free meson production. The inelastic background is subtracted empirically by fitting the inelastic data by empirical polynomial fits.  The good angular and energy resolution achieved by this apparatus are illustrated in Fig. \ref{fig:obs} for the invariant mass ($\simeq$ 1.2\%) which allowed identification of the produced $\pi^{0}$ mesons with a high signal/background ratio.  The energy of the emerging pions is also measured with good resolution ($\simeq$ 1.8\%) in HYCAL.  This resolution ($\simeq$ 90 MeV) does not  allow an experimental determination that the pion production is coherent, since it allows for the possibility of nuclear excitation.  However, at small angles this coherent production has been estimated to be small\cite{Gevorkyan:2009}. In addition any residual nuclear excitation has been empirically subtracted by extrapolating the inelastic background from the measured background at higher inelasticities, as shown in Fig. \ref{fig:obs}.

\begin{figure}
\begin{center}
\includegraphics[height=0.7\textwidth,width=0.45\textwidth] {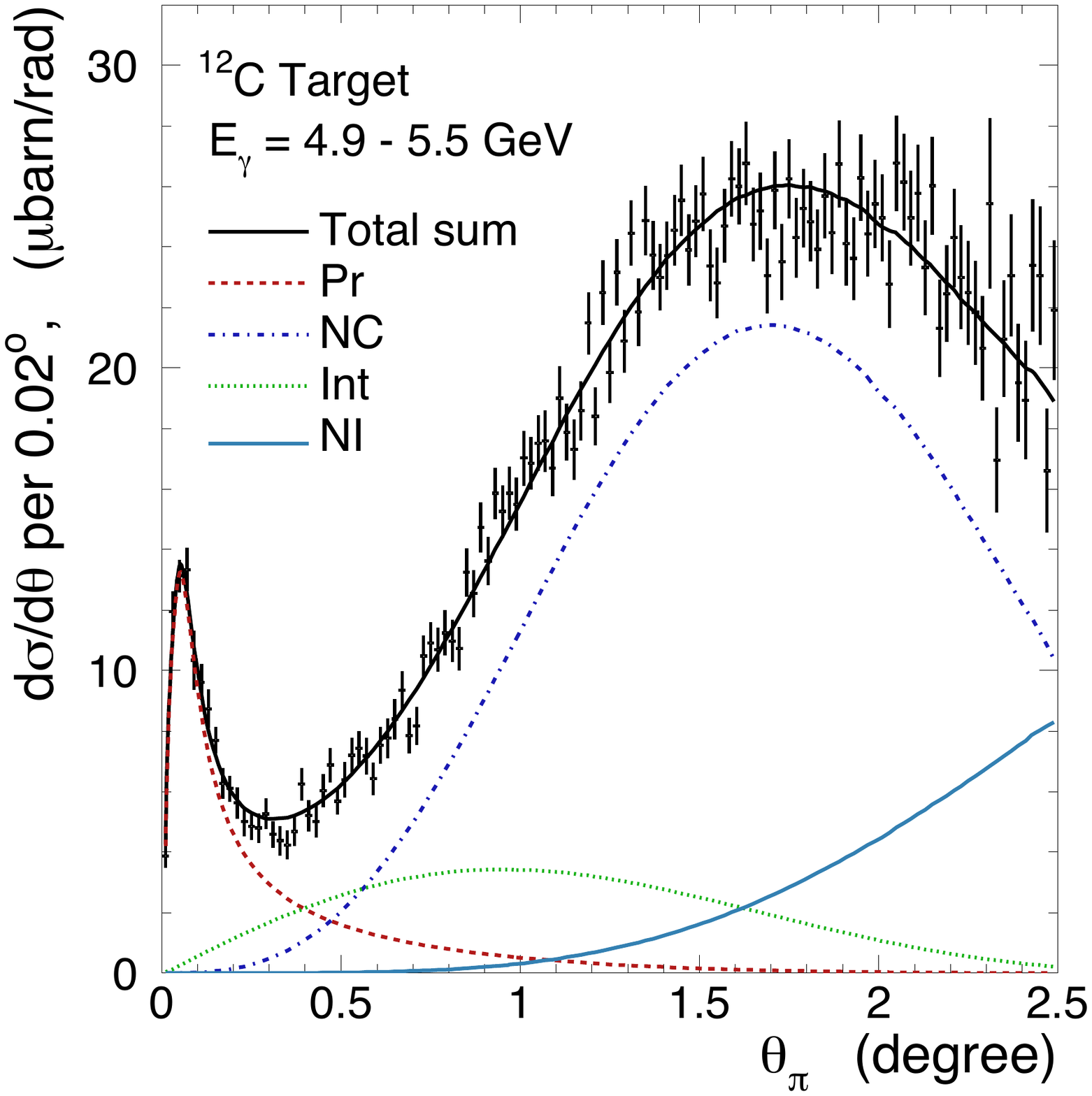}
\includegraphics[height=0.7\textwidth,width=0.45\textwidth] {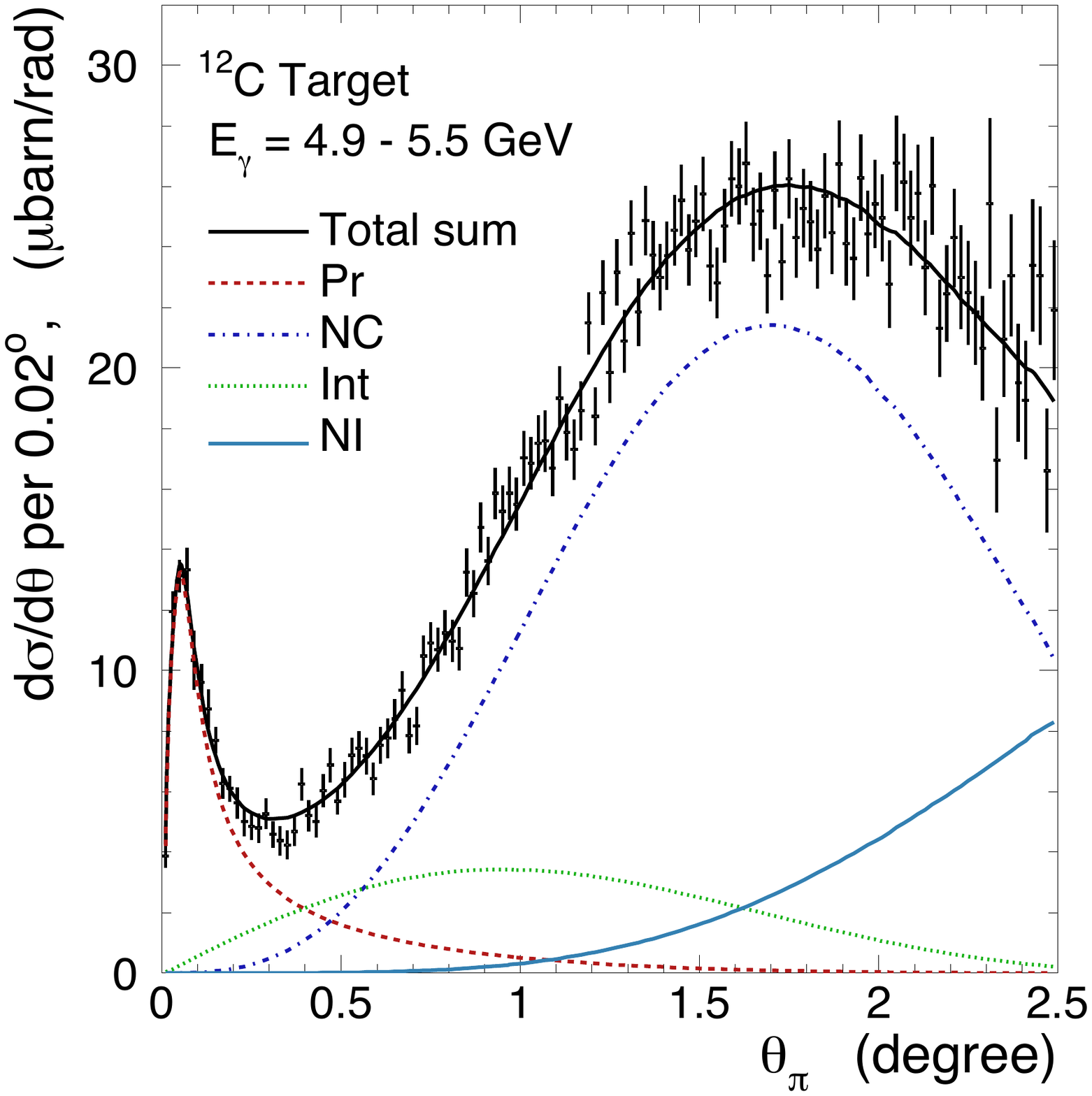}
\end{center}
\caption{Cross sections $d \sigma/d \theta$ in $\mu$barn/rad versus the lab pion angle for C (left)and Pb(right) at an average photon energy of 5.2 GeV:from \cite{Larin:2010}.
The individual contributions were obtained by a fit to the data (see text for discussion). The Primakoff contribution(Pr) peaks $\simeq 0.02^{0}$ , the strong nuclear coherent (NC) contribution(red dotted curve) peaks $\simeq1.6^{0}$  in C and  $\simeq 0.8^{0}$  in Pb with a smaller secondary maximum$\simeq1.8^{0}$. The interference (Int) contribution (green dotted curve) peaks $\simeq  0.9^{0},0.3^{0}$  in C and Pb. The nuclear incoherent (NI)  contribution(solid blue curve) rises gently with $\theta_{\pi^{0}}$.  }
\label{fig:dsig-PrimEx}
\end{figure}

The cross sections for forward angle $\pi^{0}$ photoproduction were measured on C and Pb targets with 4.9-5.5 GeV tagged photons (having an average energy of 5.2 GeV).  The resulting experimental cross sections for C and Pb are shown in Fig. \ref{fig:dsig-PrimEx}.  The data were fit by varying the magnitude of each of the four contributions of  Eq. \ref{eq:diff_cross}---Primakoff, strong, interference, and incoherent cross sections.  This was done by varying the four parameters $\Gamma_{\gamma \gamma}, C_{S}, C_{inc},\phi$ [see ref. \cite{Larin:2010} for their values].  The resulting cross section fits are also shown in Fig. \ref{fig:dsig-PrimEx}.  It can be seen that the large forward peak $\simeq 0.02^{0}$ is dominated by the Primakoff effect, which allows the value of $\Gamma_{\gamma \gamma}$ to be accurately extracted.  An important test of the dominance of the Primakoff mechanism in the small angle region is that the magnitude of this peak scales with the nuclear charge  as $Z^{2}$.  Both the predicted position of the Primakoff peak and its separation from the strong $\pi^{0}$ production peak are essential in the interpretation of the data. The fact that the theoretical cross sections are in such good agreement with the data provides confidence that this separation has been done accurately.

If there were no final state interaction the strong (nuclear coherent) peak would scale in cross section as $A^{2}$ ($A$ = atomic number)  and as $A^{2/3}$ when the mean free path of the outgoing pion is significantly smaller than the nuclear radius. In the PrimEx experiment it scales closer to the latter case ($\simeq A^{0.9}$) and makes the relative magnitudes of the strong relative to the Primakoff peaks smaller in Pb than in C.  The angle for which the the strong cross section peaks is smaller in Pb than in C, due to its larger radius (which increases $\simeq A^{1/3}$). The strong-Primakoff interference cross section is the only nuclear contribution near the Primakoff peak ($\simeq 0.02^{0}$) and enters at the few \% level.  Its strength reflects the positions and magnitudes of the strong and Primakoff cross sections.  Thus the values of the radiative width $\Gamma_{\gamma \gamma}$ obtained from the  C and Pb data pose a stringent test on the model-dependence of the result.  Consistent results for $\Gamma_{\gamma \gamma}$ were obtained from the data for these two targets, supporting the idea that the small (few \%) nuclear effects being subtracted under the Primakoff peak are well described by the theoretical calculations whose magnitudes are fit to the larger angle data.

The largest two systematic errors of the experiment are 1.0\% for the photon flux and 1.6\% for the yield extraction. The uncertainty in the flux is caused by instabilities in the photon beam and detection.  The error in the yield extraction is primarily due to uncertainties in the background subtraction, as illustrated in Fig. \ref{fig:obs}.  The total error of 2.1\% was obtained by summing all of the errors in quadrature, since there are no known correlations between them.  As stated above, this is the smallest systematic error for a photon-induced cross section that we are aware of.  The ability to accurately measure cross sections with this apparatus was tested by measurements of the Compton effect and pair production, which are accurately predicted. The measurements agree with theory to $\leq$ 2\%, which is better than the systematic error for the Primakoff effect.

\begin{figure}
\begin{center}
\includegraphics[height=1\textwidth,width=1\textwidth] {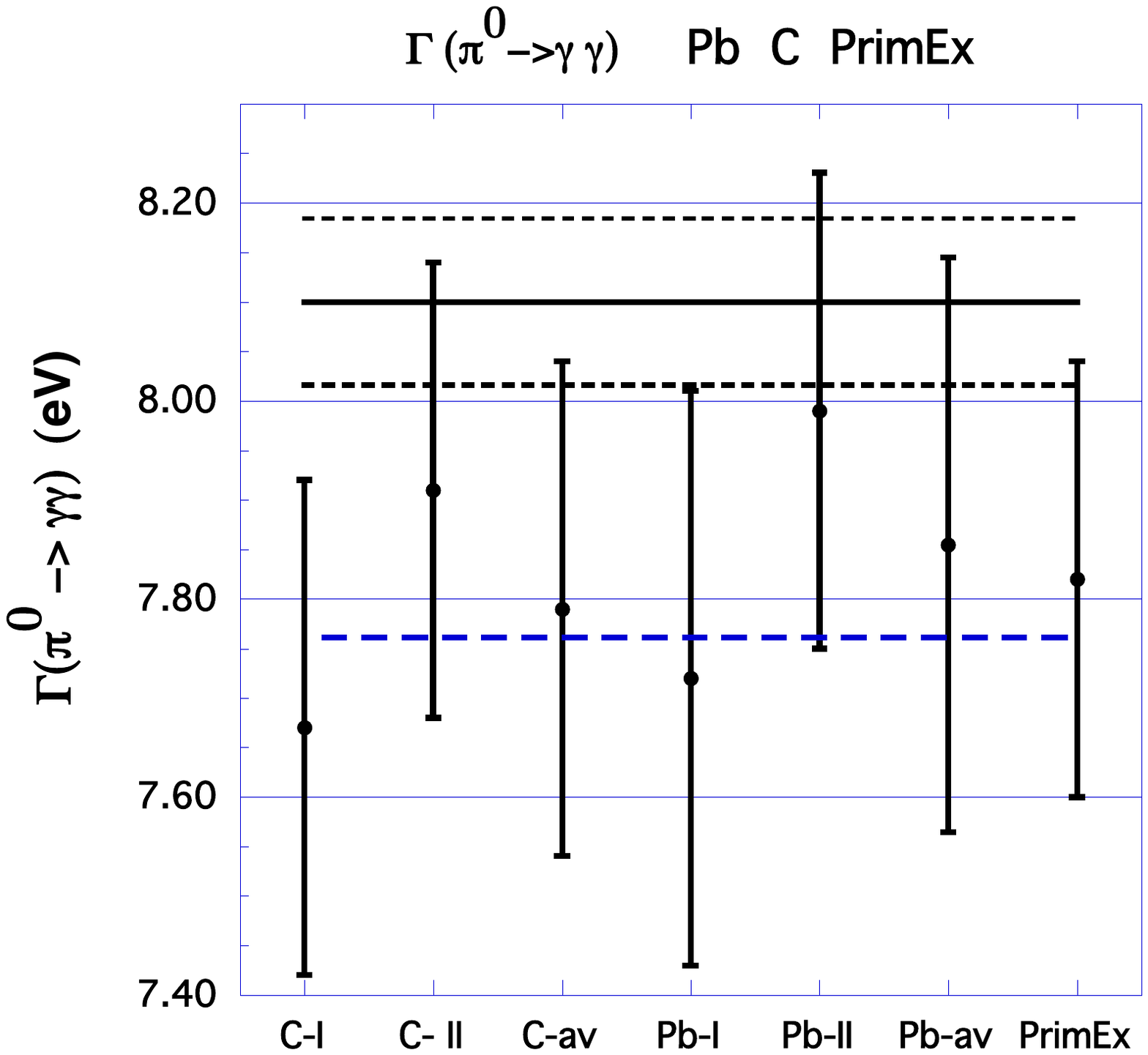}
\end{center}
\caption{The $\Gamma(\pi^{0} \rightarrow \gamma \gamma)$ widths measured by the PrimEx experiment\cite{Larin:2010} for the C and Pb targets, along with the chiral predictions\cite{bgh,kam,anm}. The errors are   statistical and systematic added in quadrature. The  averages for C and Pb are shown separately as well as the average for both targets. Since the errors for both analyses are approximately the same the numerical and weighted averages are almost equal. The lower horizontal line is  the prediction of the LO chiral anomaly\cite{bej,sdl} ( $\Gamma(\pi^{0} \rightarrow \gamma \gamma) = 7.760\,\,{\rm eV}, \tau(\pi^{0}) = 0.838\times 10^{-16}$ s). The upper horizontal lines  are  the HO chiral predictions\cite{bgh,kam,anm}($\Gamma(\pi^{0} \rightarrow \gamma \gamma)= 8.1\,\,{\rm eV}, (\tau(\pi^{0})= 0.80\times 10^{-16}$ s with its estimated  1\% error.)  }
\label{Gamma-C-Pb}
\end{figure}

The PrimEx data were independently analyzed by several groups in the collaboration. This included the event selection and independent development of software, apart from sharing the same photon flux routines which were independently checked by measurements of the pair production and Compton cross sections. The results for the $^{12}$C and $^{208}$Pb targets are shown in Fig. \ref{Gamma-C-Pb}. It is gratifying that these two independent analyses agree with each other within the errors. Perhaps even more important is the fact that the width extracted from the C and Pb targets agree within errors. This verifies the $Z^{2}$ dependence of the Primakoff cross section at the few \% level and is a strong indication that all of the nuclear effects have been properly taken into account. The PrimEx result  is
$\Gamma(\pi^{0} \rightarrow \gamma \gamma) = 7.82 \pm 0.18 \pm 0.22$ eV, where the first error is statistical and the second is systematic. Combining them in quadrature gives a total error of 2.8\%. This result for $
\Gamma_{\gamma \gamma}$ is within one standard deviation of the theoretical prediction\cite{bgh,kam,anm} and most of the results of previous measurements\cite{PDB} as shown in  Fig. \ref{fig:width}.  A summary of the present status of the $\pi^{0}$ lifetime is presented in Sec.\ref{Subsection:Summary}.

The Primakoff experiment represents a modern effort to reduce the experimental error in $\Gamma(\pi^{0} \rightarrow \gamma\gamma)$ to the 1\% level which is the present theoretical accuracy. Because the first experiment did not succeed in obtaining its precision goal of 1\%, the PrimEx collaboration proposed and executed a follow-up experiment in the fall of 2010.  Its goal (as stated in the proposal) is to reduce the error to 1.4\%, with the primary source in the systematic error being the photon intensity calibration.  Targets of $^{28}$Si and $^{12}$C were employed and the statistics were significantly improved relative to the original PrimEx run.  The data analysis is presently in progress.

%% file: section_End.tex
\section{Related Measurements}\label{Section: Related-Measurements}
There are two related extensions of the physics of the $\pi^{0} \rightarrow \gamma \gamma$ rate.  One is to consider the decay rates of the $\eta$ and $\eta^{'}$ pseudoscalar mesons. The other is to consider when one or two of the photons are off shell, i.e. at a finite value of the four momentum transfer carried by one of the two photons. This could be  either space-like ($q^{2} < 0$) or time-like($q^{2} > 0$). The space-like transitions  are accessed by
the  $e^{+}e^{-} \rightarrow e^{+}e^{-}\gamma^{*}\gamma^{*}\rightarrow e^{+}e^{-}P$  reactions ($P= \pi, \eta,\eta^{'}$) when  one of the  leptons in the final state are detected at non-zero  angles.  Such an experiment has been carried out at DESY for $\pi^{0},\eta,\eta^{'}$ production by the CELLO collaboration for $q^{2}$ values from approximately -0.3 to -3 GeV$^{2}$\cite{Behrend:1990}. The results were compared to the dipole form factor $F(q^2) = 1/(1-q^2/\Lambda^2)$ where $\Lambda$ is fit to the data. To first approximation the three form factors could be fit with $\Lambda \simeq m_{\rho}$ (the $\rho$ meson mass), although for the best fits $\Lambda$ did differ for each meson. These differences can be explained by ChPT calculations\cite{Ametller:1992}.  The results for the transition radius of the $\pi^{0}$ is 0.65$\pm$ 0.03 fm \cite{Behrend:1990}, close to the RMS radius of the charged pion.
These results for the transition radius are  model dependent since they are performed at relatively high values of $q^{2}$. A new measurement for the $\pi^{0}$ transition form factor at low $q^{2}$ values has been proposed at Frascati\cite{Babusci:2011} with the KLOE-2 $e^{+}e^{-}$ colliding beam detector. It is also possible to probe the low $q^{2}$ region by studying virtual Primakoff meson production $ e A \rightarrow e^{'} \pi^{0} A$\cite{Hadjimichael:1989}.

For the time-like region low  $q^{2}$ measurements have been performed by observing the $\pi^{0} \rightarrow \gamma e^{+}e^{-}$ reaction using neutral pions produced in the $\pi^{-}p \rightarrow \pi^{0}p$
reaction with stopped` pions\cite{Farzanpay:1992, Drees:1992}. By expanding the transition form factor at low $q^{2}$ these results are consistent with those measured in the space-like region, but with larger errors. The radiative corrections to this process have been worked out in detail in anticipation of more accurate measurements\cite{Kampf:2006}.
\newpage
At the present time the measurements of the $\Gamma (\eta, \eta^{'} \rightarrow \gamma \gamma)$ rates have  significantly larger errors than those for the $\pi^{0}$\cite{PDB}. These rates have come from $e^{+}e^{-}$ experiments. For the $\eta$ meson the Particle Data Book lists four measurements carried out between 1985 and 1990. They are all in good agreement and the resulting average is $\Gamma(\eta \rightarrow \gamma \gamma) = 0.510 \pm 0.026(5.1\%)$ keV\cite{PDB}. For the $\eta^{'}$ meson the Particle Data Book lists eight measurements carried out between 1988 and 1998. They are also in good agreement and the resulting fit is $\Gamma(\eta^{'} \rightarrow \gamma \gamma) = 4.30 \pm 0.15(3.5\%)$ keV\cite{PDB}. \footnote{As discussed in Sec. VII.A the averaging method used by the particle data group give suspiciously small errors. For $\Gamma(\eta^{'} \rightarrow \gamma \gamma)$ there are eight measurements, all using $e^{+}e^{-}$ collisions with total uncertainties which range from 7 to 27\%.  The results are in reasonable agreement with a $\chi^{2}/DOF \simeq 1$.  The overall Particle Data Book average has a 4.4\% error. This is an example of how the estimated error in the average can is reduced when there are a large number of experiments. The most accurate experiment gives $\Gamma (\eta^{'} \rightarrow \gamma \gamma) = 4.17 \pm 0.10 (2.4\%) \pm 0.27(6.5\%) $ where the first error is statistical and the second is systematic (the dominant error). It is difficult to see how the error in the average can be significantly smaller than that of the systematic error of the most accurate experiment. The problem is that it is assumed that there are no correlations between the systematic and statistical errors so that they can be added in quadrature and averaged as if they were  statistical. This  is strictly valid for the situation when the errors are statistical.  For N measurements of approximately the same accuracy the resulting error decreases as  $1/\sqrt{N}$. This is not appropriate when there are significant systematic errors. }

There has only been one Primakoff measurement for the $\eta$. This has resulted in a significantly smaller value of  $\Gamma(\eta \rightarrow \gamma \gamma) = 0.324 \pm 0.046(14.2\%)$ keV \cite{Browman:1974a} and it was not used in the Particle Data Book fit\cite{PDB}. This experiment was performed at Cornell with  bremsstrahlung beams with end points between 5.8 and 11.5 GeV with five targets between Be and U. It is of interest  that this was the same group that performed such an accurate measurement for the $\pi^{0}$ lifetime\cite{Browman:1974}. The reason that Primakoff measurements of heavier mesons is much more difficult than for the $\pi^{0}$ can be understood on the basis of the Primakoff effect discussed in Subsection \ref{subsection:Primakoff}. From Eq. \ref{eq:diff_cross} it can be seen that the differential cross section for the Primakoff effect for the $\eta$ meson is proportional to $k^{4} Z^{2}\Gamma (\eta  \rightarrow  \gamma \gamma)/m_{\eta}^{7} $. If we note that the width $\Gamma (\eta  \rightarrow  \gamma \gamma) \propto m_{\eta}^{3} $ then the Primakoff cross section decreases $\propto 1/m_{\eta}^{4}$, so that it is considerably smaller than for the $\pi^{0}$. In addition the Primakoff peak position $\theta_{max} = m_{\eta}^{2} /(2 k^{2})$, which is considerably larger than for the $\pi^{0}$. The combination of these two effects means that the nuclear interference is considerably more of a problem for the $\eta$ than for the $\pi^{0}$. This can be seen by looking at the figures for the yields for $\eta$ production\cite{Browman:1974a}. In a recent publication the nuclear incoherent effects have been re-evaluated and the conclusion is that the width measured by the Cornell group was changed to  $\Gamma(\eta \rightarrow \gamma \gamma) = 0.476 \pm 0.062(13.0\%)$ keV\cite{Rodrigues:2008}, in reasonable agreement with the PDB fit\cite{PDB} based on $e^{+}e^{-}$ two photon experiments. Although this reanalysis is a significant improvement of the treatment of the incoherent nuclear $\eta$ production, it was not meant to replace a new experiment. The PrimEx collaboration has an approved experiment to measure the Primakoff $\eta$ production cross section from the proton with the upgraded 12 GeV JLab facility\cite{Gasparian:2010}, with the goal of reaching 2.2\% accuracy.

Improving the precision of the $\eta$ and $\eta^{'}$ two photon decay rates is important for several reasons. It is important in determining the $\eta,\eta^{'}$ mixing which enters into their lifetime calculations. In addition $\eta$ and $\eta^{'}$ mixing with the $\pi^{0}$ is predicted to increase $\Gamma(\pi^{0} \rightarrow \gamma \gamma)$ by 4.5\%\cite{bgh,kam,anm}. It is also important since all of the decay rates are based on the well known branching ratios and on the two gamma decay rates\cite{PDB}. Determination of the isospin breaking $\eta \rightarrow 3 \pi$ decay rate can provide an independent determination of the mass difference of the up and down quarks $m_{d} -m_{u}$. Finally there is the more speculative issue of the nature of the $\eta^{'}$ meson, which has too large a mass to be a Nambu-Goldstone Boson, but is so in the large $N_{c}$ limit. The question of its gluonic content has been a long standing issue. Finally we note that the masses and mixing of the $\eta,\eta^{'}$ mesons have been recently calculated on the lattice\cite{Christ:2010}.

\section{Summary and Outlook}\label{Section: Summary and Outlook}
\subsection{Summary}\label{Subsection:Summary}
\begin{figure}
\begin{center}
\includegraphics[height=1\textwidth,width=1\textwidth] {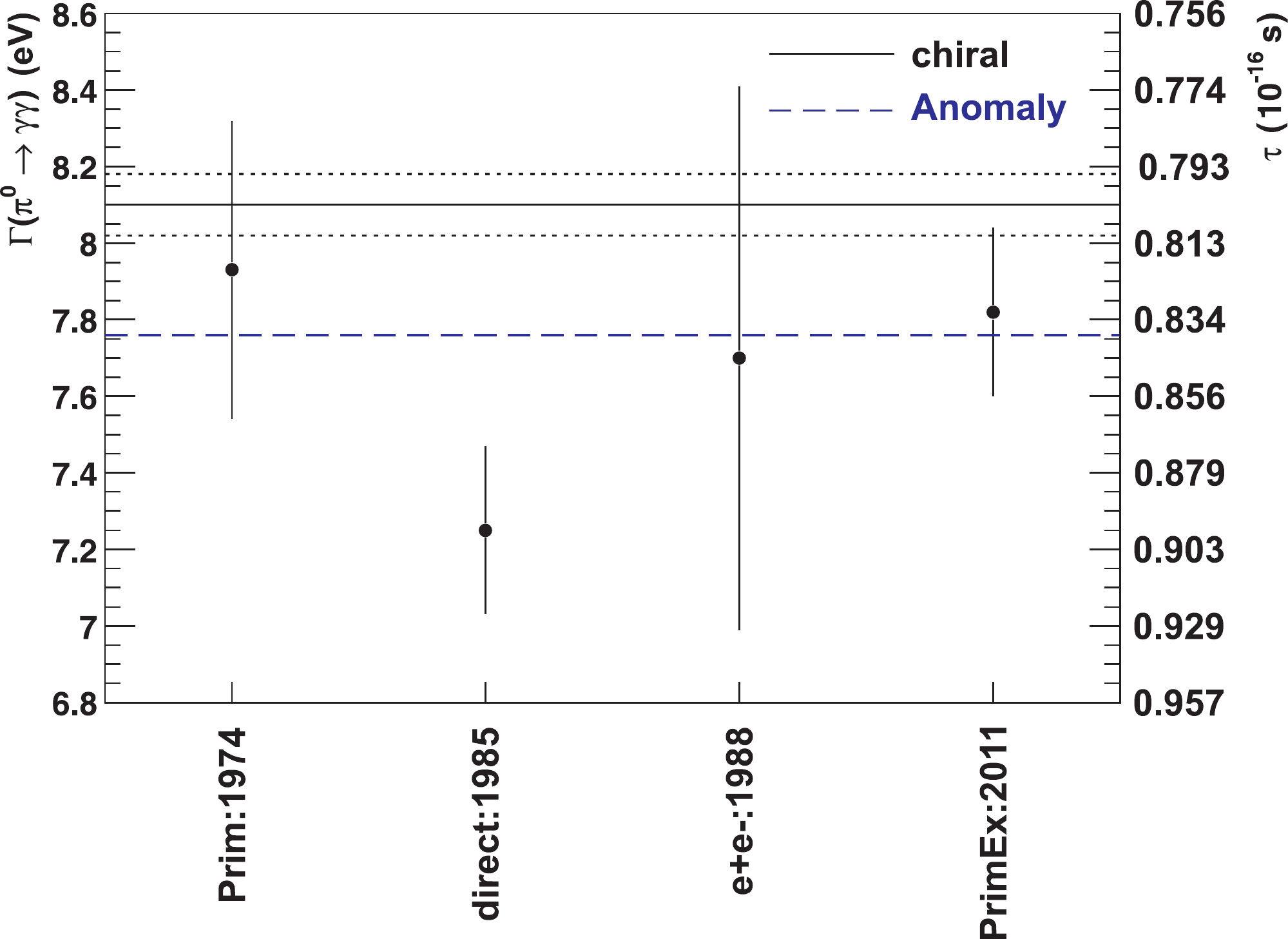}`
\end{center}
\caption{$\pi^{0} \rightarrow \gamma \gamma$ decay width
in eV (left scale) and  $\tau(\pi^{0})$, the mean $\pi^{0}$ lifetime in units of $10^{-16}$ s.(right scale) measured by the most accurate experiments. These include the 1974 Cornell Primakoff measurement\cite{Browman:1974}, the 1985 direct measurement at CERN\cite{Atherton:1985}, the 1988 $e^{+}e^{-}$ experiment at DESY\cite{Williams:1988}, the 2011 PrimEx experiment at JLab\cite{Larin:2010}. The lower dashed line is  the LO prediction of the chiral anomaly\cite{bej,sdl} ($\Gamma(\pi^{0} \rightarrow \gamma \gamma) = 7.760\,\,{\rm eV}, \tau(\pi^{0}) = 0.838\times 10^{-16}$ s). The upper solid line is  the HO chiral prediction and the dotted lines show the estimated 1\% error\cite{bgh,kam,anm}($\Gamma(\pi^{0} \rightarrow \gamma \gamma)= 8.10\,\,{\rm eV}, \tau(\pi^{0})= 0.80\times 10^{-16}$ s). For the relationship between $\Gamma(\pi^{0} \rightarrow \gamma \gamma)$ and $\tau(\pi^{0})$ see Eq. \ref{eq:Gamma-tau}.   }
\label{Gamma-pi0-summary}
\end{figure}

In this section we presented an up to date summary of the present experimental status of the $\Gamma(\pi^{0} \rightarrow \gamma \gamma)$ decay width, and this is summarized in Fig. \ref{Gamma-pi0-summary} and Table \ref{table:Gamma-summary}. The four most accurate experiments  include the 1974 Cornell Primakoff measurement\cite{Browman:1974}, the 1985 direct measurement at CERN\cite{Atherton:1985}, the 1988 $e^{+}e^{-}$ experiment at DESY\cite{Williams:1988},  and the 2011 PrimEx experiment at JLab\cite{Larin:2010}.  Following the recommendation that was made in section \ref{sec:PDB}, several of the previous Primakoff measurements which were performed at lower energies, and analyzed with imprecise theoretical models are not included. We also do not use the measurement  of radiative pion decay  $\pi^{+} \rightarrow e^{+}\nu \gamma$\cite{Bychkov:2009} which is related by a simple isospin rotation to the amplitude for $\pi^0\rightarrow\gamma\gamma$ (see section \ref{sec:pi-beta} for further discussion) because of its large (11\%) experimental error. An additional problem is that using this method to determine $\tau(\pi^{0})$ requires an isospin breaking correction which has not been made. This is of particular  importance since the chiral corrections\cite{bgh,kam,anm} to the anomaly prediction are isospin breaking. 

As we reach a new level of precision for the $\pi^{0}$ lifetime measurement the situation is not completely satisfactory. The two most accurate experiments are not in agreement. The direct experiment\cite{Atherton:1985} differs from the PrimEx result\cite{Larin:2010}  by 0.57 eV (7.5\%) which is 2.5 $\sigma$ (or 1.8 times their combined sigmas). Although this does not reach the level of a serious disagreement, this clearly needs further investigation.
In this circumstance, in our opinion,  it is difficult to make an accurate  average. Following the procedure used by the Particle Data Group\cite{PDB} one obtains an average value of $<\Gamma(\pi^{0} \rightarrow \gamma \gamma)> = 7.60 \pm 0.14 \,(1.9\%) $ eV.  This procedure, however, makes the questionable assumption that the statistical and systematic errors(added in quadrature) can be reduced by combining experiments. In this case we obtain an error of 1.9\% in a situation where  the two most accurate experiments have $\simeq$ 3.0\% total errors and disagree by 7.5\%. The RMS error is 0.30 eV (4.0\%) which is more reasonable.( See the discussion of $\Gamma (\eta, \eta^{'} \rightarrow \gamma \gamma)$ in Sec. VI.)\footnote{Note added in proof. The 2012 particle data book average, which includes the PrimEx experiment and has followed our suggestions about which of the older Primakoff experiments not to use (Sec. IV), gives an average value of $\Gamma(\pi^0\to\gamma\gamma) =7.63 \pm 2.1\%eV  (\tau(\pi^{0}) = (0.852 \pm 2.1\% )\times 10^{-16}\,{\rm s.}$) \cite{Beringer:2012}. The small difference with our average is because the PDG includes the results of the $\pi^{+} \rightarrow e^{+}\nu \gamma$ reaction\cite{Bychkov:2009}. }

The weighted error is very sensitive to the values of the published  errors since the weight assigned to each measurement =$ 1/\sigma^{2}$. As was discussed in section \ref{subsection:direct} we are concerned about the fact that the CERN direct measurement\cite{Atherton:1985} did not measure the $\pi^{0}$ momentum spectrum and therefore may have underquoted their errors. As an exercise if we increase their error by $\sqrt{2}$ so that the weight given to that experiment is reduced by a factor of two, the weighted average increases by 0.09 eV (1.2\%). We do not advocate doing this since a great deal of careful work went into this result\cite{Atherton:1985,Millikan:1985}.  We only want to show the sensitivity of the weighted average to the individual errors. We do believe that further experimental work should be done.

In summary three of the four experiments are in agreement with both the LO axial anomaly and the HO chiral predictions. They are clearly not of  sufficient accuracy to demonstrate the 4.5\% increase predicted in the width by the HO chiral corrections and to fully test them.

\begin{table}
\caption{Most Accurate Measurements of $\Gamma(\pi^{0} \rightarrow \gamma \gamma)$ and $\tau(\pi^{0}$).
For their relationship  see Eq. \ref{eq:Gamma-tau}.}
\label{table:summary}
\begin{center}
{\small
\begin{tabular}{||c|c|c|c||}\hline\hline
 & & &  \\
Reaction&Reference& $\Gamma(\pi^{0} \rightarrow \gamma \gamma)$ eV   & $\tau(\pi^{0})/10^{-16}$ s  \\
 & & &   \\ \hline
 & &  &  \\
Primakoff:1974&\cite{Browman:1974} & 7.92(0.44) & 0.821(0.044)  \\
 & & &  \\ \hline
direct:1985 & \cite{Atherton:1985}  & 7.25(0.22) & 0.897(0.028)\\
 & & &  \\ \hline
$e^{+}e^{-}$:1988 &\cite{Williams:1988}   & 7.70(0.71)& 0.845(0.078)\\
 & & &   \\ \hline
Primakoff:2011&  \cite{Larin:2010} & 7.82(0.22)& 0.832(0.023)\\
    & & &\\ \hline\hline
\end{tabular} }
\end{center}
\label{table:Gamma-summary}
\end{table}

\subsection{Outlook}\label{Subsection:Outlook}
The $\pi^{0} \rightarrow \gamma \gamma$ decay is perhaps the best example of a process that proceeds primarily via the chiral anomaly\cite{bej,sdl}, which is an essential component of QCD.  The possibility of making a precise measurement exists due to spontaneous chiral symmetry breaking, which makes the $\pi^{0}$ the lightest hadron, and consequently its  primary decay mode is into two gamma rays.  The chiral anomaly represents breaking by the electromagnetic field of the symmetry associated with the third component of the axial current\cite{bej,sdl}. The $\pi^{0}$ decay provides the most sensitive test of this phenomenon of symmetry breaking due to the quantum fluctuations of the quark fields in the presence of a gauge field.

The LO (chiral anomaly) prediction  for the $\pi^{0}$ lifetime has no adjustable parameters and is exact in the chiral limit---$(m_{u}, m_{d}, m_{\pi})\rightarrow 0$.  The HO (chiral) corrections involve isospin breaking and mix the $\eta, \eta^{'}$ mesons into the $\pi^{0}$; consequently, such corrections are proportional to $ m_{u}- m_{d}$ and  predict a $ (4.5 \pm 1.0)$\% increase to the $\pi^{0}$ decay rate\cite{bgh,kam,anm} relative to the LO anomaly calculation. The theoretical error arises from uncertainties in the low energy constants and is not due to higher orders in the chiral expansion. Therefore it is unlikely that the present theoretical error can be significantly reduced using chiral perturbation theory.  On the theoretical side, further progress probably requires lattice calculations. A first lattice calculation of $\eta, \eta^{'}$ mixing has recently been reported\cite{Christ:2010}.

As has been stressed in this review, we believe that the fundamental nature of the $\pi^{0}$ lifetime which has been calculated to HO in QCD to an accuracy of 1\% inspires a corresponding experimental effort to measure it to this precision.
The theory {\it is} consistent with experiment at the present level of accuracy. However, at the present time theory is ahead of experiment in that the estimated theoretical error of only 1\% in the chiral calculations\cite{bgh,kam,anm} is significantly smaller than the experimental uncertainty.

Considering the fundamental nature of the subject,
it is very important to have modern experiments performed with all three techniques at the 1\% level.  The PrimEx experiment has a quoted accuracy of 2.8\% \cite{Larin:2010} and the direct experiment at CERN 3.1\%\cite{Atherton:1985}.  However, the difference between the central values of these results is 7.5\% . This discrepancy clearly needs to be resolved if further progress is to be made. The PrimEx group has had a second run at JLab using $^{12}$C and $^{28}$Si targets with improved systematics. Using the analysis techniques developed for the experiment the goal is to reduce the error to $\leq$ 2\%. We also know of plans to remeasure the direct experiment with the Compass experiment at CERN\cite{Friedrich:2011}.  Finally, there exist plans to remeasure the $\pi^{0}$ lifetime at Frascati\cite{Babusci:2011} and at Belle\cite{Denig:2011}. This latter effort can also measure the rates for the $\eta$ and $\eta^{'}$ as well as the $q^{2}$ dependence associated with Dalitz decay.  We look forward to future developments.

\begin{center}
{\bf Acknowledgements}
\end{center}

The work of AMB work was supported in part by the U.S. Department of Energy and that of BRH by the National Science Foundation.  For discussions of the chiral anomaly, the loop corrections, and the chiral corrections we would like to thank J. Bijnens, J. Donoghue, J. Gasser, J. Goity, R. Jackiw, H. Leutwyler, and U. Mei{\ss}ner.  For discussions about the direct measurement we would like to thank J. Cronin. For their help with the figures we would like to thank D. McNulty and S. Schumann. For their comments on the manuscript we thank  V.R. Brown, R. Jackiw, H. Leutwyler, R. Miskimen, and L. Rosenson.  AMB would like to thank his PrimEx colleagues  for their collaboration. 